\documentclass{aa}

\usepackage{booktabs}
\usepackage{subcaption}
\usepackage{xcolor}
\usepackage{graphicx}
\usepackage[export]{adjustbox}
\usepackage{diagbox}
\usepackage{slashbox}
\usepackage{adjustbox}
\usepackage{multirow}
\usepackage{makecell}
\usepackage[normalem]{ulem}
\usepackage[varg]{txfonts}
\usepackage[stable]{footmisc}
\usepackage{url}
\usepackage{hyperref}
\newcolumntype{M}[1]{>{\centering\arraybackslash}m{#1}}

\begin{document} 

    \title{Photodissociation of aliphatic PAH derivatives under relevant astrophysical conditions\thanks{The data set associated with this work can be found under zenodo (DOI: 10.5281/zenodo.4792255).}}
   \author{A. Marciniak\inst{1} \and
           C. Joblin\inst{1}\fnmsep\thanks{Corresponding author.} \and
           G. Mulas\inst{1,2} \and
           V. Rao Mundlapati\inst{1} \and
           A. Bonnamy\inst{1} 
     }

   \institute{Institut de Recherche en Astrophysique et Plan\'{e}tologie (IRAP),                 Universit\'{e} de Toulouse (UPS), CNRS, CNES, 9 Avenue du Colonel Roche,            F-31028 Toulouse, France \\ 
             \email{christine.joblin@irap.omp.eu}
        \and
             Istituto Nazionale di Astrofisica -- Osservatorio Astronomico di Cagliari, Via della Scienza 5, I-09047 Selargius (CA), Italy
         \\
             }

   \date{Received 00 Month 2021 / Accepted 00 Month 2021}

% \abstract{}{}{}{}{} 
% 5 {} token are mandatory
 
\abstract
% context heading (optional), leave it empty if necessary 
{The interaction of polycyclic aromatic hydrocarbons (PAHs) with vacuum ultraviolet (VUV) photons triggers the emission of the well-known aromatic infrared bands (AIBs), but other mechanisms, such as fragmentation, can be involved in this interaction. Fragmentation leads to selection effects that favor specific sizes and structures. }
% aims heading (mandatory)
{Our aim is to investigate the impact of aliphatic bonds on the VUV photostability of PAH cations in a cryogenic and collisionless environment with conditions applicable for photodissociation regions (PDRs).}
% methods heading (mandatory)
{The studied species are derived from pyrene (C$_{16}$H$_{10}$) and coronene (C$_{24}$H$_{12}$) and contain aliphatic bonds either in the form of methyl or ethyl sidegroups or of superhydrogenation. Their cations are produced by laser desorption ionization and isolated in the cryogenic ion cell of the PIRENEA setup, where they are submitted to VUV photons of 10.5\,eV energy over long timescales ($\sim$1000\,s). The parent and fragment ions are mass-analyzed and their relative intensities are recorded as a function of the irradiation time. The fragmentation cascades are analyzed with a simple kinetics model from which we identify fragmentation pathways and derive fragmentation rates and  branching ratios for both the parents and their main fragments.}
% results heading (mandatory)
{Aliphatic PAH derivatives are found to have a higher fragmentation rate and a higher carbon to hydrogen loss compared to regular PAHs. On the other hand, the fragmentation of PAHs with alkylated sidegroups forms species with peripheral pentagonal cycles, which can be as stable as, or even more stable than, the bare PAH cations. This stability is quantified for the main ions involved in the fragmentation cascades by the comparison of the fragmentation rates with the photoabsorption rates derived from theoretical photoabsorption cross sections. The most stable species for which there is an effective competition of fragmentation with isomerization and radiative cooling are identified, providing clues on the structures favored in PDRs. }
% conclusions heading (optional), leave it empty if necessary 
{This work supports a scenario in which the evaporation of nanograins with a mixed aliphatic and aromatic composition followed by VUV photoprocessing results in both the production of the carriers of the 3.4\,$\mu$m AIB by methyl sidegroups and in an abundant source of small hydrocarbons at the border of PDRs. An additional side effect is the efficient formation of stable PAHs that contain some peripheral pentagonal rings. Our experiments also support the role of isomerization processes in PAH photofragmentation, including the H-migration process, which could lead to an additional contribution to the 3.4\,$\mu$m AIB.}


\titlerunning{Photodissociation of aliphatic PAH derivatives under relevant astrophysical conditions}
\authorrunning{A.~Marciniak et al.}
    
   \maketitle
%
%-------------------------------------------------------------------

\section{Introduction}

The interaction of vacuum ultraviolet (VUV) photons with polycyclic aromatic hydrocarbons (PAHs) plays a key role in the evolution of photodissociation regions (PDRs). As part of interstellar dust, PAHs absorb VUV photons and reemit the absorbed energy in the infrared (IR), leading to emission in the aromatic infrared bands (AIBs) \citep{leger1989}. Some of the absorbed photons are also expected to lead to ionization and photodissociation, which chemically alters the astroPAH population. The impact of these processes has been modeled in PDRs associated with star and planet formation \citep{visser2007, montillaud2013,andrews2015}, leading to the conclusion that only large PAHs with a typical carbon number, $N_\mathrm{C}$, of $\gtrsim$50~atoms can survive. These large species therefore appear to be the most likely carriers of the AIB emission. This conclusion also applies to the diffuse interstellar medium despite its much weaker VUV photon flux. This is due to the fact that the gas density is also much lower, which keeps the competition between rehydrogenation by reactivity with H atoms and dehydrogenation by photodissociation at the same level as in the denser bright PDRs \citep{montillaud2013}.

The description of PAH dissociation in chemical models remains simplified and does not include much molecular diversity. The evolution of dissociation rates with size is calculated from the density of vibrational states using statistical models. Only compact PAH structures, whose dissociation pathways involve hydrogen (H, H$_2$) loss, are considered. One of the motivations behind these models is to evaluate the contribution of PAHs to the formation of H$_2$ in PDRs \citep{andrews2016, castellanos2018_h2}. Still, molecular diversity has to be considered, at some point, in these models. This diversity results from formation and destruction pathways, which, in PDRs, are mainly driven by VUV photoprocessing. Studies on the spatial evolution of the AIB spectra in extended PDRs show that free PAHs are produced in the VUV-irradiated cloud layers by the evaporation of very small grains \citep{rapacioli2005, pilleri2012}. \cite{pilleri2015} conclude that these very small grains have a mixed aromatic and aliphatic composition and that PAHs with attached aliphatic sidegroups are produced by their evaporation. In particular, methyl (-CH$_3$) sidegroups are good candidates to account for the 3.4\,$\mu$m emission band, which is a satellite of the 3.3\,$\mu$m AIB attributed to aromatic CH bonds \citep{shan1991,joblin1996, pauzat1999, yang2016}. Methylated PAHs are also abundant in carbonaceous chondrites \citep{basile1984, elsila2005, sabbah2017}. In Murchison, the concentration of methyl-pyrene (CH$_3$-C$_{16}$H$_9$) is found to be lower but on the same order as that of pyrene (C$_{16}$H$_{10}$). The presence of alkylated PAHs provides insights into the underlying formation process, suggesting lower formation temperatures than for regular PAHs \citep{blumer1975} and an efficient alkyl-addition mechanism \citep{gavilan2020, santoro2020}.

Another class of PAHs that contain aliphatic CH bonds, namely superhydrogenated PAHs with H atom excess on the peripheral C atoms (also called hydro-PAHs), could be the carriers of the 3.4\,$\mu$m emission band \citep{bernstein1996,mackie2018}. Evidence for these species, however, remains controversial. From a spectroscopic point of view, these species would better account for the 3.4\,$\mu$m band compared to methyl-substituted PAHs \citep{steglich2013, maltseva2018}. From a stability point of view, though, chemical models conclude that PDRs are too hostile an environment for these superhydrogenated PAHs
\citep{andrews2016,montillaud2013}. In the laboratory, \cite{jochims1999} have performed photoion mass spectrometry experiments to measure the appearance energies of the H-loss fragment for several small methylated PAHs of sizes up to CH$_3$-C$_{14}$H$_9$, as well as a couple of dihydro-PAHs. They conclude that these species have a lower photostability compared to regular PAHs, with dihydro-PAHs being even less stable than methylated PAHs.
This article constitutes the sole results obtained so far on the VUV photoprocessing of PAHs that contain methyl sidegroups. For small dihydro-PAHs, more recent studies have shown the presence of CH$_3$- and C$_3$H$_5$-loss channels in competition with the H-loss channel \citep{west2014_dihydro, diedhiou2020}. These channels differ from the C$_2$H$_2$ loss that can be observed in regular PAHs. Experiments performed on the photodissociation of superhydrogenated pyrene derivatives upon UV-visible multiphoton excitation confirmed the loss of fragments containing odd carbon numbers \citep{wolf2016}. 

The chemical models mentioned above \citep{visser2007, montillaud2013,andrews2015} use unimolecular rates to describe the interaction with VUV photons that includes dissociation and radiative cooling. These unimolecular rates are directly calculated using statistical models, for instance to derive the IR emission rate, or derived with statistical unimolecular dissociation models from the analysis of experimental data, for example breakdown curves from measurements of photoelectrons and photoions in coincidence (PEPICO) compiled by \cite{west2018}. Using these rates, the models then treat the competition between the different processes at play in PDRs. In PEPICO experiments, the typical timescale to observe fragmentation is 0.1\,ms or less. In astrophysical environments, it can be much longer and depends on the competition with the slowest process, namely IR emission, which involves timescales of seconds or more \citep{joblin2020}. By accessing these long timescales in the laboratory, one could therefore aim to derive rates that quantify the kinetics of the fragmentation of a given PAH in a given VUV radiation field. This kinetics of fragmentation is directly governed by the competition between dissociation and radiative cooling.
In this case, the derived rates only depend on the VUV photon flux and can be appropriately scaled in models. This, however, excludes the case of multiple photon absorption, which, although rare in astrophysical conditions, is expected to play a dominant role in the dissociation of large PAHs \citep{montillaud2013}.

We recently coupled the cryogenic PIRENEA setup \citep{joblin2002} with a 10.5\,eV VUV source in order to study the photofragmentation of PAH cations in conditions that are relevant for PDRs. We focus here on cations of aliphatic PAH derivatives, which have been subject to fewer studies relative to standard PAHs. The long experimental timescale achieved in the experiment allows us to study not only the fragmentation kinetics of the parent ions but also that of subsequent generations of fragments, simulating the photoprocessing of these species in PDRs. We derive fragmentation maps and rates from the analysis of the kinetic curves recorded during the fragmentation cascades. For the major involved cations, we compare the fragmentation rates with photoabsorption rates deduced from calculations of the photoabsorption cross sections using time-dependent density-functional theory (TD-DFT). In addition, we identify species with the lowest fragmentation rates, which implies an efficient contribution of other relaxation mechanisms such as radiative cooling and isomerization. For these species, we derive characteristic rates for the radiative cooling. Finally, we draw a budget of the small neutral fragments produced in the photodissociation cascades, with special focus on the carbon- to hydrogen-loss ratio.

The manuscript is organized as follows. We describe the methods in Sect.~\ref{sec:meth}. In Sect.~\ref{sec:res}, we present the results and analyze them thanks to considerations on the molecular parameters involved in VUV photoprocessing. In Sect.~\ref{sec:astro_implications}, we discuss the application to astrophysical conditions and implications for the evolution of aliphatic PAH derivatives in PDRs.
We conclude in Sect.~\ref{sec:con}.

\section{Methods}\label{sec:meth}

\subsection{Studied species}
\begin{figure}
     \centering
     \begin{subfigure}[b]{0.1\textwidth}
         \centering
         \includegraphics[width=\textwidth]{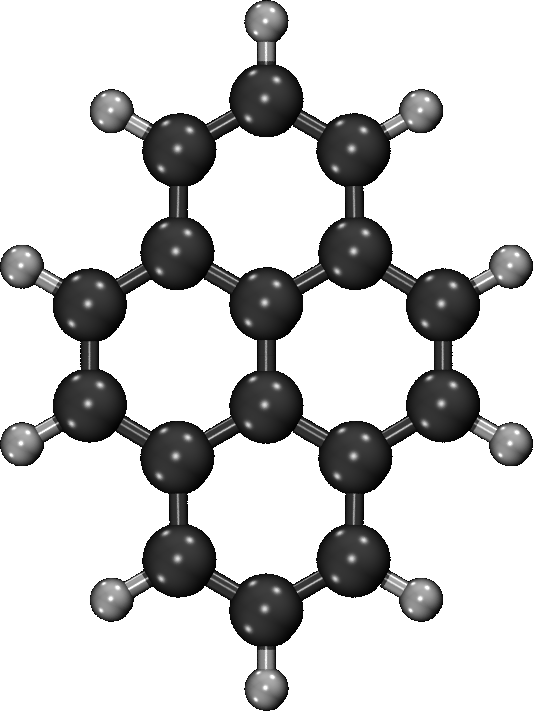} 
         \caption{}
     \end{subfigure}
     \hfill
     \begin{subfigure}[b]{0.08\textwidth}
         \centering
         \includegraphics[width=\textwidth]{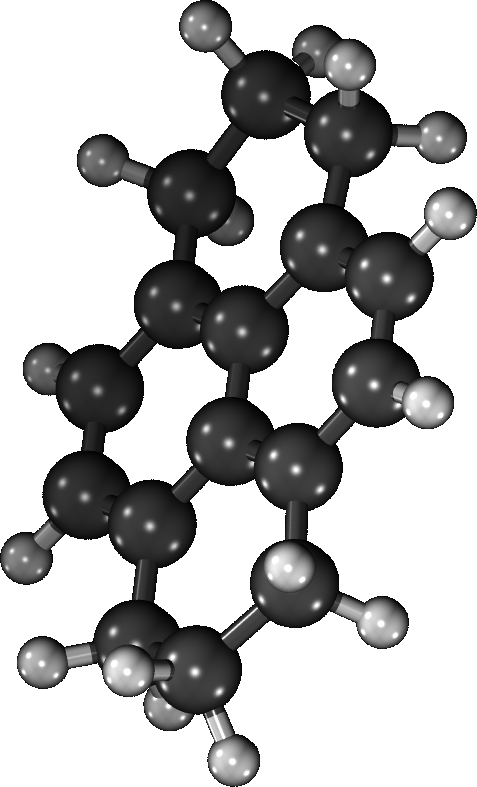}
         \caption{}
     \end{subfigure}
     \hfill
     \begin{subfigure}[b]{0.1\textwidth}
         \centering
         \includegraphics[width=\textwidth]{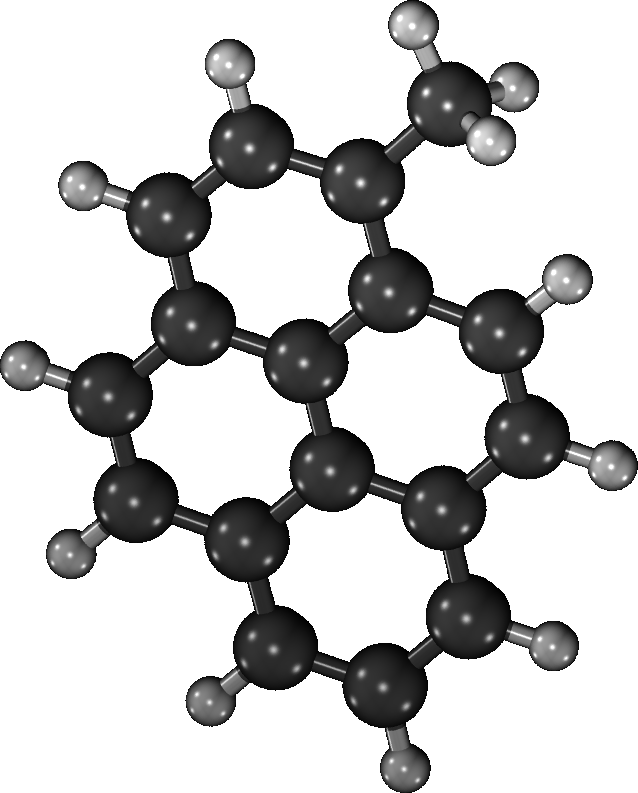}
         \caption{}
     \end{subfigure}
     \hfill
     \begin{subfigure}[b]{0.12\textwidth}
         \centering
         \includegraphics[width=\textwidth]{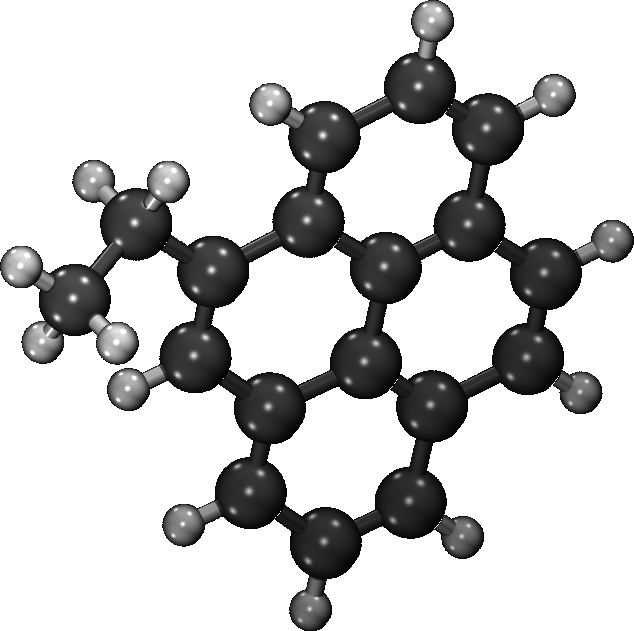}
         \caption{}
     \end{subfigure}
     \hfill
     \begin{subfigure}[b]{0.13\textwidth}
         \centering
         \includegraphics[width=\textwidth]{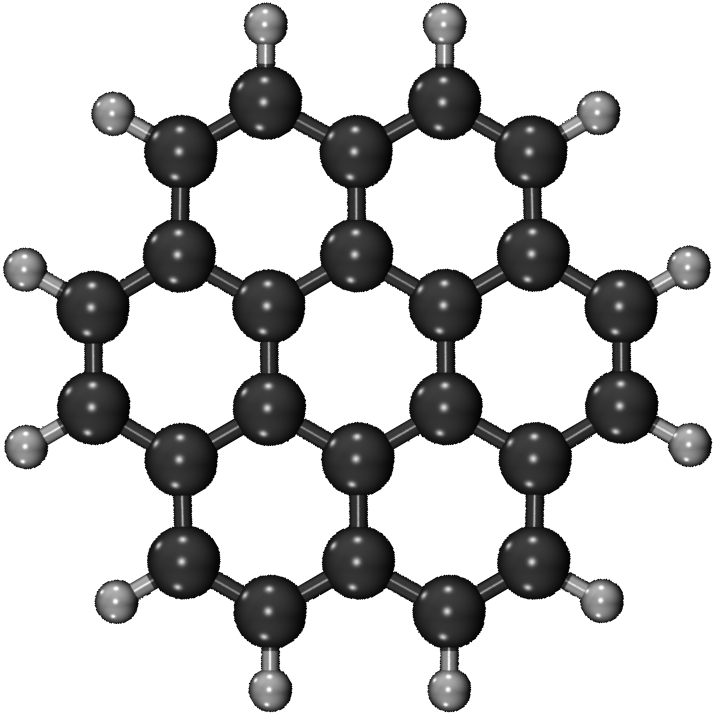}
         \caption{}
     \end{subfigure}
     \hfill
     \begin{subfigure}[b]{0.14\textwidth}
         \centering
         \includegraphics[width=\textwidth]{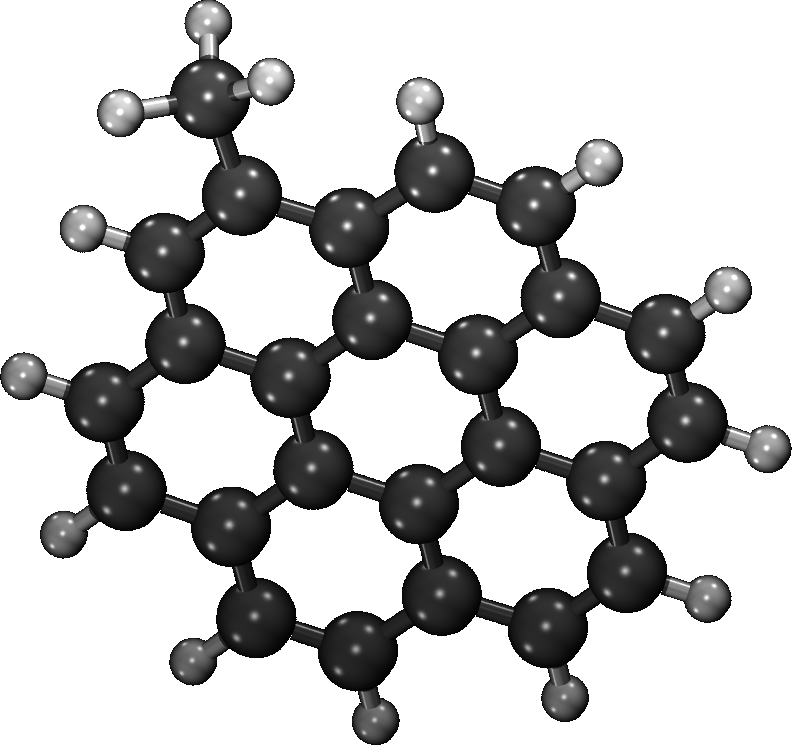}
         \caption{}
     \end{subfigure}
          \hfill
     \begin{subfigure}[b]{0.14\textwidth}
         \centering
         \includegraphics[width=\textwidth]{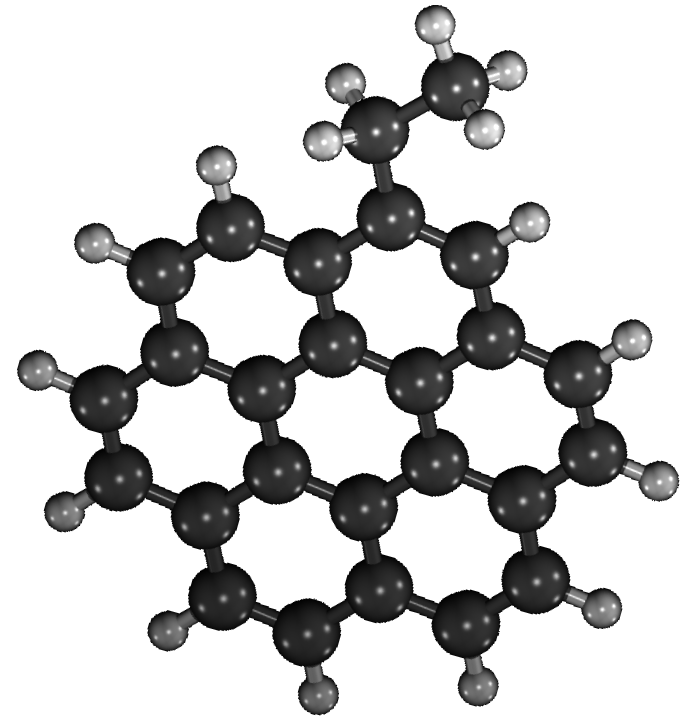}
         \caption{}
     \end{subfigure}
        \caption{Cationic molecular structures, optimized with the \textsc{Gaussian16} package at the B3LYP/6-31g(d,p) level of theory, of (a)  pyrene (C$_{16}$H$_{10}^+$, Pyr$^+$), 1,2,3,6,7,8-hexahydro-pyrene (C$_{16}$H$_{16}^+$, H$_6$-Pyr$^+$), 1-methyl-pyrene (C$_{17}$H$_{12}^+$, MePyr$^+$), 4-ethyl-pyrene (C$_{18}$H$_{14}^+$, EtPyr$^+$), coronene (C$_{24}$H$_{12}^+$, Cor$^+$), methyl-coronene (C$_{25}$H$_{14}^+$, MeCor$^+$), and ethyl-coronene (C$_{26}$H$_{16}^+$, EtCor$^+$).}
        \label{fig:molstruct}
\end{figure}

We studied seven cationic PAHs based on the availability of their neutral precursors in our laboratory: pyrene (C$_{16}$H$_{10}^+$, Pyr$^+$), 1,2,3,6,7,8-hexahydro-pyrene (C$_{16}$H$_{16}^+$, H$_6$-Pyr$^+$), 1-methyl-pyrene (C$_{17}$H$_{12}^+$, MePyr$^+$), 4-ethyl-pyrene (C$_{18}$H$_{14}^+$, EtPyr$^+$), coronene (C$_{24}$H$_{12}^+$, Cor$^+$), methyl-coronene (C$_{25}$H$_{14}^+$, MeCor$^+$), and ethyl-coronene (C$_{26}$H$_{16}^+$, EtCor$^+$). Their structures are depicted in Fig.~\ref{fig:molstruct}. Four PAH samples, namely pyrene, 1-methyl-pyrene, coronene, and hexahydropyrene, are from Sigma-Aldrich ($> 97$\% purity). The others, 4-ethyl-pyrene, methyl-coronene, and ethyl-coronene, were synthetized by E. Clar and kindly provided by L. d’Hendecourt. The IR spectra of the used substituted coronene species were previously published \citep{jourdaindemuizon90, joblin1996}.
In our experiment, the PAH cations are released and ionized in the gas phase using desorption and ionization by an Nd:YAG laser (fourth harmonics, $\lambda = 266$ nm) from an amorphous PAH deposit on an aluminum substrate. The deposit is made by a drop by drop evaporation of a solution containing the PAH into toluene (or ethanol:toluene). 

\subsection{Experimental method}
\begin{figure*}[htbp]
    \centering
    \includegraphics[width=0.9\textwidth]{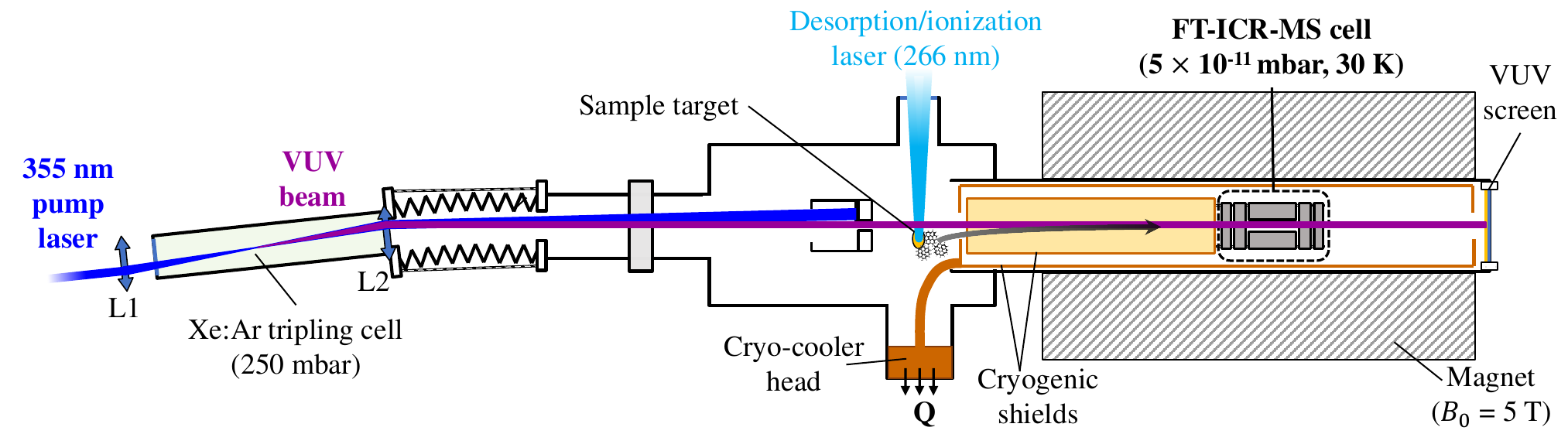} 
    \caption{Scheme of the PIRENEA setup coupled with the VUV laser source. The latter generates the VUV pulses by using a pulsed 355\,nm Nd:YAG laser focused in a Xe:Ar tripling gas cell. The VUV beam pointing stability is monitored on the VUV screen. The cationic species are produced in gas phase by desorption-ionization of a sample target with a 266\,nm laser and they are stored in the cryogenic ion cyclotron resonance cell (ICR). Then, they interact with the VUV beam and the photoproducts are measured via nondestructive Fourier transform ICR mass spectrometry (FT-ICR-MS).}
    \label{fig:setup}
\end{figure*}

The experiment was performed in the cryogenic ion trap PIRENEA, 'Piège à Ion pour la Recherche et l’Étude de Nouvelles Espèces Astrochimiques', which is dedicated to the study of the photophysics and spectroscopy of PAH cations and related species in interstellar conditions \citep[e.g.,][]{joblin2002,useli2010,west2014_dihydro,zhen2016_C78H26}.
The setup has been recently upgraded and its full functionalities will be described in a coming publication. We focus here on the coupling with a table-top VUV pulsed light source as illustrated in the scheme depicted in Fig.~\ref{fig:setup}. The VUV source consists of a cell of a Xenon:Argon gas mixture (in a number ratio 1:11) which allows the frequency tripling of a nanosecond 355\,nm Nd:YAG laser ($h\nu_\mathrm{UV} = 3.5$ eV) into a 118 nm VUV radiation ($h\nu_\mathrm{VUV} = 10.5$\,eV) as shown by \cite{lockyer1997}. At the end of the VUV cell, the generated VUV beam and the remaining 355\,nm beam are split angularly by the edge of an MgF$_2$ lens (L2) used as a prism. The 355\,nm beam is then blocked by an absorber letting only the VUV beam to propagate toward the ion trap. The distance between the L1 and L2 lenses is adjusted in order to collimate the VUV beam on a diameter size of $\sim1.5$\,mm. We used 355\,nm pulses of 10\,mJ energy at a repetition rate of 10\,Hz, which allowed us to generate about $2.5 \times 10^{11}$ VUV photons per second, which corresponds, in our experimental conditions, to a VUV photon flux of $1.4_{-0.4}^{+1.4} \times 10^{13}$\,photon$~$s$^{-1}~$cm$^{-2}$ (see Appendix \ref{app:VUV_source} for details about this estimation).

Upon shutter opening, the VUV beam crosses the cloud of PAH cations
that have been stored in the cryogenic ion trap held at a temperature of 30\,K and at a pressure of $5\times10^{-11}$\,mbar. The intensity of the desorption-ionization laser that is used to produce the cations is adjusted to minimize fragmentation and optimize the quality of the ion signal, which is related to the quality of the ion cloud. Trapping in PIRENEA is achieved using an ion cyclotron resonance (ICR) cell, which allows us to perform nondestructive Fourier transform ICR mass spectrometry (FT-ICR-MS), whose principle is explained in \cite{marshall1998}. 
Briefly, it consists of exciting resonantly the ions in their modified cyclotron frequency and observing the image current of the rotating ions. The spectrum providing the mass-to-charge ratio ($m/z$) is retrieved from the Fourier transform of the acquired signal.
Exciting resonantly into cyclotron motions can also be conveniently used to eject selectively specific masses, such as potential fragments generated during the desorption-ionization step and isotopic ($^{13}$C containing) species. Following these ejections, one can then start the VUV irradiation on pure $^{12}$C parent ions. This isolation step was not performed in the case of EtCor$^+$ as described in Appendix~\ref{app:roadmap_building_explanation}.
Finally, a helium (He) gas pulse is injected 5\,s after the ion selection and 15\,s before the VUV irradiation. Collisions of the trapped ions with He atoms cooled by collisions with the cryogenic walls is found to improve ion cooling  \citep{stockett2019}.

\begin{figure}[htbp]
    \centering
    \includegraphics[width=.5\textwidth]{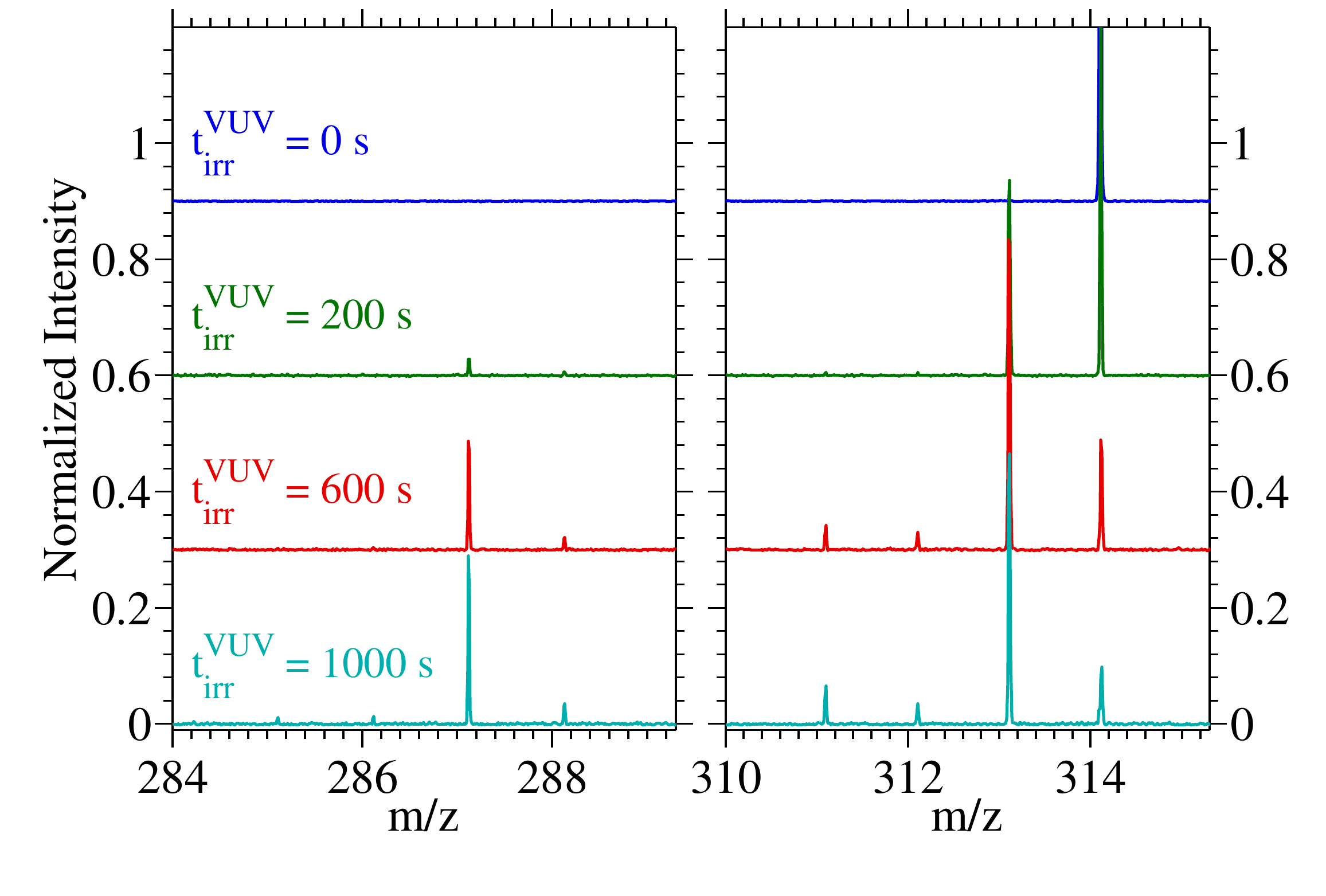}
    \caption{VUV photoprocessing of MeCor$^+$. Mass spectra acquired without VUV irradiation (blue) and at three VUV irradiation times: 200\,s (green), 600\,s (red), and 1000\,s (cyan). MeCor$^+$ is observed at $m/z = 314$ and the observed fragments are located at $m/z = 313, 312, 311, 288, 287, 286$, and $285$. For $t_\mathrm{irr}^\mathrm{VUV}= 0$\,s, the mass spectrum is truncated to a normalized intensity of 0.3 since only the parent cation is observed.}
    \label{fig:MS_MeCor}
\end{figure}

Our measurements consist of acquiring the mass spectra of the VUV photoprocessed ion cloud stored in the ICR cell as a function of VUV irradiation time. In this study, for the sake of clarity, we report values of $m/z = X$ that are truncated to the nearest integer and designate any corresponding ion by the notation '$M_X$'. As an example, we present in Fig.~\ref{fig:MS_MeCor} mass spectra recorded in the case of MeCor$^+$. We can observe the decrease in the parent cation signal while several fragment channels are growing with the increase in VUV irradiation time. The treatment of these spectra and the method used to analyze the kinetic evolution of the fragmentation channels are explained in Sect.~\ref{sec:meth_data_analysis}. After each measurement the cell is emptied and a fresh cloud of cations is injected and exposed to VUV photons. We want to emphasize that many diagnostics are used in our setup (laser power meters, VUV beam pointing monitoring, pressure gauges, etc.) in order to get similar experimental conditions day-to-day. This control of the operating conditions is crucial for the comparison of the photodissociation kinetics between all the investigated species (see Sect.~\ref{sec:res}). 

\subsection{Experimental conditions} \label{sec:meth_expcond}

Two main statements define our experimental conditions (justified below): (i) the measurements are comparable between species, (ii) the ions are cold when they absorb a VUV photon, which means that the VUV photoprocessing is purely sequential. We also stress that no dication channel was observed for all the studied species irradiated by 10.5\,eV photons. To our knowledge, the ionization energies of the methylated, ethylated, and superhydrogenated cations, which are studied here, are not reported in the literature. For Pyr$^+$ and Cor$^+$, \cite{zhen2016} have derived experimental appearance energies of $11.7~\pm~0.1$~\,eV and $10.95~\pm~0.05$~\,eV, respectively. As shown by the study of \cite{jochims1999}, regular neutral PAHs have, on average, an ionization energy 0.2 eV higher than their methylated derivatives. The presence of an aliphatic sidegroup is therefore not expected to lead to a strong shift in the ionization energy and this appears in line with the absence of a dication signal in our measurements.

As stated previously, our setup contains many diagnostics to check the stability of the experimental conditions. 
In particular, the spatial overlap between the VUV beam and the ion cloud has to be, as much as possible, similar between different measurements. We optimized this overlap by finding the maximum fragment production of Pyr$^+$ while moving the VUV beam position. The VUV beam position was observed thanks to a screen imaged by a camera (see Fig.~\ref{app:fig_VUV_calib} in the appendix). This allowed us to control the beam pointing and to reproduce it within a $\sim 100~\mu$m accuracy. The decrease in the parent peak intensity toward zero at long irradiation times suggests that the entire ion cloud can be reached by VUV photons. In these conditions, we can consider that the experimental VUV absorption rate is $k_\mathrm{abs}^\mathrm{exp} = \sigma_\mathrm{abs} \phi_0$, where $\sigma _\mathrm{abs}$ is the photoabsorption cross section and $\phi_0$ is the VUV flux. Since experimental photoabsorption cross sections are not available for the studied species, we used theoretical values (see Appendix \ref{app:sigma_PAHs}). Using the estimated VUV photon flux, we could then evaluate that $k_\mathrm{abs}^\mathrm{exp}$ equals to few $10^{-3}$s$^{-1}$. This indicates that the mean timescale between the absorption of two VUV photons is a few hundred seconds.
This time is enough for the excited ions to relax radiatively between the sequential absorption of two VUV photons. Numerical simulations show that a Cor$^+$ excited by a $\sim$10\,eV photon, if not dissociating, would relax most of its internal energy in less than 5\,s \citep{joblin2020}.
We are therefore able to experimentally simulate the VUV photoprocessing of PAHs in a cold and collisionless environment, which makes our conditions close to those found in PDRs. The observed sequential fragmentation cascades are bounded by our experimental conditions given by the VUV photon flux ($\sim 10^{13}$\,photon~s$^{-1}$~cm$^{-2}$) and the total irradiation time ($\sim 1000$\,s). A higher VUV flux or a longer irradiation time would allow us to access to more generations of daughter fragments.

\subsection{Data analysis}\label{sec:meth_data_analysis}

\begin{figure}[htbp]
    \centering
    \includegraphics[width=0.49\textwidth]{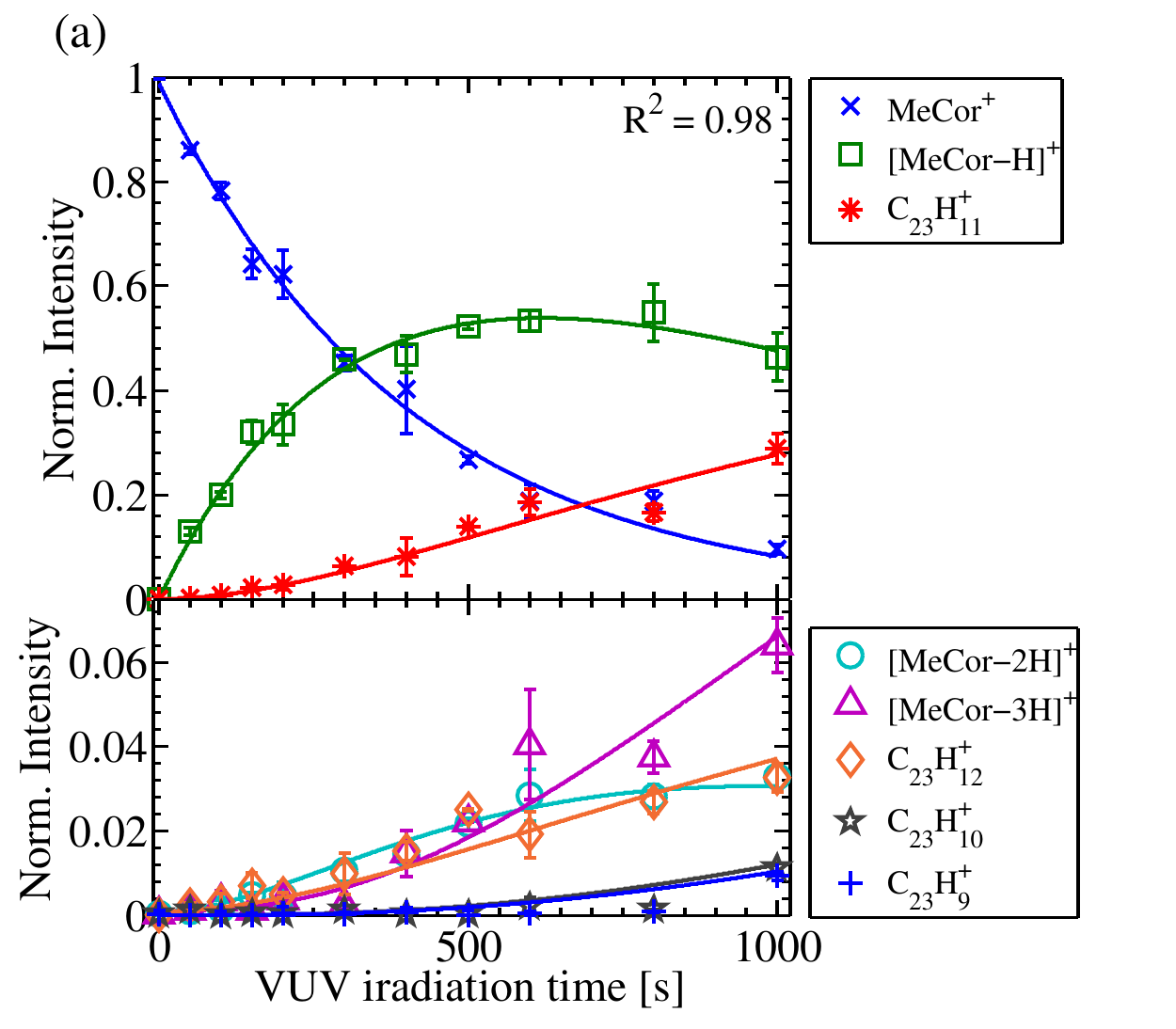} 
    \includegraphics[width=0.49\textwidth]{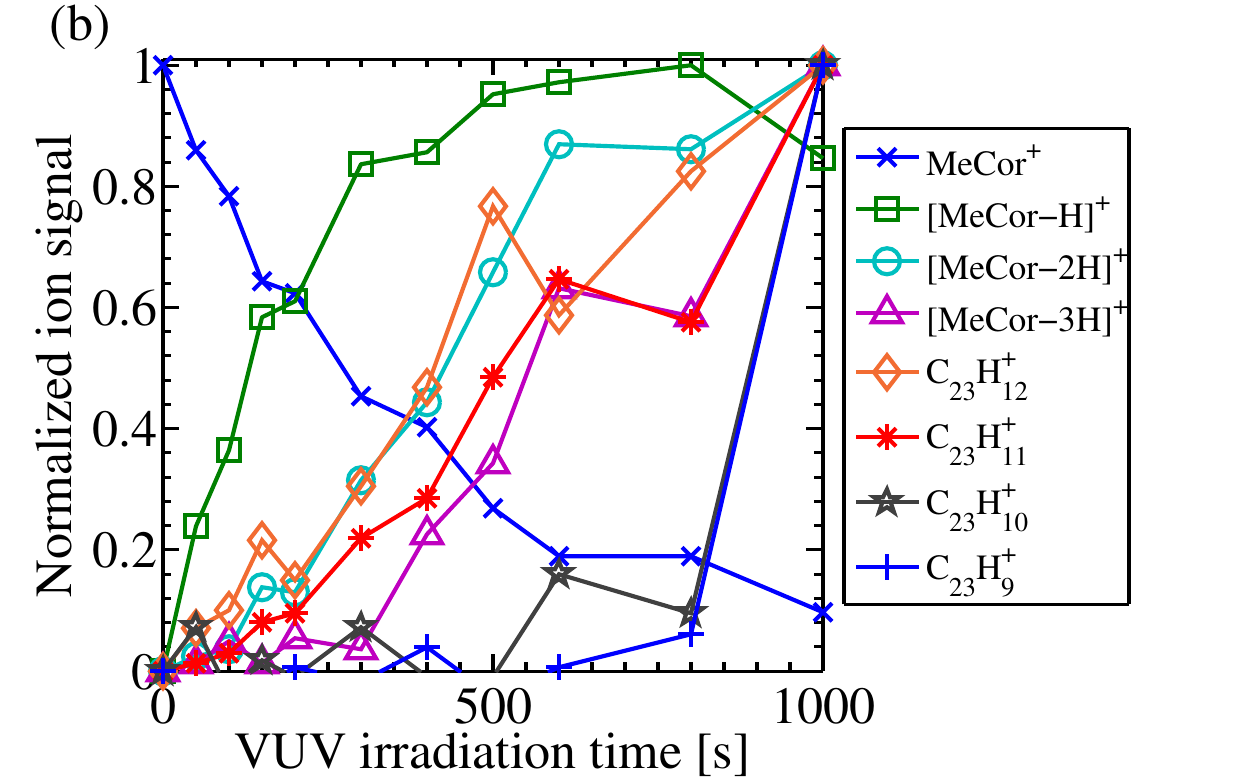}
    \caption{VUV photoprocessing kinetic curves of MeCor$^+$ and its fragments. (a) Normalized intensity of each channel (defined by Eq.~\ref{eq1}) for VUV irradiation times up to 1000\,s. The top panel corresponds to the main channels while the bottom panel is a zoom on the minor channels. The solid curves correspond to the fitting functions that were derived with the procedure described in Sect.~\ref{sec:meth_data_analysis}. (b) Normalized-to-one kinetic signal of each ion channel.}
    \label{fig:MeCor_kinetic_evol}
\end{figure}

From the measurement of the mass spectra as a function of VUV irradiation time, we retrieved kinetic curves which were analyzed to build a fragmentation map including rates between species.
As observed in the mass spectra of the MeCor$^+$ in Fig.~\ref{fig:MS_MeCor}, the high resolution combined with the ejection of the $^{13}$C isotopes permits each fragment channel to be separated without any additional analysis. Also, the high sensitivity of the technique enables us to detect ion signals that have a contribution inferior to 1\% of the total ion signa, for instance $M_{285}$ at $t_\mathrm{irr}^\mathrm{VUV}= 1000$\,s in Fig.~\ref{fig:MS_MeCor}. To further improve the quality of the spectra, a background signal is removed. Besides, each spectrum is normalized by the sum of all the detected ion peaks at $M_i$ (parent + fragments) such as: 
\begin{equation}\label{eq1}
I_{M_i}^\mathrm{norm.}=\frac{I_{M_i}}{\sum_{M_j}^{} I_{M_j}}~.
\end{equation}
This normalization procedure allows us to correct for variations in the initial parent peak intensity. This is illustrated by the small error bars on the data points in Fig.~\ref{fig:MeCor_kinetic_evol}~(a), which corresponds to the normalized intensity of the MeCor$^+$ fragment channels as a function of the VUV irradiation time. This example shows the depopulation of the parent cation toward the main fragments [MeCor-H]$^+$ and C$_{23}$H$_{11}^+$, as well as other minor fragments ([MeCor-2H]$^+$, [MeCor-3H]$^+$, C$_{23}$H$_{12}^+$, C$_{23}$H$_{10}^+$, and C$_{23}$H$_{9}^+$). 
We can also plot the fragment kinetic curves by normalizing them by their maximum, as shown in Fig.~\ref{fig:MeCor_kinetic_evol}~(b). This allowed us to identify which channel is firstly populated by the VUV photoprocessing and to get a first guess of the parent-daughter fragmentation network for each species. For instance, Fig.~\ref{fig:MeCor_kinetic_evol}~(b) clearly shows that the inflection point of the [MeCor-H]$^+$, [MeCor-2H]$^+$, and [MeCor-3H]$^+$ kinetics are shifted in time with respect to one another. This is consistent with a sequential population of these channels, which means, in this case, a sequential hydrogen (H) loss from MeCor$^+$ to [MeCor-3H]$^+$. 

With this knowledge we can build fragmentation maps, which describe the observed fragmentation cascades. For each map, the connections between the different species are described by a system of differential equations whose solution functions can fit the kinetic curves of all the detected ion channels. It reads: 
\begin{equation}\label{eq2}
\left\{
    \begin{array}{cc}
       (1^\mathrm{st}) & \displaystyle \frac{\mathrm{d}I_\mathrm{M_\mathrm{parent}}}{\mathrm{d}t}(t)= 
       \displaystyle -\sum_{M_i < M_\mathrm{parent}}{k_\mathrm{frag}^{M_\mathrm{parent}\rightarrow M_i} I_{M_\mathrm{parent}}(t)} \\
                & \vdots\\
       (i^\mathrm{th}) & \displaystyle \frac{\mathrm{d}I_{M_i}}{\mathrm{d}t}(t)=  
       \sum_{M_n > M_i}{k_\mathrm{frag}^{M_n \rightarrow M_i} I_{M_n} (t)} - \sum_{M_j < M_i}{k_\mathrm{frag}^{M_i \rightarrow M_j} I_{M_i}(t)} \\
    \end{array}
\right.~,
\end{equation}
where $ I_{M_i}(t)$ is the population of an ion channel $M_i$ as a function of the VUV irradiation time $t$ and $k_\mathrm{frag}^{M_i\rightarrow M_j}$ is the fragmentation rate from fragment $M_i$ to fragment $M_j$. The system takes into account population terms ($+k_\mathrm{frag}^{M_n\rightarrow M_i} I_{M_n}(t)$) and depopulation terms ($-k_\mathrm{frag}^{M_i\rightarrow M_j} I_{M_i}(t)$), which are implemented by us in agreement with the behavior of the observed channel kinetics and data reported in previous studies (more details in Appendix~\ref{app:roadmap_building_explanation}). The fitting procedure then adjusts the fragmentation rates in order to obtain the best fitting function for each ion channel. If some crucial terms are missing in the system (Eq.~\ref{eq2}), this results in a set of fitting functions that are partially (or completely) diverging from the measured ion channel kinetics. On the opposite, if too many terms are implemented, some fragmentation rates reach the lower bound used in the minimization procedure, indicating that they are not needed to fit the data. For each studied species, we have tried several sets of population and depopulation terms, all deduced by educated guesses concerning the plausible fragmentation pathways. Our aim was to end with a system of differential equations to build a fragmentation map, which includes a minimal set of relevant terms and provides a satisfactory quality of fit even for minor channels. For instance, we show the quality of this fitting procedure for MeCor$^+$ in Fig.~\ref{fig:MeCor_kinetic_evol}~(a) (fitting curves for all the other species are displayed in Fig. \ref{app:fig_kinetics_all_PAHs}). By extracting the specific fragmentation rates ($k_\mathrm{frag}^{M_i\rightarrow M_j}$), we can define the total fragmentation rate of an ion $M_i$ ($k_\mathrm{frag}^{M_i}$) and the corresponding branching ratios of each of its daughter fragments ($R^{M_i\rightarrow M_j}$) by:

\begin{equation}\label{eq3&4}
k_\mathrm{frag}^{M_i}=\sum_{M_j<M_i}^{} k_\mathrm{frag}^{M_i\rightarrow M_j} ~~ , ~~ R^{M_i\rightarrow M_j}=\frac{k_\mathrm{frag}^{M_i\rightarrow M_j}}{k_\mathrm{frag}^{M_i}} ~,
\end{equation}
where the $M_j$ are the daughter fragments of $M_i$.
These values are displayed in the retrieved fragmentation maps (see Figures~\ref{fig:maps_bare_PAHs}, \ref{fig:Me_Et_PAH_roads}, and \ref{fig:H6-pyrene_road}) and discussed in Sec.~\ref{sec:res}. We stress that the construction of these fragmentation maps using the system of differential equations given by Eq.~\ref{eq2} is a complex multivariable problem that does not always have an unique acceptable solution. Indeed, if $N$ is the number of detected fragments plus parent, a sequential fragmentation could imply up to $N \times (N-1)/2$ fragmentation rates between all the detected fragments. A map of this kind, built without any other input, would surely succeed in fitting the data but very little uncorrelated information could be extracted. On the opposite, we constructed each map by using the observed parent-daughter relations (see Fig.~\ref{fig:MeCor_kinetic_evol}~(b)), logical relations between masses of the fragments, and educated guesses coming from previous studies dealing with the (photo)fragmentation of these species. This allowed us to obtain satisfactory fitting curves with less than 25\% of the possible paths. This procedure is described in Appendix~\ref{app:roadmap_building_explanation}, where we detail the logic and the encountered complexities for each species. Our degree of confidence in the retrieved maps depends on certain fragment channels whose peculiar kinetics largely constrains the possible paths. We finally made use of these maps to draw a global picture of the VUV photoprocessing of the studied PAH cations (see Sects.~\ref{sec:res_bare_PAH}-\ref{sec:res_alphatic_PAH}) and to quantify the relative stability of the different ions produced during the dissociation cascade (see Sect.~\ref{sec:cooling}).

\subsection{DFT and TD-DFT calculations}\label{sec:meth_(TD-)DFT calculations}

A number of species potentially present (or formed) in the experiment have been studied theoretically. We optimized their structures using the density functional theory \citep[DFT,][]{dreizler2012} as implemented in the \textsc{Gaussian16} quantum chemistry package \citep{g16}. These calculations were performed with the B3LYP exchange-correlation functional \citep{Becke1993} and the 6-31g(d,p) basis set \citep{ditchfield1971a, hariharan1973a, hehre1972a}, employing the resolution of identity approximation \citep{weigend1998} as applicable. A harmonic vibrational analysis was performed, with the same level of theory, at all the optimized geometries, to make sure they are really minima and not saddle points of the electronic potential energy surface which represent transition states.\\
Subsequently, we used the \textsc{Octopus} implementation \citep{octopus2020} of TD-DFT in real time and real space \citep{yabana1996} to evaluate the complete electronic photoabsorption spectrum of each species. These calculations, following calibrations performed in previous works \citep{malloci2004, malloci2007, wenzel2020}, were carried out with a simulation box defined by the union of spheres centered on each atom of the given molecule, each with a 8~\AA\ radius. \textsc{Octopus} represents all physical quantities such as Kohn-Sham wavefunctions, electron density, etc., in a discrete grid in the simulation box, whose spacing was chosen as 0.18~\AA. In the \textsc{Octopus} simulations we employed the local spin density approximation for the exchange-correlation functional \citep{dirac1930,perdew1981}, and we used the frozen-core approximation, representing core electrons by standard pseudo-potentials \citep{kleinman1982}. Atom positions were kept fixed during the time evolution, so that the resulting spectra represent vertical excitations, neglecting vibronic structure. This combination of simulation box size, grid spacing, and exchange-correlation functional was previously shown to provide good numerical convergence \citep{wenzel2020} and an overall acceptable agreement with available spectra of PAHs \citep{malloci2004}. \\
Since the photoabsorption spectra are derived from a numerical Fourier transform of the electric dipole moment following an initial Dirac-delta perturbation, they show an artificial broadening. This corresponds to the minimum frequency that can be adequately sampled from a Fourier transform of a function with a finite length. All of our real-time simulations cover $\sim$26\,fs, yielding an energy resolution, in the resulting spectra, of $\sim$0.15\,eV. This is already better than the accuracy of TD-DFT in predicting the energy of excited states, which is usually on the order of 0.3\,eV at best \citep[see e.g.][]{laurent2013}

\section{Results and discussion}\label{sec:res}

In this section, we present the parent cation kinetic curves as well as the associated fragmentation maps resulting from the fragmentation cascades. The presented maps provide the fragmentation rates and the branching ratios (Eq.~\ref{eq3&4}), which were extracted from the fitting procedure (all fits are in the supplementary Fig.~\ref{app:fig_kinetics_all_PAHs}). In our experimental conditions, the values of the fragmentation rates depend on the VUV photon flux but the branching ratios do not (cf. Sect.~\ref{sec:meth_expcond}). We stress that the signal-to-noise ratio of the latest detected fragments is lower than others, due to the overall loss of trapped ions over time and the need of longer VUV irradiation time to observe the last fragments in the cascade. This implies that the extracted $k_\mathrm{frag}^{M_i}$ and $R^{M_i\rightarrow M_j}$ values of the fragmentation map ending channels have higher uncertainties. We also display the lowest-energy structures of some relevant ions in these maps. 
We first present and discuss the results on the VUV photoprocessing of bare cationic PAHs (see Sect.~\ref{sec:res_bare_PAH}). Then, we reveal what is happening when aliphatic bonds are at play in these PAHs, by detailing the similarities and the differences observed in the parent cation kinetics and the retrieved maps (see Sect.~\ref{sec:res_alphatic_PAH}). We finally discuss the involved molecular parameters and mechanisms, in order to better interpret our results (see Sect.~\ref{sec:discuss_vuv_processing}).

\subsection{Fragmentation of the bare PAH cations}\label{sec:res_bare_PAH}

\begin{figure}[htbp]
    \centering
    \includegraphics[width=0.45\textwidth]{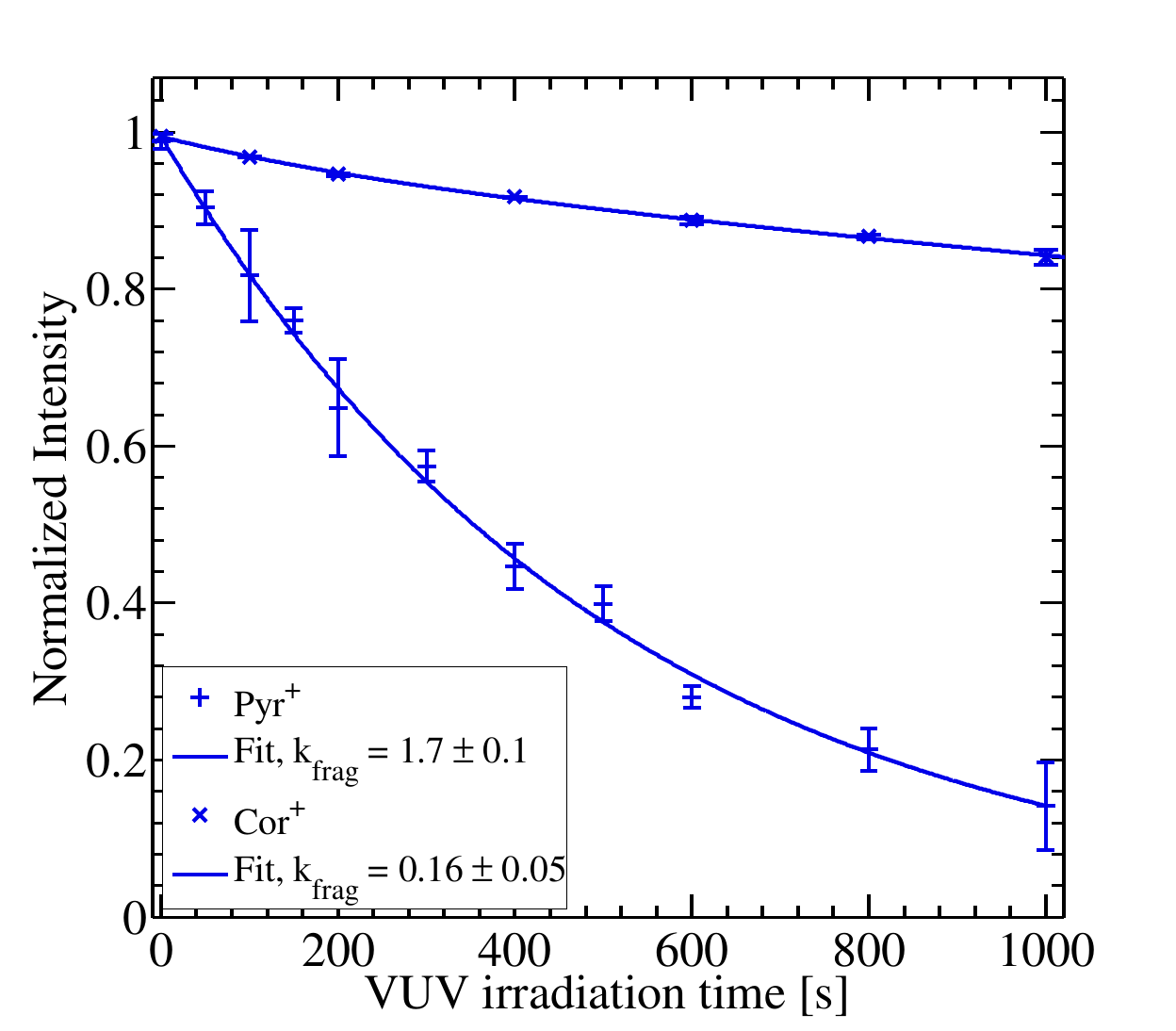} 
    \caption{Normalized intensity of Pyr$^+$ and Cor$^+$ as a function of the VUV irradiation time. The fitting curves and the extracted fragmentation rates (see method in Sect.~\ref{sec:meth_data_analysis}) are displayed for each species with $k_\mathrm{frag}$ in $10^{-3}$s$^{-1}$.}
    \label{fig:parent_kinetic_evol_bare_PAHs}
\end{figure}

We compare the kinetic curves of the bare PAH cations as a function of the VUV irradiation time in Fig.~\ref{fig:parent_kinetic_evol_bare_PAHs}. The displayed fragmentation rates were extracted from the fitting procedure, which is described in Sect.~\ref{sec:meth_data_analysis}, and they show that the VUV photofragmentation of Pyr$^+$ is much faster ($1.9\times10^{-3}$\,s$^{-1}$) than the one of Cor$^+$ ($0.16\times10^{-3}$\,s$^{-1}$).

\begin{figure*}[htbp]
    \centering
    \includegraphics[width=0.95\textwidth]{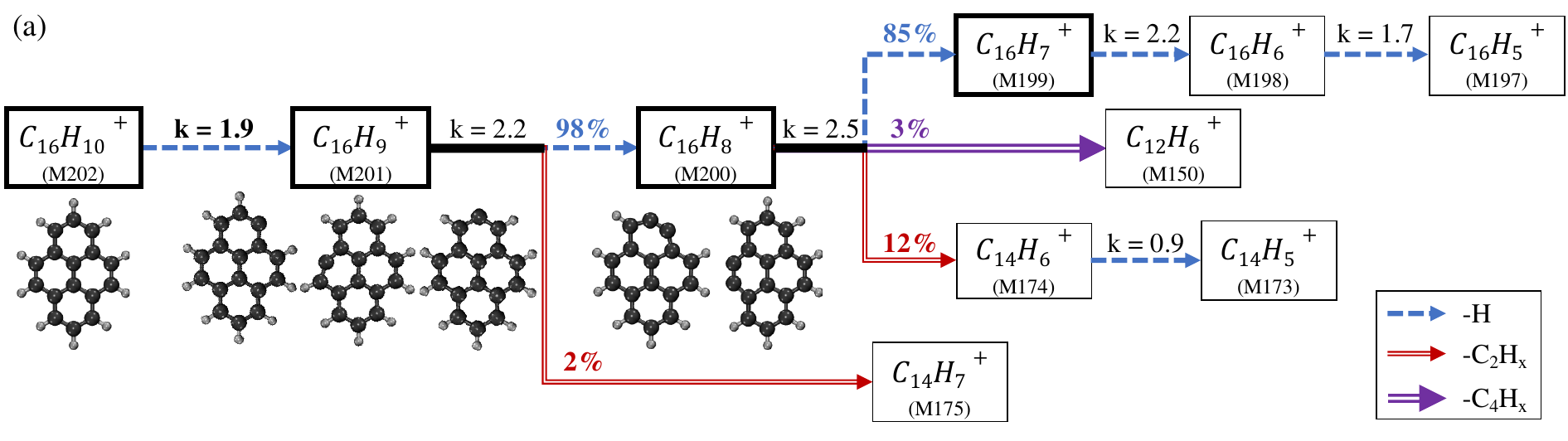}
    \includegraphics[width=0.89\textwidth]{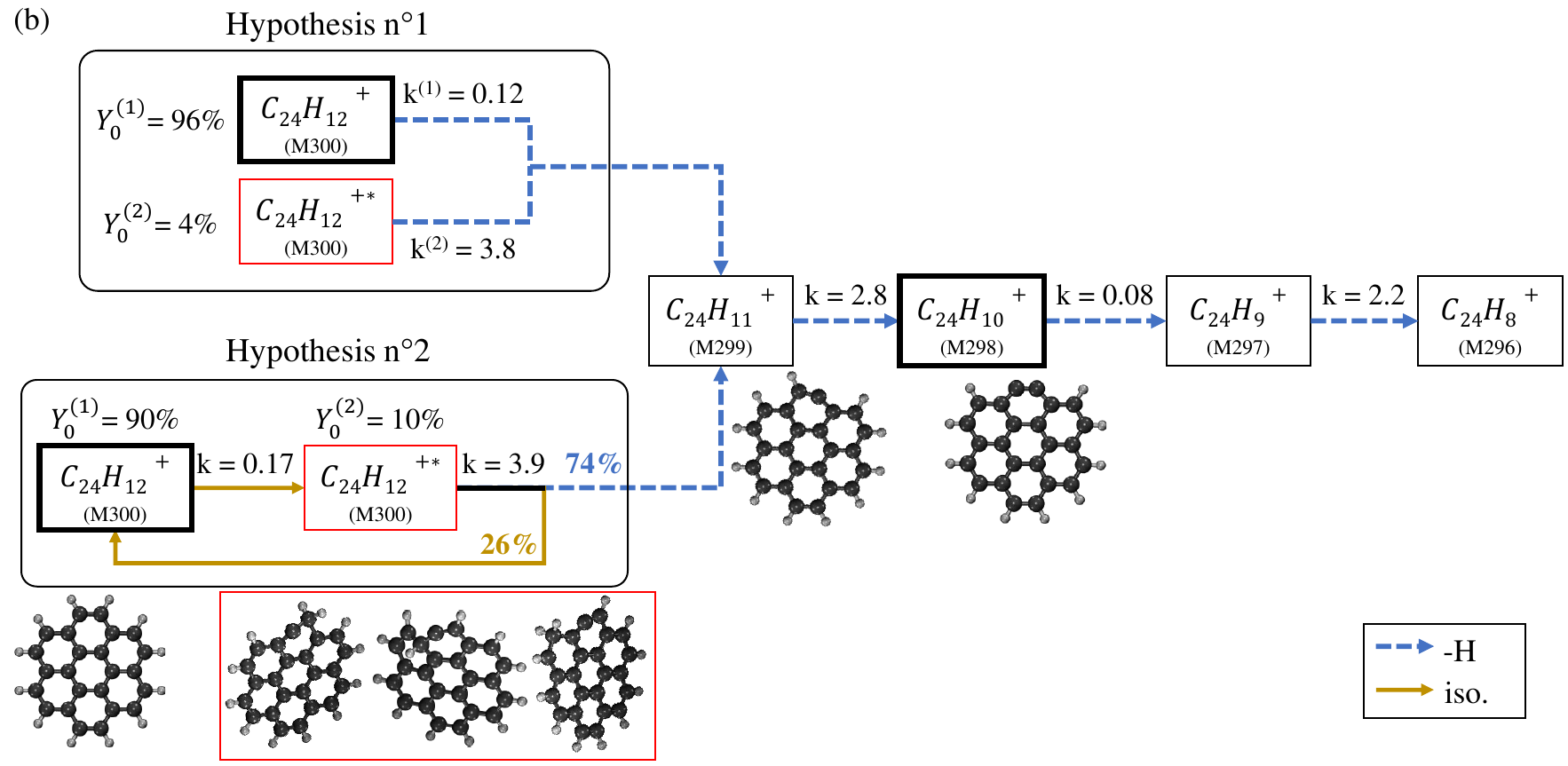}
    \caption{VUV photofragmentation map of (a) Pyr$^+$ and (b) Cor$^+$. Each arrow represents a specific fragment loss or an involved mechanism triggered by a VUV photon absorption (a legend is provided for each map). The fragmentation rates $k_\mathrm{frag}$ (here $k$ for simplification) are displayed in units of $10^{-3}$ s$^{-1}$. When needed, the branching ratios are also displayed. The optimized molecular structures are depicted for the parent cations and the key fragments, as well as their expected isomers. Bold boxes correspond to the major ion channels (yield higher than 10\% during the kinetics). The case of Cor$^+$ is peculiar since the fitting requires the contribution of a population of isomers (red box). Two plausible scenarios connecting the parent to the isomers are shown.}
    \label{fig:maps_bare_PAHs}
\end{figure*}

The Pyr$^+$ and Cor$^+$ fragmentation maps (Fig.~\ref{fig:maps_bare_PAHs}~(a-b)) show that their main fragmentation path corresponds to a sequential H loss: up to five H atoms for Pyr$^+$ and four H atoms for Cor$^+$, this number being limited by the irradiation time as discussed above. For this sequential H-loss branch of the map, the retrieved fragmentation rates are all comparable in the case of Pyr$^+$ (about $2\pm0.2\times10^{-3}$~s$^{-1}$) while they exhibit strong variations for Cor$^+$ (from $0.1\times10^{-3}$~s$^{-1}$ to $2.9\times10^{-3}$~s$^{-1}$). Besides, carbon-loss channels are present in the Pyr$^+$ map, whereas they are absent from the Cor$^+$ map.

The fragmentation cascade of Pyr$^+$ was previously studied with PIRENEA using the continuous irradiation of a UV-visible Xe arc lamp (photon energy range of [1--5]\,eV). In these conditions a very minor channel of C$_2$H$_2$ loss was observed for Pyr$^+$ and a more significant one for C$_{16}$H$_8^+$ \citep{west2014_pyr}. The results obtained with 10.5\,eV photons present some similarities but also differences with these earlier results. In particular, no C$_2$H$_2$ loss channel is retrieved for Pyr$^+$ but only for C$_{16}$H$_9^+$ and C$_{16}$H$_8^+$.

The Cor$^+$ map is mostly consistent with the sequential H loss reported in earlier measurements with PIRENEA using the Xe arc lamp and discussed in \cite{montillaud2013} but also with other studies \citep{castellanos2018_lab, west2018}. In line with these results, it shows that species containing an odd number of hydrogens, such as C$_{24}$H$_{11}^+$ and C$_{24}$H$_9^+$ are easier to dissociate compared to the ones containing an even number of hydrogens. A point of interest that was reported by \cite{castellanos2018_lab} is the foreseen role of isomers in the production of [Cor-H]$^+$. We found that a good fit of the [Cor-H]$^+$ kinetic curve, in particular the initial steep rise (see supplementary Fig.~\ref{app:fig_kinetics_all_PAHs}~(e)), requires the contribution of two precursors: Cor$^+$ itself and another one that we label Cor$^{+*}$. As shown by \cite{trinquier2017_H_shifted_isomers,trinquier2017_ring_alteration}, a number of isomers can be formed upon Cor$^+$ activation, mostly resulting from H-migration (some examples are depicted in Fig.~\ref{fig:maps_bare_PAHs}~(b)). The resulting formation of an aliphatic bond would energetically favor the photofragmentation at the next absorbed VUV photon. As discussed in Appendix \ref{app:roadmap_building_explanation}, we could not obtain a satisfactory fit of the kinetic curves without including an initial population of Cor$^{+*}$ (4 to 10\% of the total parent ion abundance), which would therefore result from the desorption-ionization process. Two hypothesis are presented in Fig.~\ref{fig:maps_bare_PAHs}~(b) concerning the Cor$^{+*}$ production. In the first one, no isomer is produced during the VUV irradiation; its population expires then quickly and only contributes to the steep rise of the [Cor-H]$^+$ kinetics. In the second hypothesis, Cor$^{+*}$ is produced upon VUV irradiation of Cor$^+$ population. In this second hypothesis, a repopulation path goes back to the initial Cor$^+$ reservoir with a probability of 26\%. This may correspond to isomers that are more stable than those resulting from H-migration. They could correspond to the vinylidene or ethynyl isomers calculated by \cite{trinquier2017_ring_alteration}. The two presented schemes provide a comparable fit of the data and similar values of k$_\mathrm{frag}$ in the corresponding maps. They represent extreme cases and the actual scheme is likely to lay in between. We could have expected a similar behavior for [Cor-2H]$^+$ but it was not possible to evidence it because our experiment is not optimized to detect such subtle effects on daughter species.

\subsection{Fragmentation of the aliphatic PAH derivatives}\label{sec:res_alphatic_PAH}

\begin{figure}[htbp]
    \centering
    \includegraphics[width=0.45\textwidth]{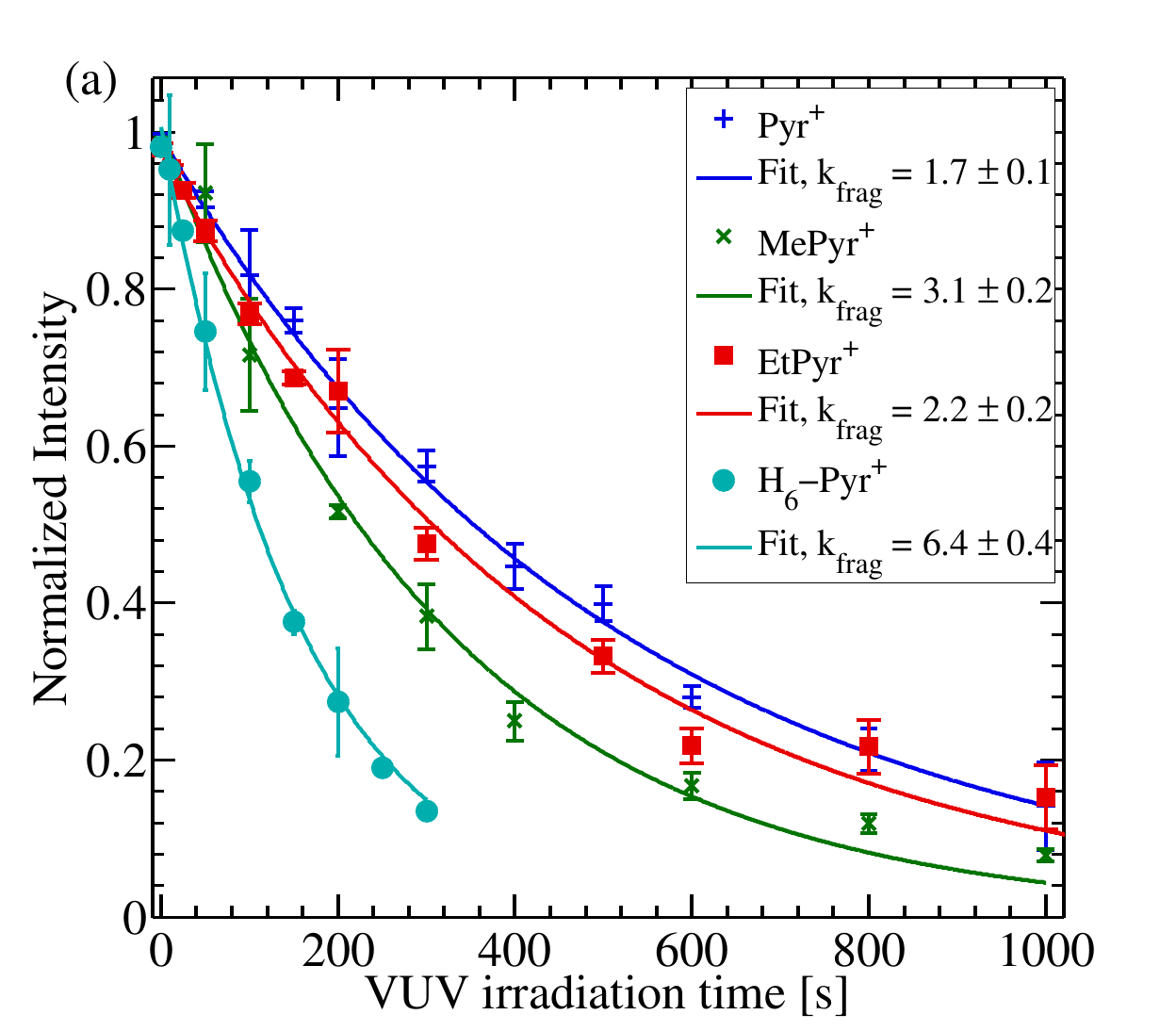} 
    \includegraphics[width=0.45\textwidth]{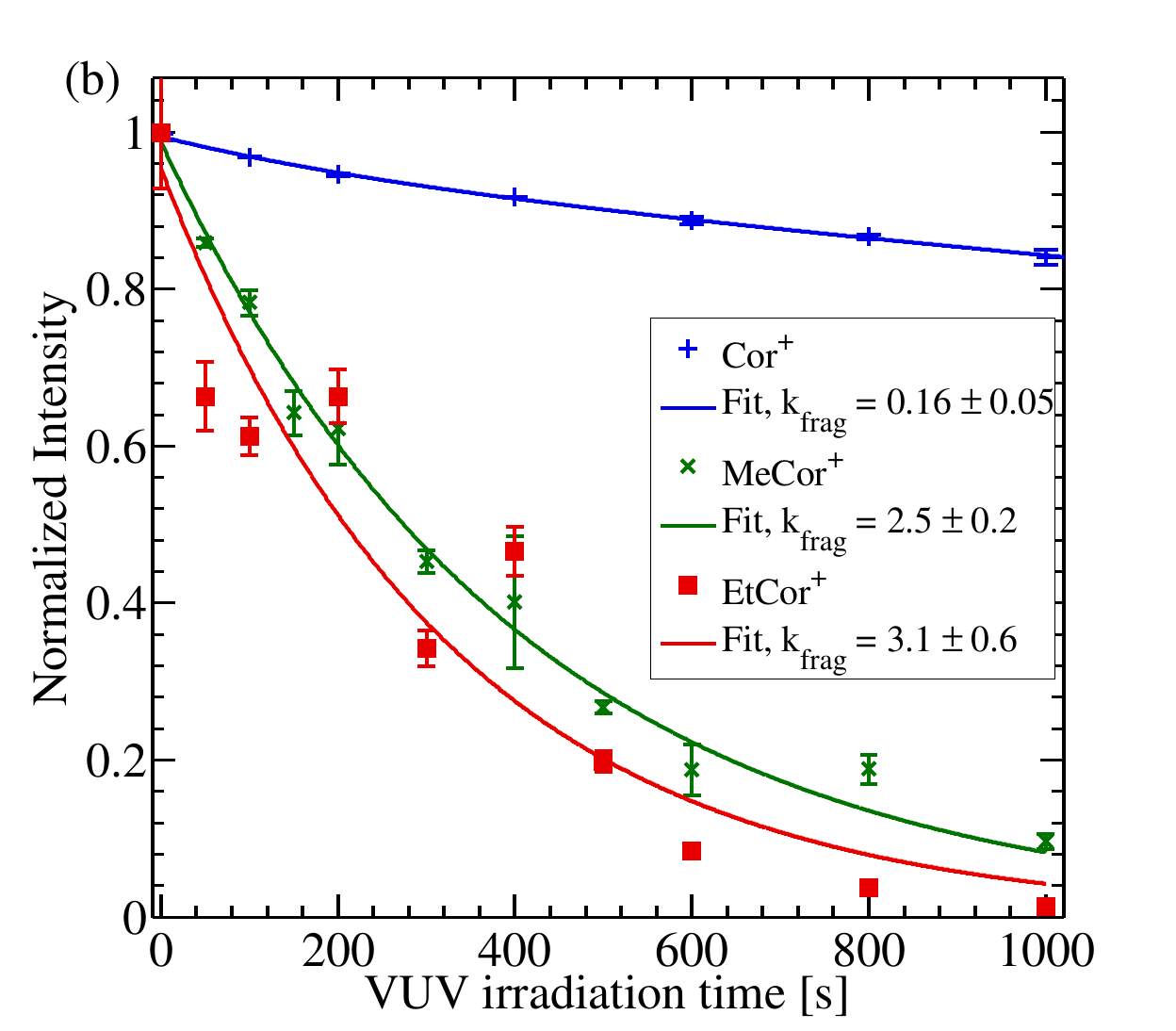}
    \caption{Normalized intensity as a function of the VUV irradiation time of the studied parent cations: (a) pyrene-like species and (b) coronene-like species.
    The fitting curves and the extracted fragmentation rates (see method in Sect.~\ref{sec:meth_data_analysis}) are displayed for each species with $k_\mathrm{frag}$ in $10^{-3}$s$^{-1}$.}
    \label{fig:parent_kinetic_evol}
\end{figure}

The fragmentation kinetic curves of the aliphatic PAH derivatives are compared with those of the bare PAH parent cations in Fig.~\ref{fig:parent_kinetic_evol} by separating, for the sake of clarity, pyrene-like species (Fig.~\ref{fig:parent_kinetic_evol}~(a)) from coronene-like species (Fig.~\ref{fig:parent_kinetic_evol}~(b)). The retrieved fragmentation maps are reported in Figs.~\ref{fig:Me_Et_PAH_roads} and ~\ref{fig:H6-pyrene_road} for alkylated PAHs and H$_6$-Pyr$^+$, respectively. The results are discussed below for both types of species.

\subsubsection{Methylated and ethylated PAHs}
\begin{figure*}[htbp]
    \centering
    \includegraphics[width=0.9\textwidth]{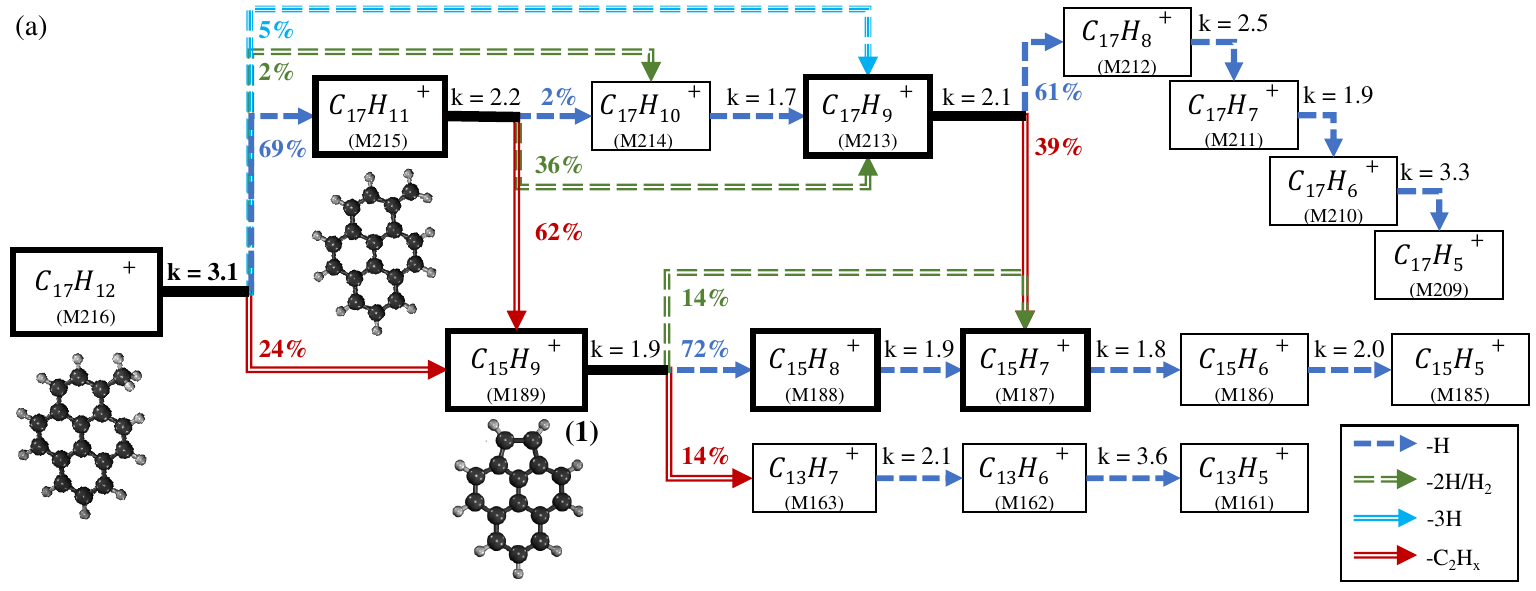}
    \includegraphics[width=0.89\textwidth]{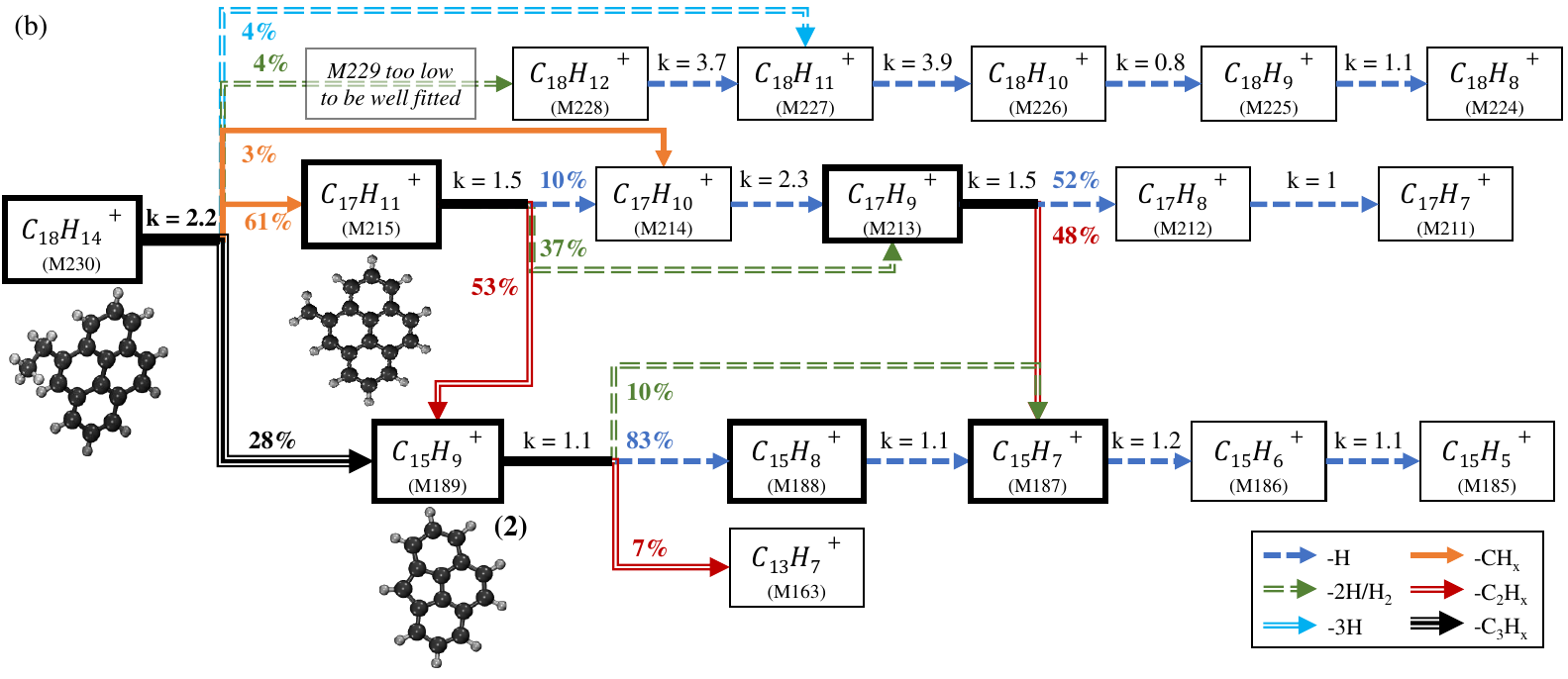}
    \includegraphics[width=0.7\textwidth]{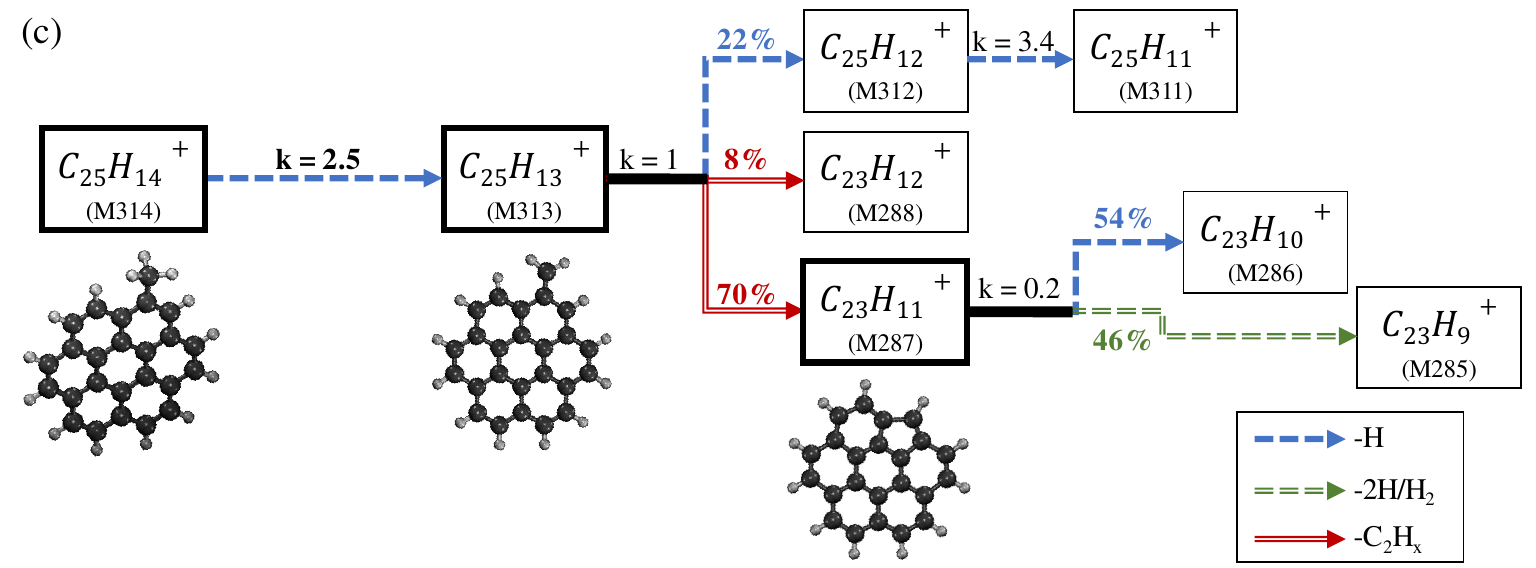}
    \includegraphics[width=0.71\textwidth]{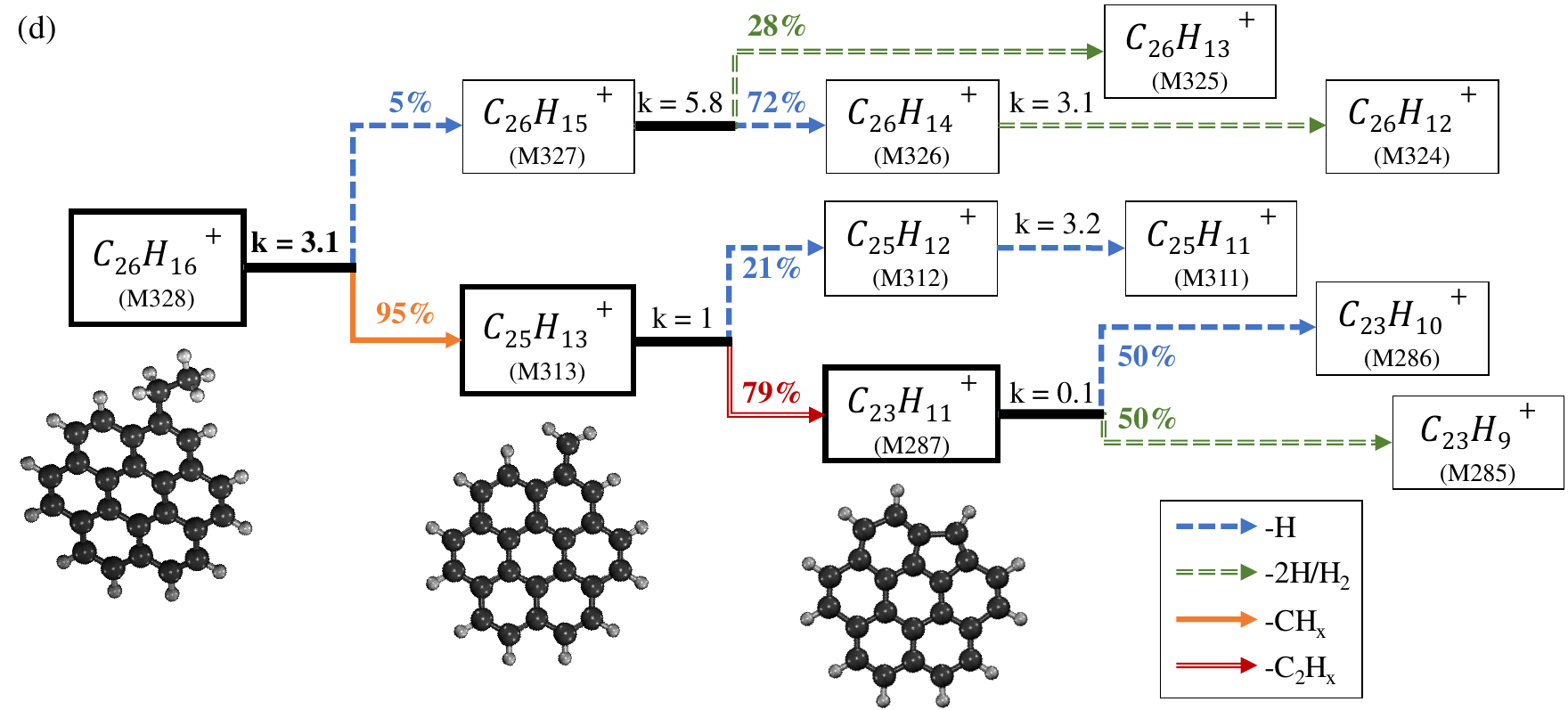}
    \caption{VUV photofragmentation maps of (a) MePyr$^+$, (b) EtPyr$^+$, (c) MeCor$^+$, (d) EtCor$^+$. Caption similar to Fig. \ref{fig:maps_bare_PAHs}.}
    \label{fig:Me_Et_PAH_roads}
\end{figure*}

The extracted fragmentation rates are similar for methylated and ethylated PAHs, in the range of $2.7~\pm~0.4\times10^{-3}$\,s$^{-1}$ (cf. Fig.~\ref{fig:Me_Et_PAH_roads}). Although higher, these values remain close to the fragmentation rate of Pyr$^+$ ($1.9\times10^{-3}$\,s$^{-1}$) but they completely differ from the one of Cor$^+$ ($0.16\times10^{-3}$\,s$^{-1}$). The retrieved fragmentation maps are shown in Fig.~\ref{fig:Me_Et_PAH_roads}. They exhibit strong similarities. For methylated PAHs, the first fragmentation step mainly consists in a single H loss (69\% for MePyr$^+$ and 100\% for MeCor$^+$) while, for ethyl-PAH cations, the first step mainly consists in a CH$_3$ loss (61\% for EtPyr$^+$ and 95\% for EtCor$^+$). In both cases, these steps lead to fragments that have the same mass, namely C$_{17}$H$_{11}^+$ ($M_{215}$) and C$_{25}$H$_{13}^+$ ($M_{313}$) for pyrene and coronene derivatives, respectively. The structure of C$_{17}$H$_{11}^+$ has been the focus of several studies, including \cite{kokkin2014}, \cite{rapacioli2015}, \cite{jusko2018}, and \cite{west2018_MePyr}, the last two demonstrating that the isomer with a methylene group, as drawn in Fig.~\ref{fig:Me_Et_PAH_roads}, is preferentially formed. We also expect the formation of such a group following the dissociation of the ethyl group. However, in the case of EtPyr$^+$, the methylene group is formed at a different position (named C$_4$) than in MePyr$^+$ for which it is at the C$_1$ position. This leads to different isomer structures, as displayed in Fig.~\ref{fig:Me_Et_PAH_roads}~(a) and (b).
In the case of MeCor$^+$ (resp. EtCor$^+$), we further observe that the parent fragmentation rate is $\sim$~16 (resp. $\sim$~20) times larger than the one of Cor$^+$, which reinforces the idea that the H loss (resp. CH$_3$ loss) comes from the alkyl group and that a methylene-coronene structure is expected for C$_{25}$H$_{13}^+$ (see Fig.~\ref{fig:Me_Et_PAH_roads}~(c-d)). The structure of this ion is the same in both maps, due to the symmetry of the coronene molecule. The associated fragmentation rates and branching ratios are found to be similar (as expected), except for the minor C$_{23}$H$_{12}^+$ channel, which could not be extracted in the EtCor$^+$ case due to experimental limitations (see Appendix~\ref{app:roadmap_building_explanation}).  

Another common fragment is noticeable for the pyrene-like species: C$_{15}$H$_{9}^+$ ($M_{189}$). For MePyr$^+$ (resp. EtPyr$^+$), it is produced either by a direct fragmentation of the parent, with a branching ratio of 24\% (resp. 28\%), or by a secondary fragmentation of the methylene-pyrene cation, through C$_2$H$_2$ loss with a branching ratio of 62\% (resp. 53\%). For coronene-like species, this channel is echoed by the common fragment C$_{23}$H$_{11}^+$ ($M_{287}$), which comes from a secondary fragmentation of the methylene-coronene with a branching ratio of $\sim$70\%. These C$_2$H$_2$-loss fragments of methylene-PAH ions result in an odd number of C atoms and most likely contain a five-membered ring, as earlier proposed in the case of methylene-pyrene fragmentation \citep{kokkin2014,jusko2018}. The structure of these species are displayed in Fig.~\ref{fig:Me_Et_PAH_roads}. The five-membered ring position of C$_{15}$H$_{9}^+$ depends on the position of the alkyl group, leading to isomer (1) having the pentagonal ring on the long axis of the pyrene (Fig.~\ref{fig:Me_Et_PAH_roads} (a)) and isomer (2) having it on the short axis of pyrene (Fig.~\ref{fig:Me_Et_PAH_roads} (b)). The fragmentation maps show that the two isomers have different fragmentation rates, $1.9\times10^{-3}$\,s$^{-1}$ for isomer (1) compared to $1.1\times10^{-3}$\,s$^{-1}$ for isomer (2). The branching ratio of the fragments ($\Sigma{(H,2H)}/C_2H_2$) also differs to some extent, with values of 86\%:14\% and 93\%:7\% for isomer (1) and (2), respectively. In addition, C$_{13}$H$_7^+$ is an ending product for isomer (2) while it keeps fragmenting for isomer (1).

Finally, alkylated species exhibit routes involving sequential H losses. The first one corresponds to $M_{215} \rightarrow M_{214} \rightarrow M_{213} \rightarrow M_{212} \rightarrow M_{211} $ for pyrene-based species and to $M_{313} \rightarrow M_{312} \rightarrow M_{311}$ for coronene-based species. In the case of pyrene-based species, there is an efficient competition with C$_2$H$_2$ loss for C$_{17}$H$_{11}^+$ and C$_{17}$H$_{9}^+$ due to the formation of the pentagonal ring. Still, overall the dissociation proceeds comparably to pyrene. The comparison with the bare PAHs is even more obvious in the case of coronene-based species for which we notice that the following specific fragmentation rates: $k_\mathrm{frag}^{M_{313}\rightarrow M_{312}}=R^{M_{313}\rightarrow M_{312}} k_\mathrm{frag}^{M_{313}}\approx 0.2\times10^{-3}$\,s$^{-1}$ and $k_\mathrm{frag}^{M_{312}\rightarrow M_{311}} \approx 3\times10^{-3}$\,s$^{-1}$, show the same trend as the values obtained along the Cor$^+$ map, which suggests that channels involving the CH aromatic bonds are also involved in these fragmentation steps.
A second route of sequential H losses is observed in the case of ethylated species, although it is relatively minor, at the level of a few percent. It likely leads to the formation of ethynyl-substituted PAHs, starting from C$_{18}$H$_{10}^+$ and C$_{26}$H$_{12}^+$ in the case of EtPyr$^+$ and EtCor$^+$, respectively. Ethynyl-substitued PAHs are expected to be as stable as bare PAHs \citep{rouille2015, rouille2019}. The fragmentation rate derived for C$_{18}$H$_{10}^+$, $k_\mathrm{frag}^{M_{226}} = 0.8\times10^{-3}$\,s$^{-1}$ suggests that these species could even be more stable than the bare PAH ($k_\mathrm{frag}^{M_{202}} = 1.9\times10^{-3}$\,s$^{-1}$) upon VUV irradiation. A longer irradiation time would have been necessary to observe the fragmentation of C$_{26}$H$_{12}^+$ in our experiment.

\subsubsection{The case of H$_6$-Pyr$^+$}

\begin{figure}[htbp]
    \centering
    \includegraphics[width=0.5\textwidth]{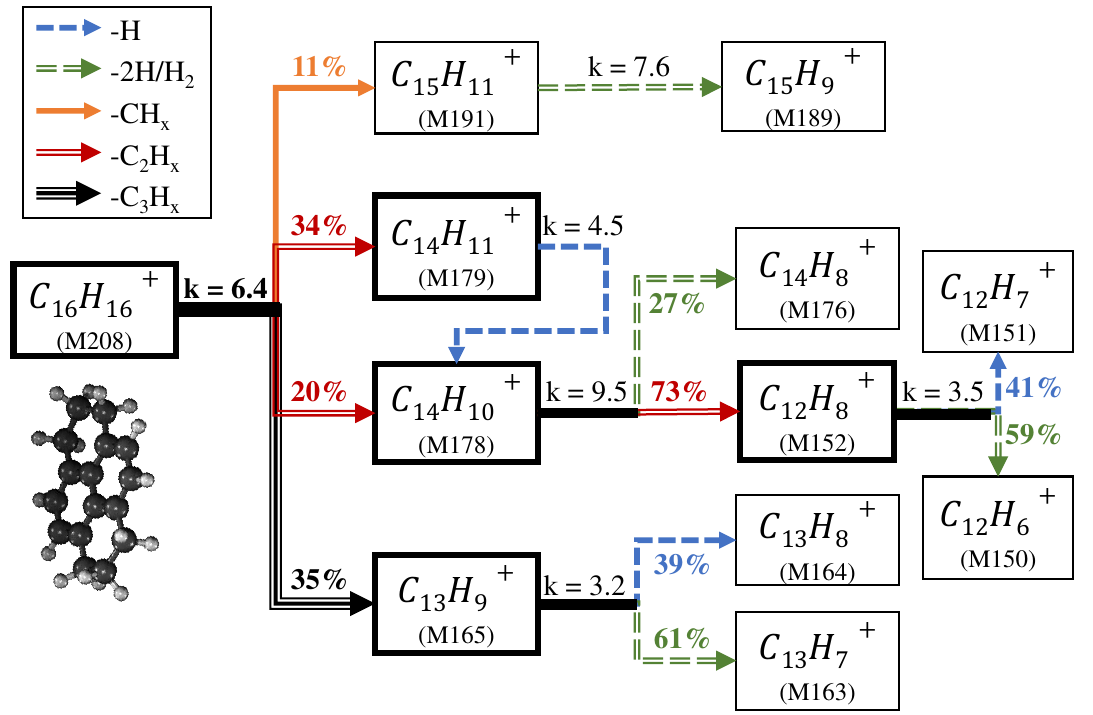}
    \caption{VUV photofragmentation map of H$_6$-Pyr$^+$. Caption similar to Fig. \ref{fig:maps_bare_PAHs}.}
    \label{fig:H6-pyrene_road}
\end{figure}

H$_6$-Pyr$^+$ is the only superhydrogenated species studied in this work. As shown by Fig.~\ref{fig:parent_kinetic_evol}~(a), H$_6$-Pyr$^+$ has a fragmentation rate about 2 to 3 times higher than other species ($k_\mathrm{frag} = 6.4\times10^{-3}$s$^{-1}$). Moreover, as displayed in the fragmentation map (see Fig.~\ref{fig:H6-pyrene_road}), the retrieved fragmentation rate of each by-product has a high value, which shows that not only the parent but also the fragments are efficiently dissociated.
We have identified four primary fragmentation channels that involve a large number of hydrogens (five to seven atoms) and some carbons (one to three atoms), namely 'CH$_5$', 'C$_2$H$_5$', 'C$_2$H$_6$', and 'C$_3$H$_7$', the quotes representing the fact that each channel can involve several neutral fragments.
We found that these pathways are consistent with available data in the literature. First, the dissociation pathways of small superhydrogenated PAHs have been explored using the analysis of imaging PEPICO experiments combined with molecular structure calculations. For small dihydro-PAHs the fragmentation was found to proceed via the loss of H and CH$_3$ \citep{west2014_dihydro}. For a higher level of hydrogenation, new channels open, which involves the loss of larger hydrocarbons, C$_2$H$_4$ and C$_3$H$_5$, as shown by \cite{diedhiou2019,diedhiou2020}. This was rationalized by these authors by invoking isomerization processes through H migration and ring opening. For the studied species, the different channels were determined to have comparable activation energies, $\sim$2\,eV or below for the H and CH$_3$ loss channels \citep{diedhiou2020}.
In addition, the fragmentation of H$_6$-Pyr$^+$ has been studied by \cite{gatchell2015} and \cite{wolf2016}, using collision induced dissociation and two-photon dissociation, respectively. The authors concluded  that the backbone fragmentation is increased because of the two carbon ring with additional H atoms, and that low dissociation channels involving CH$_3$ and C$_2$H$_4$ loss enter in competition with H-loss channels. \cite{wolf2016} in their multiple photon dissociation determined an internal energy of $\sim$6\,eV for dissociating H$_6$-Pyr$^+$. The larger energy reached in our experiment following the absorption of a 10.5\,eV photon is  therefore consistent with the loss of multiple fragments. 

\subsection{Molecular parameters involved in VUV photoprocessing}\label{sec:discuss_vuv_processing}

Several molecular parameters are at play in the VUV photofragmentation efficiency of any species, namely: (i) the VUV photoabsorption cross section, (ii) the competition with nondissociative mechanisms, and (iii) the involved dissociation energies.

The VUV photoabsorption cross sections of PAH cations have not been recorded so far. We therefore have to rely on theoretical photoabsorption cross sections computed by TD-DFT following \cite{malloci2004}.
The calculated cross sections exhibit sharp and strong resonances below 8\,eV, mostly resulting from $\pi^* \leftarrow \pi$ transitions, and a rising slope above 8\,eV, where $\sigma^* \leftarrow \sigma$ transitions dominate the photoabsorption spectrum (see supplementary Fig.~\ref{app:fig_sigma_abs}). The computed cross sections of the most symmetrical species, such as Pyr$^+$, H$_6$-Pyr$^+$, Cor$^+$ or C$_{23}$H$_{11}^+$, show large variations in the energy range around 10.5\,eV, while other species have smoother variations. This suggests that these species have more probability than the other species to have a much larger or smaller cross section at 10.5\,eV depending on whether or not the VUV photon energy is right on the peak of a strong electronic transition or not. This effect cannot be predicted since, with the used theory level, we cannot expect a better accuracy than $\pm~0.5$\,eV for the positions of the electronic transition energies. Therefore, in the following, we consider mean values for the photoabsorption cross sections, which are averaged values over 10 to 11\,eV energy range (see supplementary Table.~\ref{app:table_averaged_sigma_kabs}).
An additional point of interest concerns band broadening.
The lifetimes of excited cationic states have been measured for PAHs in this energy range; they fall in the range of a few to few tens of femtoseconds \citep{marciniak2015,herve2021}. This results in a natural band broadening in the range of 0.1 to 0.5 eV, which can be compared to the value of the FWHM of 0.15\,eV in Octopus simulations (see Sect.~\ref{sec:meth_(TD-)DFT calculations}). 
One should bear in mind, however, that bandwidths in the computed spectra are artificial and arbitrary. They could be made arbitrarily small by increasing the length of the simulations, and only their integrated band intensities have physical meaning. This means that, on the one hand, actual bands can be much broader, if they have for instance substantial vibronic structure, or undergo strong lifetime broadening, neither of which is included in theoretical vertical electronic spectra; on the other hand, if a given band is not subject to substantial physical broadening, it may be much sharper than it appears on our theoretical spectra.

It is generally accepted that the absorption of a VUV photon, if not leading to ionization, is followed by rapid internal conversion to the ground excited state with a lot of vibrational energy. This energy can then activate various processes, namely fragmentation, isomerization and radiative cooling.
In our experiment we can only track directly the fragmentation. In the absence of fragmentation the hot ion, after a potential isomerization step, turns back into a cold ion at the same mass but not necessarily with the same structure compared to the parent. This happens after a given cooling time (see Sect.~\ref{sec:meth_expcond}), which depends on the efficiency of the radiative cooling involving the emission of IR photons and eventually visible photons by the so-called recurrent fluorescence mechanism \citep{leger88,martin2015,saito20}. The competition between fragmentation and radiative cooling, which governs the stability upon photodissociation, can be indirectly studied from the comparison of the values of fragmentation rates with those of photoabsorption rates. As discussed in Sect.~\ref{sec:cooling}, Cor$^+$, [Cor-2H]$^+$, and C$_{23}$H$_{11}^+$, are the most stable species of this study.

A number of isomeric configurations have been computed in the case of dissociating PAH ions, more specifically for Pyr$^+$ \citep{simon2017, parneix2017} and [MePyr-H]$^+$ \citep{jusko2018}, which are of interest here.
Based on the analysis of the fragmentation maps, we could find only one clear case of isomerization given by Cor$^+$ (see Sect.\ref{sec:res_bare_PAH}). 
We determined that the highest fragmentation rate leading to [Cor-H]$^+$ is retrieved from a population of Cor$^{+*}$ isomer(s) and not from the standard Cor$^+$, for both presented hypotheses in Fig.~\ref{fig:maps_bare_PAHs}~(b). 
Cor$^{+*}$ is likely to be formed by H-shift leading to the formation of an aliphatic bond as discussed in \cite{trinquier2017_H_shifted_isomers} and \cite{castellanos2018_lab}. The latter authors concluded that the formation of such aliphatic sites is critical for the first fragmentation steps in PAHs. A similar process might therefore be active for other species in our study but could not be evidenced in the analysis of our data. This is mentioned in Sect.~\ref{sec:res_bare_PAH} for [Cor-2H]$^+$. We have also tested different possible isomerization schemes in relation with the C$_{17}$H$_{11}^+$ fragment. However, we could not find a solution that would satisfy our selection criteria (cf. Appendix~\ref{app:roadmap_building_explanation}.)

Finally, in the framework of statistical unimolecular dissociation the fragmentation probability depends on the value of the activation energy (AE) and on the density of vibrational states, which, at a given internal energy, depends on the molecular size.
The value of AE  corresponds to the minimal energy required to fragment the molecule in a given channel. Because of intramolecular vibrational redistribution, the fragmentation probability for a given molecular size is higher for lower values of AE. In addition, this probability becomes very unlikely when the values of AE get close to the absorbed energy. For aromatic CH bonds, a dependence of AE with the number of unpaired electrons was reported, with a value deduced from RRKM modeling of experimental PEPICO breakdown curves of 4.40\,eV for the loss of the first H, and of 3.16\,eV for the loss of the second \citep{west2018}. The authors derived a value of 2.8\,eV for aliphatic CH bonds in MePyr$^+$ \citep{west2018_MePyr} and in the range of 1.2--2.4\,eV for aliphatic CH bonds in superhydrogenated PAHs \citep{diedhiou2020}. Although these values might have some uncertainties, they constitute a consistent set of data since they were derived using the same experimental technique and analysis procedure. In particular they show that the weakest CH bonds in our study are for  H$_6$-Pyr$^+$. In the case of the dissociation of Cor$^+$ involving H-shift and the dissociation of an aliphatic CH bond, relevant values for AE have been calculated by \cite{castellanos2018_lab} and \cite{trinquier2017_H_shifted_isomers}. This leads to AE values of, typically, 3.6\,eV for the H shift and 2.2\,eV for the H loss from the formed aliphatic CH bond.

\section{Astrophysical application and implications} \label{sec:astro_implications}

\subsection{Competition between fragmentation and cooling}\label{sec:cooling}

In Sect.~\ref{sec:discuss_vuv_processing}, we discuss the different molecular parameters that can affect the lifetime of a given PAH under VUV irradiation. First, the photoabsorption cross sections vary by  typically 30\% around their mean values calculated at 10.5\,eV (see Table~\ref{app:table_averaged_sigma_kabs}). These variations are likely to be larger depending on whether there is a peak in the cross sections precisely in resonance with the VUV photon energy. This effect seems particularly important in the case of H$_6$-Pyr$^+$ for which we calculated a value of $\overline{k_\mathrm{abs}}$ of 1.54~$\times10^{-3}$~s$^{-1}$, which is about four times smaller than the value derived for $k_\mathrm{frag}$.
The second, and probably more critical parameter driving the stability is the competition between fragmentation and radiative cooling. This depends on both the AE value for a given fragmentation channel and the size of the PAH, which governs the delocalization of the vibrational energy. This effect explains why pyrene is less stable than coronene but also why aliphatic-substituted species are less stable than standard PAHs.

\begin{figure}[htbp]
    \centering
    \includegraphics[width=0.47\textwidth]{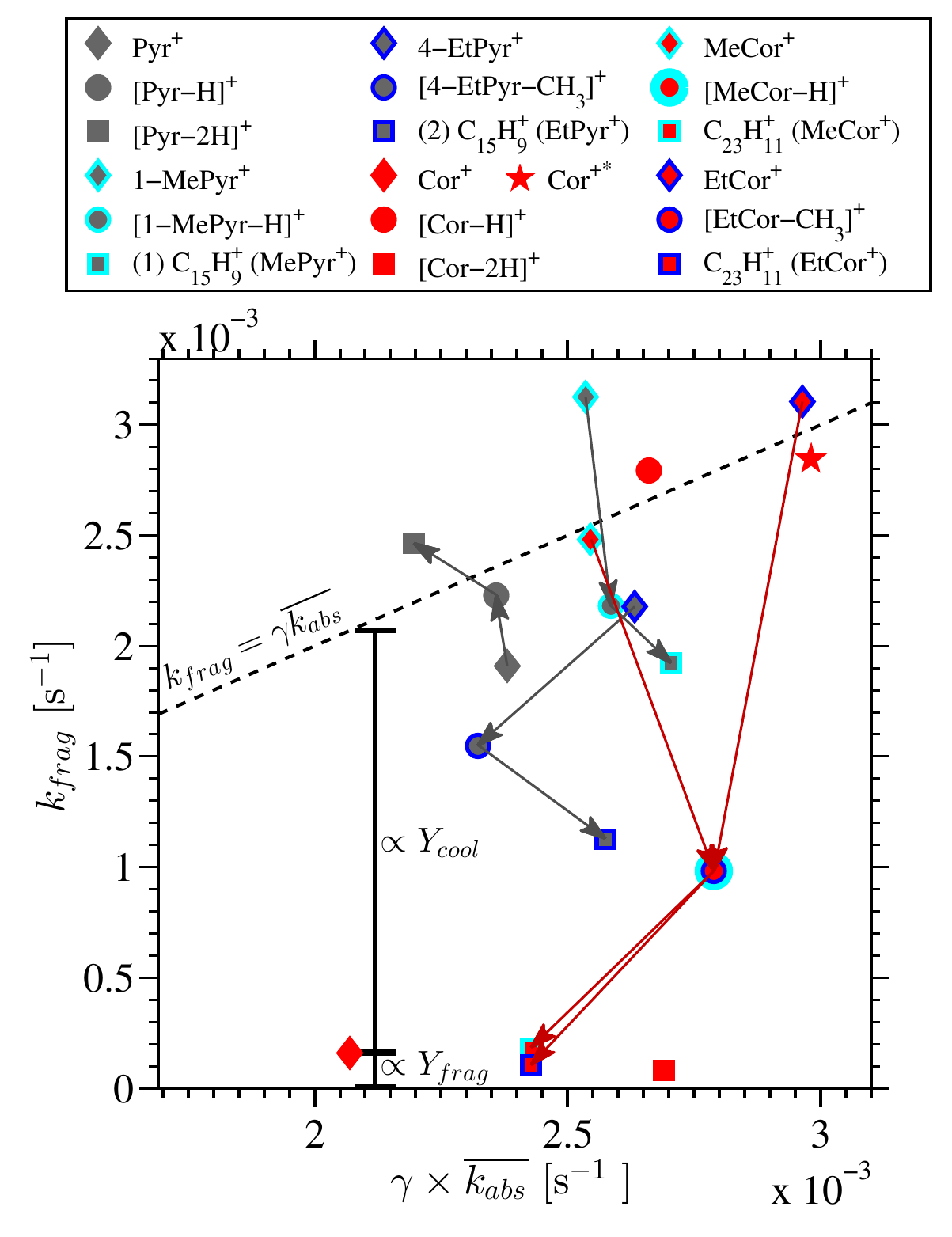}
    \caption{Fragmentation rate of the studied species (apart H$_6$-Pyr$^+$) as a function of the rescaled average absorption rate (by the $\gamma$ factor, see text for details). The pyrene-based (resp. coronene-based) species are in gray (resp. red). The cyan (resp. blue) edge indicates the methylated (resp. ethylated) species. The dashed line corresponds to $k_\mathrm{frag}=\gamma \overline{k_\mathrm{abs}}$. The fragmentation fraction ($Y_\mathrm{frag}$) and the cooling fraction ($Y_\mathrm{cool}$) are displayed for Cor$^+$ as an example. 
    For MeCor$^+$ and EtCor$^+$, the arrows point to the increased stability in the fragments formed by VUV processing.
    All the $Y_\mathrm{cool}$ values are given in Table~\ref{table:Y_cool}.}
    \label{fig:rescaled_cross_sections}
\end{figure}

To further discuss the relative stability of the different molecular families, we performed a quantitative comparison of the fragmentation rates with the photoabsorption rates. In order to take into account the uncertainty on both the theoretical photoabsorption cross sections and the VUV photon flux (see Appendix~\ref{app:VUV_source}), we empirically rescaled the photoabsorption rates as follows. We selected a group of the more fragile species, due to their smaller size and/or smaller values of AE for their fragmentation channels. This group consists of Pyr$^+$, [Pyr-H]$^+$, [Pyr-2H]$^+$, MePyr$^+$, EtPyr$^+$, [Cor-H]$^+$, MeCor$^+$, and EtCor$^+$. We exclude H$_6$-Pyr$^+$ from the analysis since its photoabsorption rate relative to the fragmentation rate is more difficult to rationalize. For the above-selected species we assumed that there is negligible competition with radiative cooling and therefore the fragmentation rates give a direct measurement of the photoabsorption rates. We could then calculate a corrective factor $\gamma$, with:
 \begin{equation}
     \gamma = \frac{1}{N} \sum_i^N \frac{k_\mathrm{frag}^{M_i}}{\overline{k_\mathrm{abs}^{M_i}} } ~~ 
    \mathrm{with}~~\left\{ \begin{array}{c}
        \overline{k_\mathrm{abs}^{M_i}}=\overline{\sigma _\mathrm{abs}^{M_i}} \times \phi_0 \\
         \\
       M_i \in \mathfrak{M}  \\
    \end{array}
    \right.~,
 \end{equation}
where $\overline{\sigma_\mathrm{abs}^{M_i}}$ is the average TD-DFT absorption cross section of the species $M_i$ reported in Table~\ref{app:table_averaged_sigma_kabs}, $\phi_0$ is the VUV flux and $\mathfrak{M}$ is the ensemble of selected species detailed above. We obtained a value of $\gamma \sim$ 2.0.
Then, we applied $\gamma$ to $\overline{k_\mathrm{abs}^{M_i}}$ for all the species for which we have computed a TD-DFT photoabsorption cross section and we plotted $k_\mathrm{frag}^{M_i}$ as a function of $\gamma \times \overline{k_\mathrm{abs}^{M_i}}$. The results are presented in Fig.~\ref{fig:rescaled_cross_sections}. 

Different groups can be identified in Fig.~\ref{fig:rescaled_cross_sections} on the basis of their yield of fragmentation, $Y_\mathrm{frag} = k_\mathrm{frag}/(\gamma \overline{k_\mathrm{abs}})$, or its complementary value, which is the yield of cooling, $Y_\mathrm{cool}= 1 - Y_\mathrm{frag} $. The first group contains the set of species that were selected to derive the $\gamma$ value and are therefore located around the $k_\mathrm{frag} = \gamma \overline{k_\mathrm{abs}}$ line. The second group gathers the most stable species, namely  Cor$^+$, [Cor-2H]$^+$, and C$_{23}$H$_{11}^+$ for which $Y_\mathrm{cool} \gtrsim 0.92$. Finally, the last group gathers the species in an intermediate regime, namely the species containing a methylene group formed from the fragmentation of alkylated PAHs, except from 1-MePyr$^+$, and the fragment C$_{15}$H$_9^+$, which is formed from the fragmentation of alkylated pyrene species. The latter ion contains a pentagonal cycle, similarly to  C$_{23}$H$_{11}^+$, which is formed in the fragmentation of alkylated coronene species. The corresponding values of $Y_\mathrm{cool}$ are provided in Table~\ref{table:Y_cool}. The arrows displayed in Fig.~\ref{fig:rescaled_cross_sections} show the remarkable trend of increasing stability upon 10.5\,eV irradiation (given by $Y_\mathrm{cool}$) from alkylated PAHs to the methylene form ([MePAH-H]$^+$, [EtPAH-CH$_3$]$^+$) and finally toward the five-membered ring structure. 

\begin{table}[htbp]
\caption{\label{table:Y_cool} Cooling fraction ($Y_\mathrm{cool}$) of the most stable species. The uncertainty is estimated to be $\pm~0.15$.}
    \centering \small{
        {\renewcommand{\arraystretch}{1.5}
\begin{tabular}{M{2.6cm}M{0.9cm}M{2.6cm}M{0.9cm}} \hline\hline
\textbf{Species}                  & $Y_\mathrm{cool}$ & \textbf{Species}               & $Y_\mathrm{cool}$ \\ \hline

[1-MePyr-H]$^+$                   & 0.16       & Cor$^+$                        & 0.92       \\

(1) C$_{15}$H$_9^+$   (MePyr$^+$) & 0.29       & [Cor-2H]$^+$                   & 0.97       \\

                                  &            &                                &            \\
                                  
[4-EtPyr-CH$_3$]$^+$              & 0.33       & [MeCor-H]$^+$                  & 0.65       \\

(2) C$_{15}$H$_9^+$   (EtPyr$^+$) & 0.56       & C$_{23}$H$_{11}^+$ (MeCor$^+$) & 0.92 \\

                                   &             &                       &            \\
                                  
                                  &            & [EtCor-CH$_3$]$^+$             & 0.65       \\
                                  
                                  &            & C$_{23}$H$_{11}^+$ (EtCor$^+$) & 0.96     \\ \hline
\end{tabular}
           }}
\end{table}

\subsection{Fragment budget}\label{sec:res_frag_budget}

We have observed that the VUV photoprocessing triggers mainly H losses for the bare cationic PAH, with possibly some C$_2$H$_2$ losses when the PAH size is small, while, for aliphatic PAH derivatives, the main fragmentation step depends on the aliphatic bonds at play. Counter-intuitively, for methyl-PAHs (resp. ethyl-PAHs), this first VUV induced fragmentation step mainly leads to an H loss (resp. a CH$_3$ loss), which opens the way to additional C$_2$H$_2$ losses, thus inducing an alteration of the PAH skeleton. For H$_6$-Pyr$^+$, the VUV photoprocessing induces carbon losses which occur all around the PAH skeleton and involve up to three carbons following the absorption of a single 10.5\,eV photon. In order to draw a detailed count of the neutral fragments that result from the cascade of fragmentation of each species, we define an effective fragmentation rate for each channel involved in the cascade. This rate takes into account the successive fragmentation steps and it is defined iteratively as: 
\begin{equation}\label{eq5}
\left\{
    \begin{array}{cc}
       (1^\mathrm{st}) & k_\mathrm{eff}^{A\rightarrow B_i}=  R^{A\rightarrow B_i} k_\mathrm{frag}^{A} \\
                & \vdots\\
       (j^\mathrm{th}) & k_\mathrm{eff}^{A\rightarrow \cdots M_i\rightarrow M_j}=  \frac{\displaystyle \left(\sum_{n}^{}  k_\mathrm{eff}^{A\rightarrow \cdots M_n\rightarrow M_i}\right) \times R^{M_i\rightarrow M_j} k_\mathrm{frag}^{M_j}}{\displaystyle \sum_{n}^{}  k_\mathrm{eff}^{A\rightarrow \cdots M_n\rightarrow M_i} + R^{M_i\rightarrow M_j} k_\mathrm{frag}^{M_j}} \\
    \end{array}
\right.~,
\end{equation}
where $A$ is the parent cation, $B_i$ is one of its direct daughter species, $n$ is the index of the $M_n$ channels that are populating the $M_i$ channel, and $k_\mathrm{frag}^{M_j}$ is the fragmentation rate of the ion $M_j$. With this definition, the $(j+1)^\mathrm{th}$ effective rate is necessarily smaller than the $j^\mathrm{th}$ one. We can then define the total production rate of the X fragment, which reads:
\begin{equation}\label{eq6}
K_\mathrm{tot}^{X-\mathrm{loss}}=\sum_{M_i-M_j=M_{X-\mathrm{loss}}}^{N_{\mathrm{channel~of~}X-\mathrm{loss}}}  k_\mathrm{eff}^{A\rightarrow \cdots M_i\rightarrow M_j}~.
\end{equation}
It corresponds to the discrete sum of the effective fragmentation rates leading to the release of a given X fragment in gas phase (for a better understanding of Eqs.~\ref{eq5} and~\ref{eq6}, an example is given in Appendix~\ref{app:Ktot_explanation}). The values of $K_\mathrm{tot}^{X-\mathrm{loss}}$ for the studied PAH parent cations are listed in Table~\ref{table:Ktot}.
For H$_6$-Pyr$^+$ the irradiation time was 300\,s and not 1000\,s as for the other studied cations.
The data provided in Table~\ref{table:Ktot} can be summarized as follows. Concerning the bare species, the photodissociation cascade of Cor$^+$ has a production rate of H atoms, which is $\sim$10 times lower than the photodissociation cascade of Pyr$^+$. The production rate of carbon fragments is marginal (only a small contribution of 7\% from the Pyr$^+$ cascade). Carbon loss becomes much more efficient for species containing aliphatic bonds and with a value of $\left(\sum C\right)/\left(\sum H\right) \gtrsim 0.3$ for both pyrene and coronene-derived species. This loss involves C$_2$H$_x$ fragments and also CH$_x$ and C$_3$H$_x$ at the first fragmentation step (cf. bold values in Table~\ref{table:Ktot}) for ethylated species and H$_6$-Pyr$^+$. Since significantly lower AE are involved in the fragmentation of the species containing aliphatic bonds (cf. Sect.~\ref{sec:discuss_vuv_processing}), the fragmentation rate of these species proceeds at the photoabsorption rate whereas for subsequent steps there is more competition with cooling (cf. Sect.~\ref{sec:cooling} and Fig.~\ref{fig:rescaled_cross_sections}). This results in a strong production of fragments in the first fragmentation step for coronene-related species ($\sim$ 65\% of produced fragments) relative to pyrene-related species ($\sim$ 30\% of produced fragments). In the next step, the dissociation of species containing a methylene group involves an additional C$_2$H$_2$ loss, which is energetically favored by the formation of the pentagonal ring. This step is observed to be size dependent (lower fragmentation rates for larger molecules) as shown in Fig.~\ref{fig:rescaled_cross_sections}. We can foresee that, for larger sizes, the C-loss channels arise mostly from the dissociation of alkylated and superhydrogenated species.

\begin{table*}[htbp]
\caption{\label{table:Ktot} Production rate ($K_\mathrm{tot}^{X-\mathrm{loss}}$ in $10^{-3}$s$^{-1}$) and corresponding fraction (\%) of the total loss for the different X fragment channels produced in our VUV irradiation conditions. The C$_N$ list below each parent species corresponds to the number of carbons that belong to the detected fragments. The values in bold represent the contribution of the first fragmentation step. The $\sum C_nH_x/\sum H,H_2$ ratio corresponds to the production rate of carbonaceous fragments (CH$_x$, C$_2$H$_x$, etc.) relative to hydrogen (H, 2H, and H$_2$) fragments. The $\sum C/\sum H$ ratio corresponds to the production rate of C over H atoms. }
    {\centering {\small
    {\renewcommand{\arraystretch}{1.5}
    \begin{tabular}{c|*6{M{1.4cm}}|M{1.3cm}M{1.3cm}}
        \hline \hline 
        
        \backslashbox{\textbf{Parent}}{\textbf{Fragment}}          & \textbf{H}         &\textbf{2H or H$_2$}&\textbf{CH$_x$}   &\textbf{C$_2$H$_x$}&\textbf{C$_3$H$_x$} &\textbf{C$_4$H$_x$} & { \textbf{$\displaystyle \frac{\sum C_nH_x}{\sum H,H_2}$}}  & \textbf{{ $\displaystyle \frac{\sum C}{\sum H}$}} \\ \hline 
       
       \makecell{\textbf{ Pyr$^+$} \\ (C$_{16}$, C$_{14}$) }                    & \makecell{ \textbf{1.9}~+~2.8 \\ \textbf{38\%}~+~55\% }                        &          - & -                            & \makecell{0.3 \\  6\%}  & -     &   \makecell{0.06 \\  1\%}  &   0.07  & 0.13 \\ 
        
        \makecell{\textbf{H$_6$-Pyr$^+$}$^{(a)}$ \\  (C$_{16}$, C$_{15}$, \\ C$_{14}$, C$_{13}$, C$_{12}$)}            & \makecell{3.1 \\ 19\%}     & \makecell{\textbf{0.7}~+~4.2 \\ \textbf{5\%}~+~24\%}         & \makecell{\textbf{0.7} \\ \textbf{5\%}}                    & \makecell{\textbf{4.4}~+~0.9 \\  \textbf{7\%}~+~6\%}  &  \makecell{\textbf{2.2} \\  \textbf{14\%}}  & -   &  1.1$^{(b)}$ & 0.35$^{(b)}$\\ 
       
       \makecell{\textbf{MePyr$^+$ } \\ (C$_{17}$, C$_{15}$, C$_{13}$)}     & \makecell{\textbf{2.2}~+~4.4 \\ \textbf{23\%}~+~46\%}  & 
       
       \makecell{\textbf{0.2}~+~0.8 \\ \textbf{2\%}~+~8\%}          & -                            & \makecell{\textbf{0.8}~+~1.1 \\ \textbf{8\%}~+~12\%} & -            & - & 0.25 & 0.29 \\  
       
       \makecell{\textbf{EtPyr$^+$} \\ (C$_{18}$, C$_{17}$, \\ C$_{15}$, C$_{13}$)}                & \makecell{3.1 \\ 46\%}                       & \makecell{0.7 \\ 10\%}  &\makecell{\textbf{1.4} \\ \textbf{21\%}} & \makecell{0.9 \\ 14\%}  & \makecell{\textbf{0.6} \\ \textbf{9\%}}   & - & 0.78 & 0.42\\ 
        
        \makecell{\textbf{Cor$^+$} \\ (C$_{24}$)}                    & \makecell{\textbf{0.16}~+~0.34 \\ \textbf{32\%}~+~68\%}                     & -          & -                            & -            &               & - & 0 & 0\\ 
        
        \makecell{\textbf{MeCor$^+$} \\ (C$_{25}$, C$_{23}$)} & \makecell{\textbf{2.5}~+~0.5 \\ \textbf{68\%}~+~13\%}   & \makecell{0.07 \\ 2\%}          & -                            & \makecell{0.6 \\ 17\%}   & -      & -     &   0.20 & 0.29 \\  
        
        \makecell{\textbf{EtCor$^+$} \\ (C$_{26}$, C$_{25}$, C$_{23}$)}               & \makecell{0.7 \\ 16\%}                       & \makecell{0.31 \\ 7\%}  &\makecell{\textbf{3.0} \\ \textbf{64\%}}  & \makecell{0.6 \\ 13\%}  & -    & - &  3.4 & 0.36\\  \hline
        \end{tabular}
      
         }}}
         {\tiny $^{(a)}$ For H$_6$-Pyr$^+$, we have considered that the first CH$_5$-loss channel corresponds to a CH$_3$ loss plus a 2H or H$_2$ loss. This has been taken into account in the calculated fractions and ratios that are displayed here. 
         
         $^{(b)}$ These ratios are calculated for a total VUV irradiation time  of $t_\mathrm{irr}^\mathrm{VUV} = 300$\,s, which is shorter than the irradiation time of 1000\,s used for the other species.}
\end{table*}

\subsection{Astrophysical implications}\label{sec:astro}
The fact that the physical conditions in our experiment mimic those found in PDRs constitutes the strong point of the present work. The produced PAH cations are put in isolated conditions in the cryogenic ICR cell of PIRENEA and submitted to the irradiation of the VUV photons with a mean time between the absorption of two VUV photons of a few hundred seconds (cf. Sect.~\ref{sec:meth_expcond}). This timescale can be compared to a timescale of several hours in the NGC 7023 NW PDR, which was studied by \cite{montillaud2013} and tens of minutes in the brighter Orion Bar \citep{joblin2020}.

A first implication of the reported measurements concerns the fragmentation rate of aliphatic C-H bonds, which could be quantified here relative to the fragmentation rate of aromatic C-H bonds. For all the studied species, all the aliphatic C-H bonds that could contribute to the 3.4\,$\mu$m emission band, whether in methyl/ethyl groups or in superhydrogenated PAHs, are removed following the absorption of the first VUV photon (cf. Figs~\ref{fig:Me_Et_PAH_roads} and ~\ref{fig:H6-pyrene_road}). Although this trend could be attenuated for larger PAH sizes, it suggests that these aliphatic C-H bonds are immediately eroded when exposed to VUV irradiation. The presence of methylated compared to ethylated species might be favored for the following reason. Since the first step of dissociation of alkylated species leads to the formation of the same species carrying methylene sidegroups, it is not excluded that methyl sidegroups could be reformed by reactivity of these species with H atoms, at gas densities that are high enough that the reaction rate is competitive compared to the VUV photodissociation rates. Assuming an optimistic reaction rate of $10^{-9}$~cm$^3$~s$^{-1}$, we can estimate a reaction rate of 2.8\,h and 17\,min for a gas density of $10^{5}$ and $10^{6}$~cm$^{-3}$, respectively. Therefore, the reaction could be competitive at the dense interface of VUV irradiated PDRs \citep{joblin2018}, which calls for a measurement of this reaction rate. In fact, the observation of the 3.4\,$\mu$m AIB, if carried by the fragile methylated CH bonds, is favored by both the possibility to reconstruct these bonds by gas-phase chemistry and that of replenishment of the alkylated PAHs by evaporation of mixed aromatic/aliphatic nanograins in dense PDR interfaces as shown by \cite{pilleri2015} and \cite{bouteraon2019}. There is another possibility that we cannot exclude to account for the 3.4\,$\mu$m band, which is related to the formation of aliphatic CH bonds due to H-migration, a process that was evidenced in the case of Cor$^+$ and invoked as a general process in the fragmentation of PAHs \citep{castellanos2018_lab}. Interestingly, an earlier theoretical study by \cite{jolibois2005} suggested that such a process, which is intimately coupled to photodissociation, could result in building emitters at 3.4\,$\mu$m. The efficiency of this process still needs to be determined.
Finally, regarding superhydrogenated PAHs, the case of H$_6$-Pyr$^+$ suggests that it is harder to maintain these species in the gas-phase, due to both some destruction of the carbon backbone with photoprocessing and a higher photoabsorption cross section. Although experiments have to be performed on larger PAHs than pyrene, superhydrogenated PAHs appear thus less promising as carriers for the 3.4\,$\mu$m AIB.

The fragmentation cascades observed upon VUV irradiation selectively form the most stable species. PAHs carrying a methylene group are found to have an increased stability relative to alkylated PAHs, yet this is not enough to favor their presence in irradiated PDRs. On the other hand, their fragments containing a pentagonal ring are at least as stable as the bare PAHs. This suggests that, in an evolutionary chemical scenario, stable PAHs resulting from either methylated or ethylated species would retain the scars of their origin in the form of one or more pentagonal rings at their edge, where the methyl or ethyl group(s) was (were) attached. For ethylated-PAHs, there is also a small route that leads to the formation of species at least as stable as the bare PAHs, which likely carry an ethynyl group (cf. Sect.~\ref{sec:res_alphatic_PAH} and Fig.~\ref{fig:Me_Et_PAH_roads}). These results are in line with earlier studies \citep{rouille2015, rouille2019}, which suggest that ethynyl-substitued PAHs are potentially good candidates for astroPAHs, if they can be formed efficiently as discussed by \cite{rouille2019}.

The formation of pentagonal rings upon energetic processing was demonstrated spectroscopically, in the case of small PAHs containing two to three rings \citep{bouwman2016, petrignani2016, deHaas2017}. The facile route toward the pentagons upon energetic processing of small PAHs led the latter authors to suggest that species composed of both pentagonal and hexagonal cycles might therefore be abundant in PDRs. Our study validates this proposal for physical conditions relevant to PDRs and for larger PAHs.
It is therefore interesting to study if these species have specific bands that would help identifying them in space \citep{galue2014, bouwman2020}. We also remark that both the pyrene and coronene derivatives carrying pentagonal rings that we studied here, show an overall IR activity systematically stronger than their respective parent species, in theoretical vibrational spectra, while they do not show large systematic differences in their calculated photoabsorption spectra at $\sim$10.5\,eV. This may contribute to their increased stability relative to the regular PAHs, as seen in the case of C$_{15}$H$_{9}^+$ (isomer (2)) relative to Pyr$^+$ (see Fig.~\ref{fig:rescaled_cross_sections}).

Finally, the UV-processing of small carbonaceous grains and PAHs has been proposed as a plausible (top-down) mechanism to account for the observed abundances of hydrocarbons (in particular C$_2$H and c-C$_3$H$_2$ seen by radiotelescopes) in the first layers of PDRs exposed to a low VUV flux \citep{pety2005, guzman2015}. The release of small hydrocarbons when hydrogenated amorphous carbon grains (a–C:H) are exposed to VUV photons has been studied in the laboratory \citep{alata2015,dartois2017}. The authors measured a photoproduction yield of 96.5\% for H$_2$ and 3\% for CH$_4$. Larger fragments C$_2$H$_x$, C$_3$H$_x$, and C$_4$H$_x$ were found to be produced with smaller abundances: 17, 7.7, and $\leq$3\% relative to CH$_4$. Still, such a production rate was found to be compatible with observations assuming continuous replenishment of fresh material in the UV-irradiated PDR. Similary, our results suggest that the production of hydrocarbons by top-down chemistry is consistent with the overall picture of evolutionary PAH chemistry involving nanograin evaporation followed by the destruction of aliphatic sidegroups, as suggested by \cite{pilleri2015}. As summarized in Sect.~\ref{sec:res_frag_budget}, CH$_x$, C$_2$H$_x$, C$_3$H$_x$ can be produced efficiently relative to H, 2H, or H$_2$ loss. In addition the involved cross section ($\sim$10$^{-16}$\,cm$^{2}$) is several orders of magnitude larger compared to the photodestruction cross section of a–C:H, which has been evaluated to be $\sim 3 \pm 0.9 \times 10^{-19}$\,cm$^{2}$ \citep{alata2014}, comparable to the value of $\sim 8 \pm 2 \times 10^{-20}$\,cm$^{2}$ more recently obtained for superdeuterated coronene films \citep{mennella2021}. This large difference can be ascribed to the fact that the process occurs in condensed phase in these studies, whereas it occurs on isolated molecules in our experiment. The dominant mechanism of fragmentation in isolated molecules is unimolecular reactions in statistical equilibrium after the molecule reaches the ground electronic state in a highly excited vibrational state (cf. Sect.~\ref{sec:discuss_vuv_processing}). This is suppressed in condensed phase, where vibrational energy is swiftly redistributed to the whole solid, on timescales much shorter than those required for fragmentation \citep{boutin1995}. This makes the mechanism reported here an especially competitive top-down mechanism in hydrocarbon chemistry, resulting in much less destruction of the precursors needed to account for the observations.

\section{Conclusion}\label{sec:con}
The weakening of the carbon backbone due to the presence of aliphatic CH bonds was initially reported by \cite{gatchell2015} for superhydrogenated pyrene species in collision with fast He atoms. Such conditions would apply to supernova shock waves. We extend here the study to the conditions of PDRs where the interaction with VUV photons is the main energetic process altering PAHs. We also consider the role of aliphatic sidegroups in addition of super-hydrogenation and include the larger PAH coronene. Overall, our experiments confirm the weakening of the carbon backbone in all cases where aliphatic bonds are present. However, in the case of alkylated PAHs, only the cycle that carries the sidegroup is affected and the fragmentation turns that cycle to a pentagonal ring, leading to the formation of a PAH that is found to be at least as stable, if not more, compared to the bare PAH. In the case of superhydrogenated PAHs, although we have more limited data, we can foresee that each additional H is susceptible to cause some damage, and increasing their number increases the probability to lose more C atoms \citep{diedhiou2020}, which severely limits the lifetime of these species in PDRs.  \cite{reitsma2014} observed that superhydrogenated coronene cations favorably lose their H atom to stabilize the backbone under a soft X-ray excitation. This conclusion is surprising but has still to be tested under VUV irradiation. In general, it would be interesting to extend our measurements to larger PAHs, which are more likely to survive in PDRs \citep{montillaud2013, andrews2015}. The production of small hydrocarbons in relation with the photoprocessing of aliphatic PAH derivatives has been evaluated and we can conclude that this process is much more competitive to contribute to hydrocarbon chemistry in PDRs relative to the processing of larger a:C-H grains. Finally, this work provides more evidence on the importance of isomerization processes upon VUV processing. Of particular interest here is the H-migration process, which might provide another way to account for the observed 3.4\,$\mu$m AIB. In our experiment it was evidenced only for Cor$^+$ but we cannot exclude that it is involved in some other cases, as suggested by \cite{castellanos2018_lab}. Determining to which extent this process generalizes as suggested in earlier studies \citep{castellanos2018_lab}, is clearly an aspect that should be tackled by further experiments.

\begin{acknowledgements}
We acknowledge L. Noguès, O. Coeur-Joly, and D. Murat for their technical support on the PIRENEA setup including its recent upgrade. We acknowledge funding from the European Research Council under the European Union's Seventh Framework Programme ERC-2013-SyG, Grant Agreement no. 610256, NANOCOSMOS. This project was also granted access to the HPC resources at the CALMIP supercomputing centre under project P20027. G. M. thanks S. Leurini, G. Malloci, and A. Bosin for making available their workstations for some of the calculations.
\end{acknowledgements}

%%% REFERENCES
\bibliographystyle{aa}
\bibliography{VUVPAH_aliph}

\begin{appendix}

\section{Additional diagnostics to estimate the VUV flux}\label{app:VUV_source}

\begin{figure}[htbp]
    \centering
    \includegraphics[width=0.5\textwidth,valign=c]{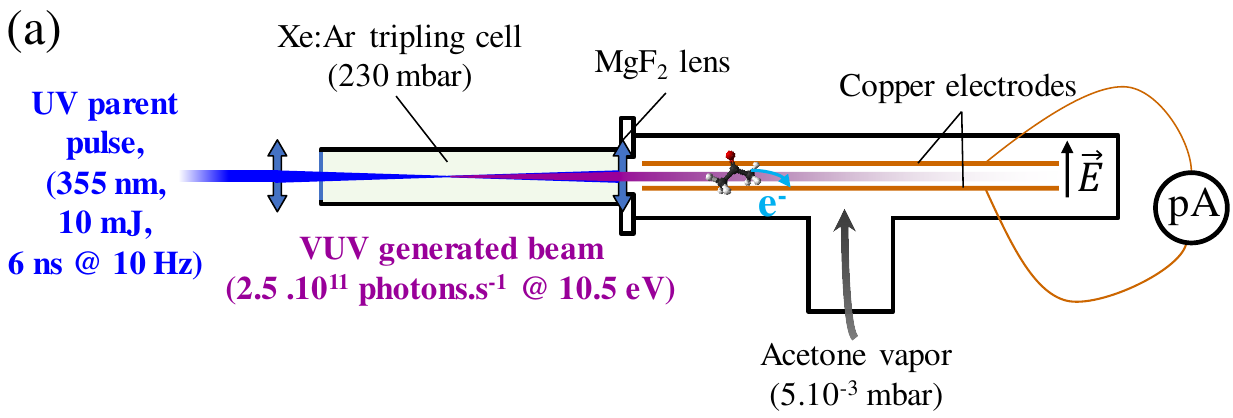} \includegraphics[width=0.33\textwidth,valign=c]{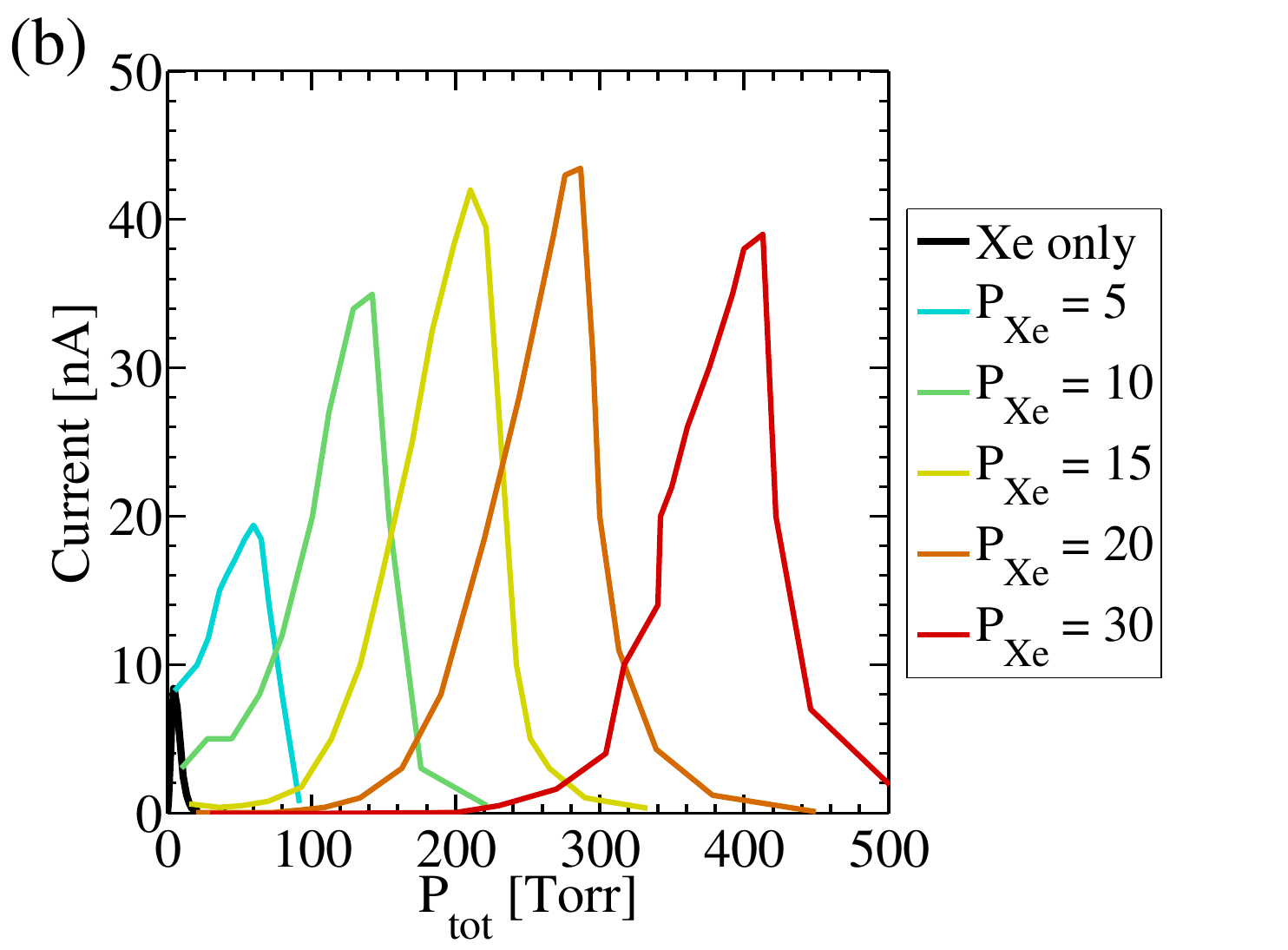} \includegraphics[width=0.14\textwidth,valign=c]{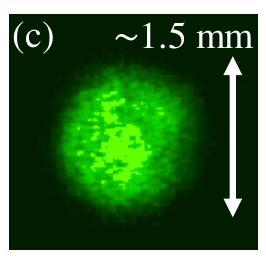} 
    \caption{Characterization of the VUV source. (a) Setup to record the number of VUV photons generated per second. (b) Calibration for 10 mJ UV pulses. (c) Image of the VUV beam profile.}
    \label{app:fig_VUV_calib}
\end{figure}

The developed VUV source is based on the frequency tripling of a 355\,nm UV-laser in a cell containing a Xe:Ar gas mixture. It has the advantage of being a table-top setup but the drawback is that the VUV photon energy is not tunable ($h\nu_\mathrm{VUV} = 10.5$\,eV). The conversion efficiency of such a system is on the order of a few $10^{-6}$ with a saturation threshold at high input energy per UV pulse. We calibrated this source with the diagnostic presented in Fig.~\ref{app:fig_VUV_calib}~(a). This latter consists in a 25 centimeter-long cell, which is filled with acetone vapor ($P_\mathrm{acetone} \approx 5.10^{-3}$~mbar), just after the tripling cell. Acetone is ionized by the VUV beam but not by the 355\,nm laser pulses. The photoemitted electrons are then attracted and collected by a copper electrode located on the side of the VUV beam and the current is measured with a pico-amperemeter. The high pressure of acetone allows us to make the hypothesis that all the VUV photons are absorbed before the end of the acetone vapor cell. The recorded current is proportional to the number of photons per second. We calibrated the VUV source by measuring the current as a function of the total pressure ($P_\mathrm{tot}= P_\mathrm{Xe} + P_\mathrm{Ar}$), for several initial Xe pressures ($P_\mathrm{Xe}$) and for an energy of 10 mJ per parent pulse (Fig.~\ref{app:fig_VUV_calib}~(b)). The optimum was found for an initial $P_\mathrm{Xe} \sim 17$\,Torr and a ratio of typically 1:11 relative to Ar. These results are consistent with the literature \citep{lockyer1997}. We could derive $N_\mathrm{VUV} = 2.5~\times~10^{11}$~photon.s$^{-1}$, a value that is certainly underestimated because the collection efficiency of photoelectrons is below 1.

After calibration, the VUV source was coupled to the PIRENEA setup. Inside PIRENEA, the VUV beam is viewed thanks to a screen that is localized after the ICR cell, at the end of the setup. This screen (coated with a solid coronene deposit) converts the absorbed VUV light into visible light. The resulting signal is imaged by a camera that allows the diameter of the beam to be measured (see Fig.~\ref{app:fig_VUV_calib}~(c)) and the beam stability to be checked in term of pointing, profile, and intensity. In particular, the radius was estimated to be $r \sim 0.75$\,mm, with a quasi 'flat-top' profile, which enables a homogeneous VUV excitation of the ion cloud. 
The flux is defined by $\phi _0 = N_\mathrm{VUV}/(\pi r^2)$ and, with the measured parameters, we obtained $\phi_0~\approx~1.4~\times~10^{13}$~photon~s$^{-1}$~cm$^{-2}$ for a 'flat-top' profile. Although great care was taken to obtain similar VUV beam conditions from day to day, some drifts and fluctuations were observed. In particular, we could observe that the beam profile was drifting and fluctuating from flat-top to Gaussian-like. This effect tends to decrease the effective radius of the VUV beam and thus to increase the VUV flux. For instance, a decrease by 20\% (150~$\mu$m) in the effective radius induces an increase of about 56\% in the VUV flux.

Both the error on the measurement of the photoelectron current and the error linked to the drifts and fluctuations of the VUV beam radius suggest that the value determined for the VUV flux is underestimated. This is somehow confirmed by the value of the coefficient $\gamma~\sim~2$ (see Sec.~\ref{sec:cooling}), which indicates that the estimated photon flux is too low by a factor of about two. Part of this factor can also be ascribed to uncertainties in the theoretical photoabsorption cross sections at 10.5\,eV. More specifically, we can estimate an interval of confidence for the measured VUV flux by taking into account (i) a minimal detection efficiency of 0.8 for the 'acetone cell diagnostics' and (ii) fluctuations on the beam radius of about 20\% (positive and negative). These combined errors then give: $\phi_0~=~1.4_{-0.4}^{+1.4}~\times~10^{13}$~photon~s$^{-1}$~cm$^{-2}$ which leads to an interval of -30\% and +96\% around $\phi_0$.

\section{Analyzing the kinetic curves and building the fragmentation maps}\label{app:roadmap_building_explanation}

\begin{figure*}[htbp]
    \centering
    \includegraphics[width=0.45\textwidth,valign=c]{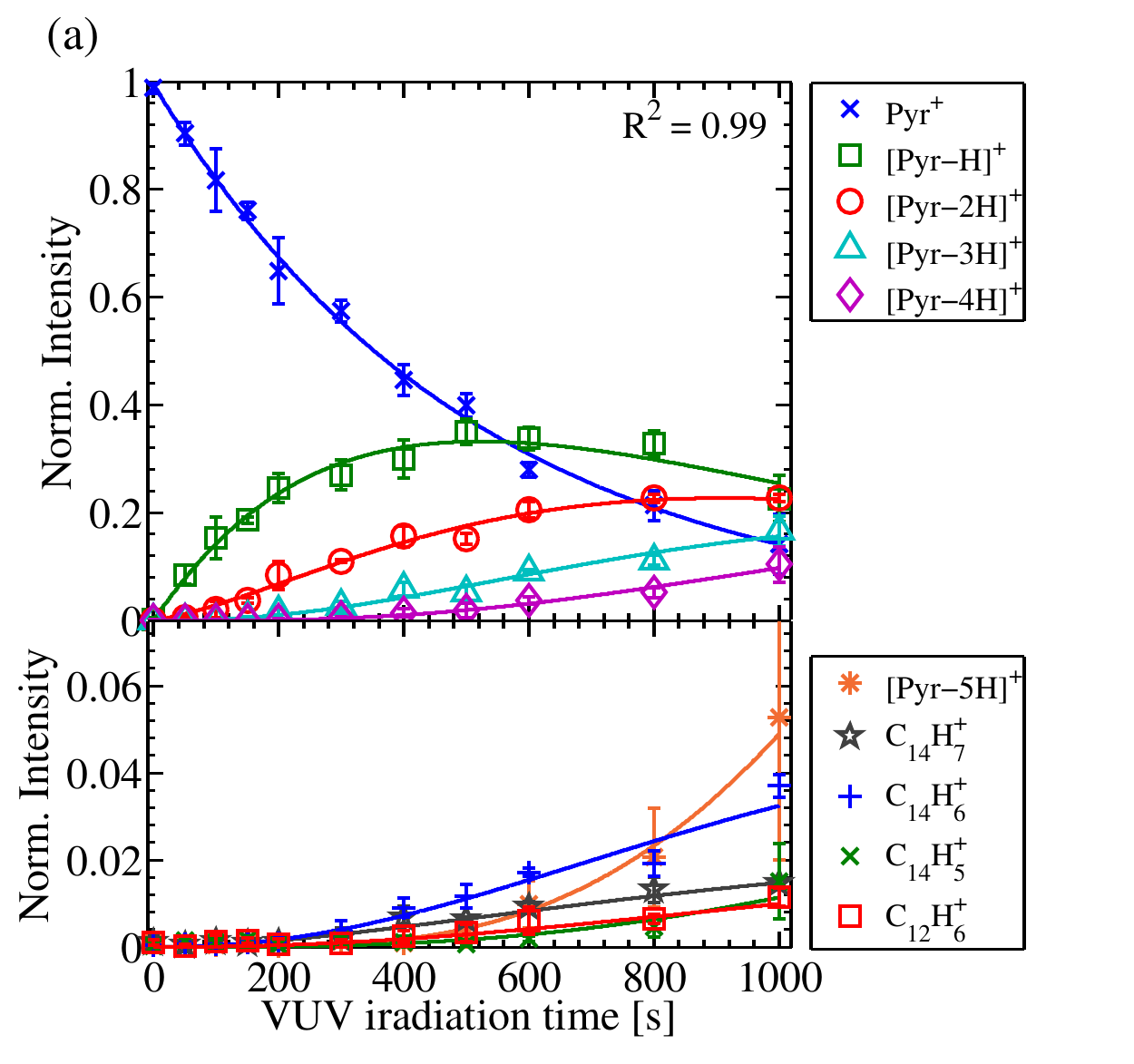} \includegraphics[width=0.45\textwidth,valign=c]{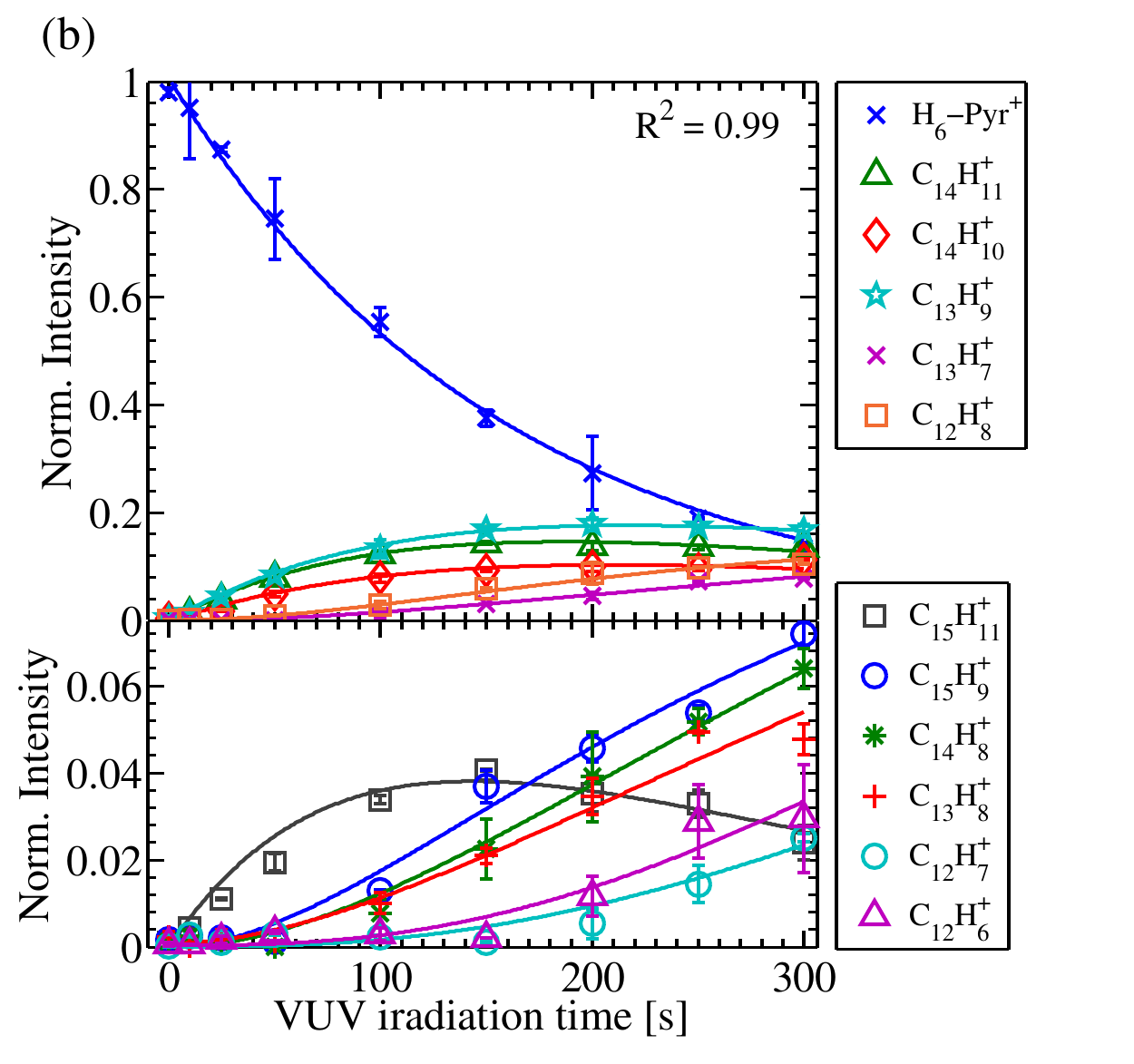} 
    \includegraphics[width=0.45\textwidth,valign=c]{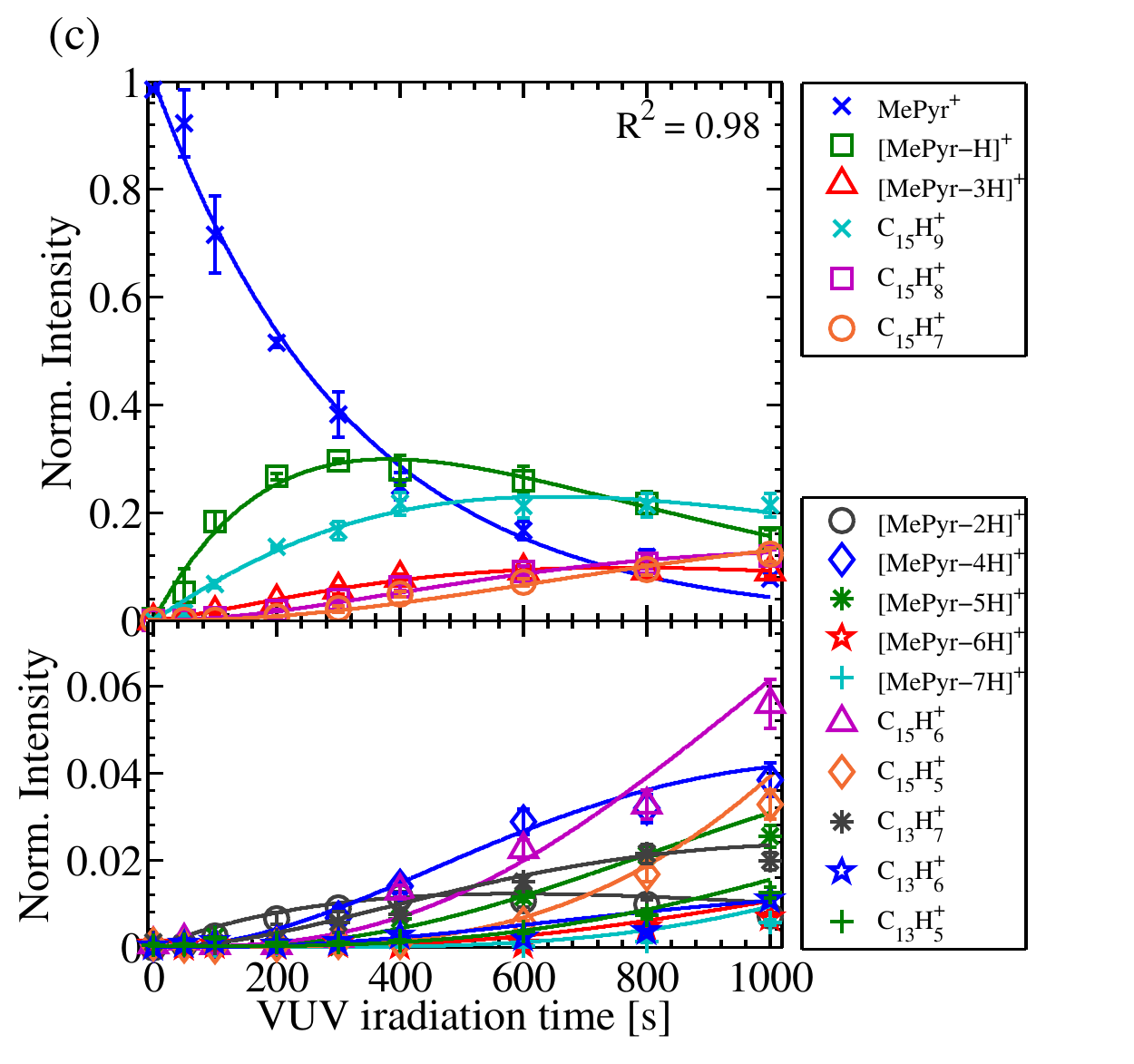}
    \includegraphics[width=0.45\textwidth,valign=c]{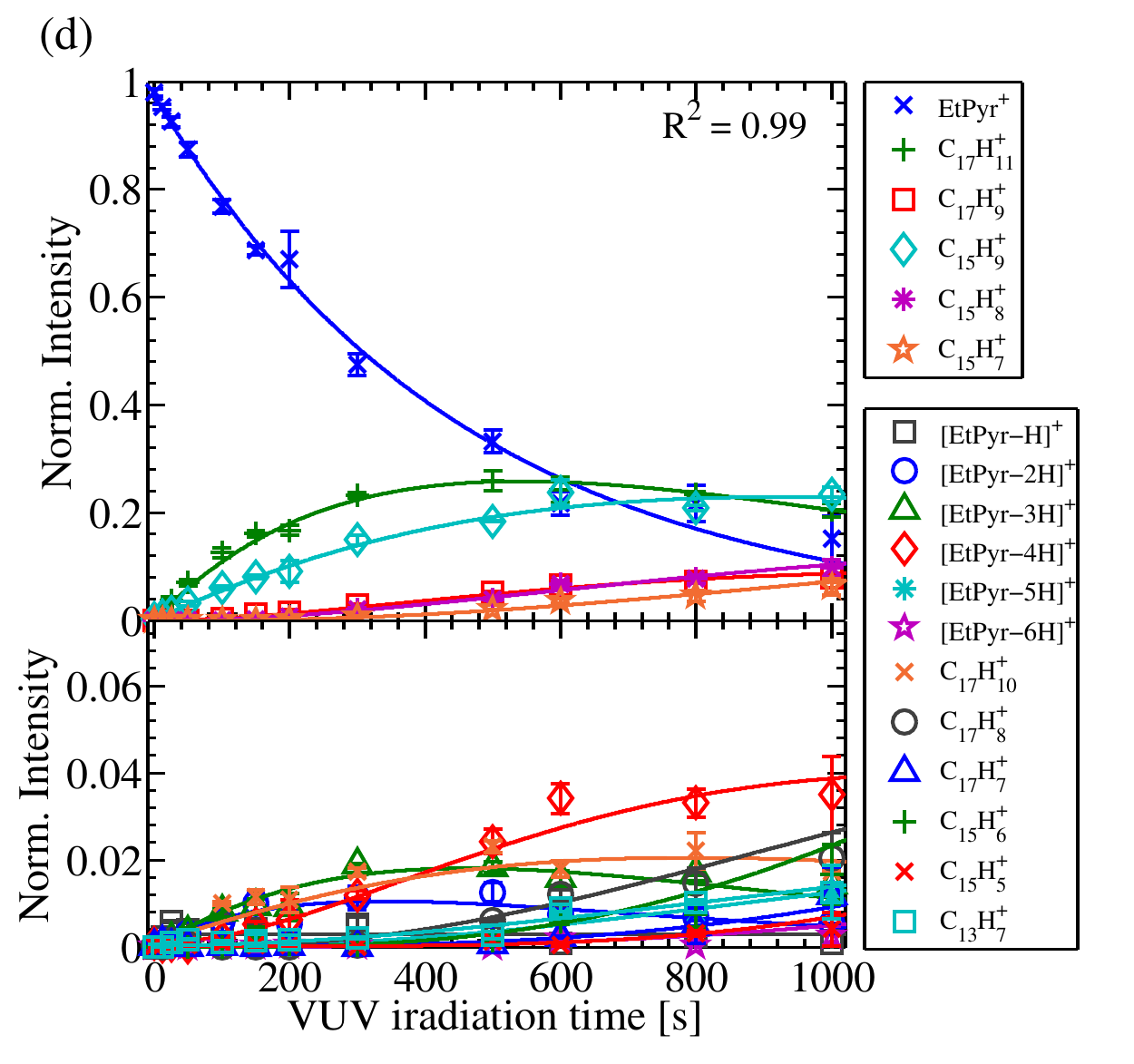}
    \includegraphics[width=0.45\textwidth,valign=c]{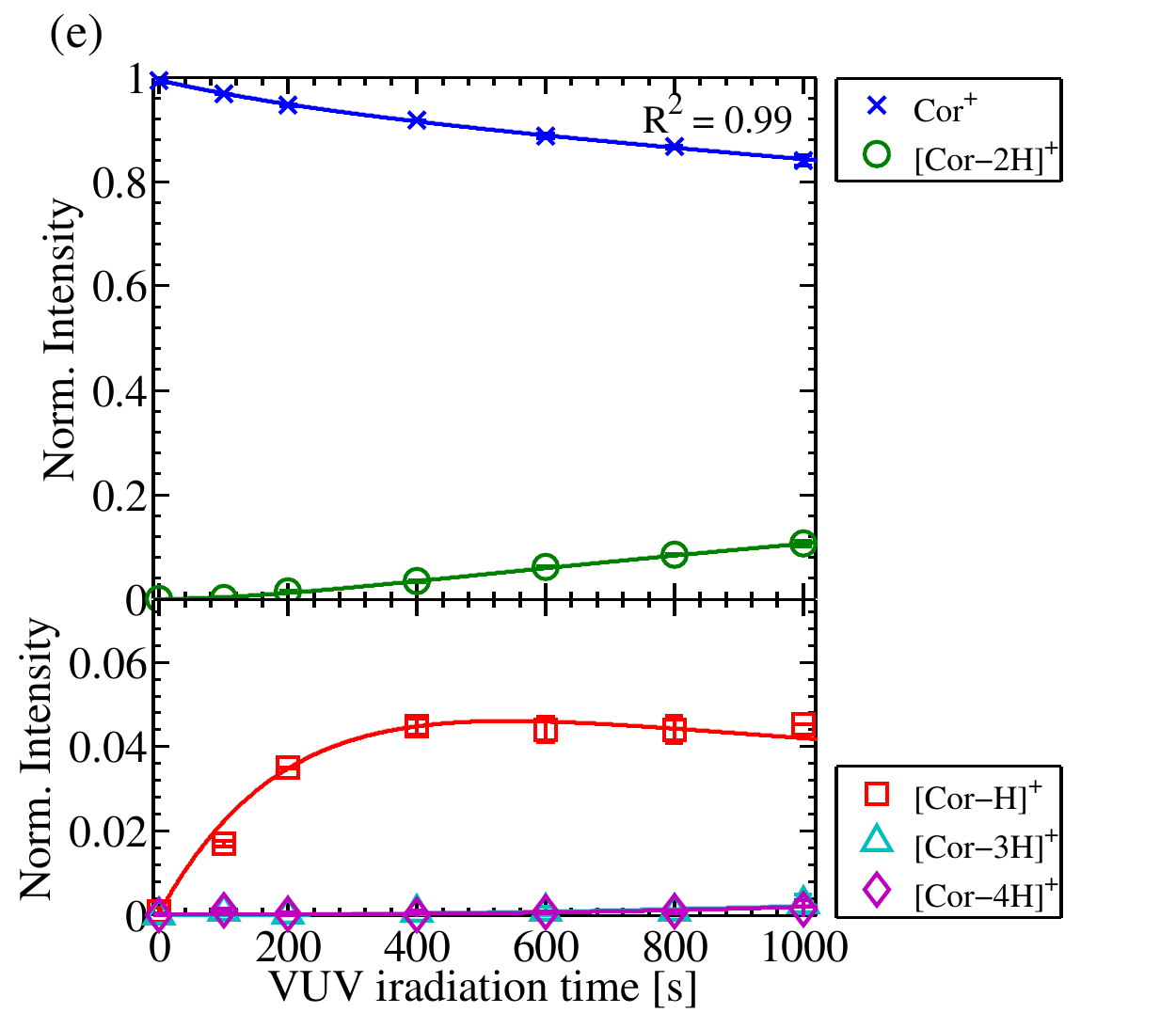}
    \includegraphics[width=0.45\textwidth,valign=c]{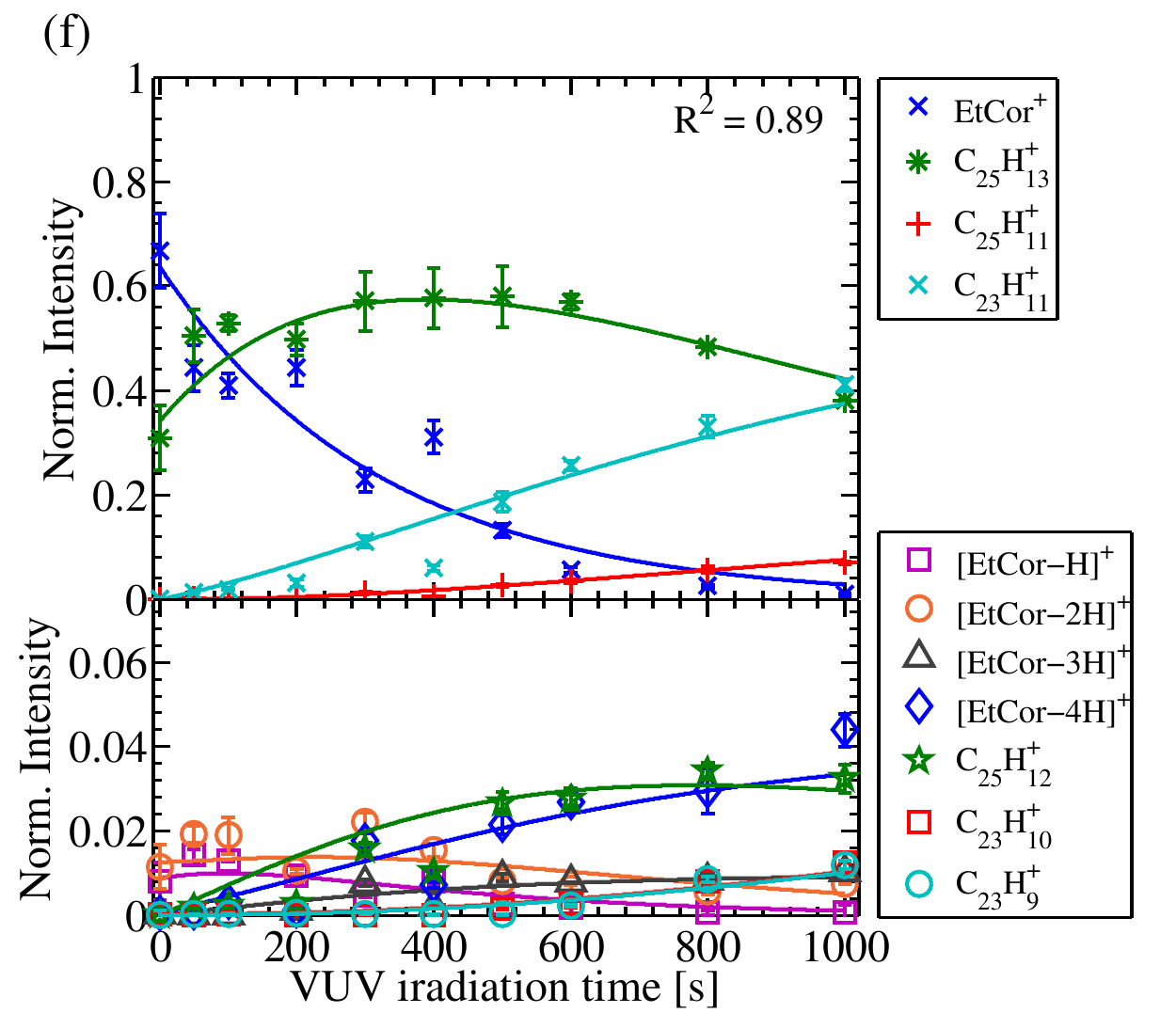}
    \caption{Kinetic curves of the studied PAH cations and their fragments: (a) Pyr$^+$, (b) H$_6$-Pyr$^+$, (c) MePyr$^+$, (d) EtPyr$^+$, (e) Cor$^+$, (f) EtCor$^+$ (MeCor$^+$ is shown in Fig.~\ref{fig:Me_Et_PAH_roads}). For each figure, the top panel shows the main channels while the bottom panel is a zoom on the minor channels. The solid curves correspond to the fitting functions that were derived with the procedure described in Sect.~\ref{sec:meth_data_analysis}. The derived photofragmentation maps are shown in 
    Figs. \ref{fig:maps_bare_PAHs},~\ref{fig:Me_Et_PAH_roads}, and \ref{fig:H6-pyrene_road}.}
    \label{app:fig_kinetics_all_PAHs}
\end{figure*}

As stated in Sec.~\ref{sec:meth_data_analysis}, the analysis of the measured kinetic curves has to be performed carefully, case by case, in order to retrieve the fragmentation rates and branching ratios included in the fragmentation maps (Figs.~ \ref{fig:maps_bare_PAHs},~\ref{fig:Me_Et_PAH_roads}, and~\ref{fig:H6-pyrene_road})
from the fit of the experimental data obtained by solving the system of differential equations (Eq.~\ref{eq2}). Through the analysis, we made sure that these maps: (i) followed the observed logic between the kinetics of the different channels, (ii) were in coherence with the results of previous studies, (iii) had solutions that fitted the data reasonably well (R$^2 \geq 0.98$, apart for EtCor$^+$ where R$^2=0.89$), and (iv) permitted parameters that are physically relevant to be extracted. The procedure consisted firstly in observing the normalized-to-one signal of all the channels together on a same graph for a studied species, such as in Fig.~\ref{fig:MeCor_kinetic_evol}~(b) for MeCor$^+$. By estimating the derivative at $t_\mathrm{VUV} = 0$ and/or the inflexion moment of the kinetic curve, we could rank the channels in term of timing of appearance. This gave a logical order to respect for the construction of the fragmentation map. In particular, for some channels, the derivative at $t_\mathrm{VUV} = 0$ is 'strongly' positive and it only decreases all along the kinetics, which means that these channels are directly populated from the parent cation. These observations reveal some groups of fragments that can be classified as 'primary', 'secondary', 'tertiary', etc. channels. We also used the relation between fragment masses to constrain the parent/daughter links. Obviously a lower mass fragment cannot populate a higher mass fragment and a fragment owning a certain number of H atoms cannot populate a fragment with a higher number of H atoms. Thanks to this kind of constraints we could build a first simple fragmentation map within the assumption that the 10.5 eV photons were absorbed sequentially by the parent and its fragments. Then, we checked that the proposed paths were in agreement with previous studies, which have reported the appearance energies of some PAH fragments or the existence of certain losses for the studied species \citep{west2014_dihydro,west2018,diedhiou2020}. If the solution functions did not fit the data well, we iteratively added new depopulation and population terms in Eq.~\ref{eq2}, guided by educated guesses. This was done within the logic that is aforementioned with the aim of getting a satisfactory fitting solution even for the weakest signals. If some terms were not necessary, they led to a negligible fragmentation rate. On the contrary, very high retrieved $k_\mathrm{frag}$ values were not acceptable because fragmentation rates cannot be faster than photoabsorption rates. A high $k_\mathrm{frag}$ indicates that some paths are missing around the channel of interest since these paths could redirect the excess of input population. In particular, when a retrieved $k_\mathrm{frag}$ exceeds 3 $\overline{k_\mathrm{abs}}$, it indicates that there may be an issue in the fragmentation map since we estimated that the rescaling factor between these two quantities is $\gamma \sim 2$. Moreover, methylated and ethylated PAH derivatives have fragments in common, thus we used common paths between the maps of these species. In the following, we give some details of the fragmentation map construction for each studied species (see Fig.~\ref{fig:maps_bare_PAHs}, ~\ref{fig:Me_Et_PAH_roads}, and \ref{fig:H6-pyrene_road}). The retrieved fitting curves are shown in Fig.~\ref{fig:MeCor_kinetic_evol} for MeCor$^+$ and in Fig.~\ref{app:fig_kinetics_all_PAHs} for the other species.

\paragraph{\textbf{Pyr$^+$.}}
The main paths of Pyr$^+$ were quite straightforward to construct as they correspond to H-loss channels for which there is a sequential appearance from Pyr$^+$ to [Pyr-5H]$^+$. The paths that connect C$_{14}$H$_7^+$, C$_{14}$H$_6^+$, and C$_{12}$H$_6^+$ are found to be uncertain. Indeed, their appearance time is around the [Pyr-3H]$^+$ appearance time, which means that they may be linked to [Pyr-2H]$^+$ or around. We also used the results obtained by \cite{west2014_pyr} to set up the paths presented in Fig.~\ref{fig:maps_bare_PAHs}~(a). The C$_{14}$H$_7^+$ fragment was not detected in \cite{west2014_pyr}, which may indicate that the path leading to this species may be uncommon. 

\paragraph{\textbf{MePyr$^+$ and EtPyr$^+$.}}
For MePyr$^+$, the [MePyr-H]$^+$, C$_{15}$H$_9^+$, [MePyr-2H]$^+$, and [MePyr-3H]$^+$ channels show a positive derivative at $t_\mathrm{VUV} = 0$ (with lower and lower values in the announced order). As these derivatives are not similar, we inferred that lower values result from a combination of additional delayed paths for instance going through a first intermediate fragment. For this reason, we added the 'transverse' paths that populate  C$_{15}$H$_9^+$, [MePyr-2H]$^+$, and [MePyr-3H]$^+$ from [MePyr-H]$^+$. We found that the channel [MePyr-2H]$^+$ was especially constraining for the fragmentation map because any wrong path around this channel was leading to an inconsistently high $k_\mathrm{frag}$ value (around 30). This issue was fixed in the proposed fragmentation map Fig.~\ref{fig:Me_Et_PAH_roads}~(a). As mentioned in the main text, we also tried to setup an isomerization scheme for the MePyr$^+$ map. In particular, we considered two populations of C$_{17}$H$_{11}^+$: one leading to the sequential H-loss route ($M_{214} \rightarrow M_{213} \cdots$) and the other one leading to the C$_{15}H_{9}^+$ fragment. This scheme was not appropriate since it could not properly fit the rising edge of the C$_{15}H_{9}^+$ kinetic curve. We also tested the impact of transverse paths between isomers and fragments, which resulted in a good fit but with some of the retrieved k$_\mathrm{frag}$ values being too high (>20) to be acceptable.

The fragmentation map of EtPyr$^+$ was built by mimicking the one of MePyr$^+$ (see Fig.~\ref{fig:Me_Et_PAH_roads}~(b)). A sequential H-loss route ($M_{230} \rightarrow M_{228} \rightarrow M_{227} \rightarrow M_{226} \rightarrow M_{225} \rightarrow M_{224}$) was needed but independent from the H-loss route that starts from C$_{17}$H$_{11}^+$ ($M_{215} \rightarrow M_{214} \rightarrow M_{213} \rightarrow M_{212} \rightarrow M_{211}$). In the case of MePyr$^+$, these H-loss channels could be connected to both C$_{17}$H$_{11}^+$ and the parent.
Similarly to MePyr$^+$, the $M_{214}$ channel of EtPyr$^+$ is a strong constraint for the map. Finally, we neglected the $M_{229}$ channel because it was leading to too high values of $k_\mathrm{frag}$, while its signal was below 0.5\% and with strong fluctuations. The final retrieved map and fitting functions remain very satisfactory without taking into account this channel.

\paragraph{\textbf{H$_6$-Pyr$^+$.}}
In the case of H$_6$-Pyr$^+$, we could define groups of fragments that have the same kinetic behavior. Four channels (C$_{15}H_{11}^+$, C$_{14}H_{11}^+$, C$_{14}H_{10}^+$, and C$_{13}H_{9}^+$) have a similar positive derivative at $t_\mathrm{VUV} = 0$, so they can be classified as primary fragments that are directly populated from the parent. Five channels (C$_{12}H_{8}^+$, C$_{15}H_{9}^+$, C$_{13}H_{8}^+$, C$_{14}H_{8}^+$, and C$_{13}H_7^+$) have similar kinetics with a quasi-null derivative at $t_\mathrm{VUV} = 0$, so they belong to the group of secondary fragments. Finally, two channels (C$_{12}H_{7}^+$ and C$_{12}H_6^+$) have the most delayed kinetics. The retrieved map, shown in Fig.~\ref{fig:H6-pyrene_road}, minimizes the number of necessary paths to obtain a satisfactory fit of all the channels.

\paragraph{\textbf{Cor$^+$.}}
The Cor$^+$ case was peculiar because we could not fit the rising edge of the $M_{299}$ channel (which saturates quickly) just by using a simple sequential depopulation relation such as $M_{300}\rightarrow M_{299}\rightarrow M_{298}\rightarrow M_{297}\rightarrow M_{296}$. Indeed, we had to add an intermediate population of cations at the same mass (C$_{24}$H$_{12}^{+*}$ in Fig.~\ref{fig:maps_bare_PAHs}~(b)), which we attribute to one or more isomers of coronene. This isomer population is justified by the studies of \citet{trinquier2017_H_shifted_isomers,trinquier2017_ring_alteration} and \citet{castellanos2018_lab}, which show that stable isomers can be formed in particular by H migration. We found several ways to include a contribution of isomers in the first fragmentation step. The two hypotheses displayed in Fig.~\ref{fig:maps_bare_PAHs}~(b) represent the two extreme cases but scenarios in between provide as good fits and comparable retrieved $k_\mathrm{frag}$ values. We found that all scenarios yield almost indistinguishable rates for all fragmentation steps from C$_{24}$H$_{11}^+$ and beyond, which are therefore robustly determined regardless. A common constrain in these scenarios is to start with a nonzero fraction (4 to 10\%) of the isomer population at $t_\mathrm{VUV}=0$\,s. This implies that at least a fraction of the isomers was produced following the laser desorption-ionization process. One of the schemes presented in Fig.~\ref{fig:maps_bare_PAHs}~(b) leads to a repopulation of the isomers upon irradiation by 10.5\,eV photons, whereas the other case does not require it. The former scenario seems more consistent, although further experiments would be needed to conclude on this point.

\paragraph{\textbf{MeCor$^+$ and EtCor$^+$.}}
In the case of MeCor$^+$, we retrieved a distinct H-loss route (as in the case of MePyr$^+$), two channels induced by a C$_2$H$_X$ loss (C$_{23}$H$_{12}^+$ and C$_{23}$H$_{11}^+$), which have comparable secondary kinetics, and two channels (C$_{23}$H$_{10}^+$ and C$_{23}$H$_{9}^+$) that are very delayed. The resulting fragmentation map (see Fig.~\ref{fig:Me_Et_PAH_roads}~(c)) is sufficient to provide a good fit of all the channel kinetic curves. We could not find evidence for an isomerization mechanism, as in the Cor$^+$ case. The weak $M_{288}$ channel (C$_2$H loss from [MeCor-H]$^+$) is present while its analog $M_{190}$ in the case of MePyr$^+$ or EtPyr$^+$ was not observed. In the case of MePyr$^+$ or EtPyr$^+$, this channel may be an intermediate fragment that directly dissociates with the remaining excess of energy, while it may cool down and lead to C$_{23}$H$_{12}^+$ for MeCor$^+$.

Similarly to EtPyr$^+$ with respect to MePyr$^+$, we built the fragmentation map of EtCor$^+$ (see Fig.~\ref{fig:Me_Et_PAH_roads}~(d)) by mimicking the MeCor$^+$ map and simply adding a sequential H-loss route that contains $M_{327}$, $M_{326}$, $M_{325}$, and $M_{324}$. We stress that the normalized intensity of the EtCor$^+$ is not starting at 1 in Fig.~\ref{app:fig_kinetics_all_PAHs}~(f) because it was not possible to isolate with a good enough S/N the parent cation from the isotopic peaks and the fragments ([EtCor-H]$^+$, [EtCor-2H]$^+$, and C$_{25}$H$_{13}^+$) produced by the desorption-ionization laser. Therefore, we irradiated the initial ion mixture and summed the isotopic channels in the analysis. The quality of the data is lower than for other species (bigger fluctuations) but, Figure~\ref{app:fig_kinetics_all_PAHs}~(f) shows that the fit is fair (R$^2 = 0.89$) and therefore the values of the retrieved fragmentation rates can be considered. Because of this experimental limitation, we could not separate the C$_{23}$H$_{12}^+$ ($M_{288}$) contribution from the C$_{23}$H$_{11}^+$ ($M_{287}$) channel. Still, the coherence of the retrieved fragmentation rates and branching ratio between the EtCor$^+$ map and the MeCor$^+$ map shows that this possibly missing path has a negligible impact.

\section{Example of $k_\mathrm{eff}$ and $K_\mathrm{tot}$ computations}\label{app:Ktot_explanation}

In order to illustrate a computation of the $k_\mathrm{eff}$ and $K_\mathrm{tot}$ parameters, we show in Fig.~ \ref{app:fig_explanation_keff} a simple fragmentation map for an hypothetical parent cation $A$ and its successive fragments $B$, $C$, $D$, $E$, and $F$. In this simple example, the following equations give the different $k_\mathrm{eff}$ and $K_\mathrm{tot}$:

\begin{equation}\label{app:eq1}
\large{
\left\{
    \begin{array}{c}
    k_\mathrm{eff}^{A\rightarrow B\rightarrow D}=  \displaystyle \frac{R^{A\rightarrow B} k_1 k_2}{R^{A\rightarrow B}k_1 +  k_2}   \\
    k_\mathrm{eff}^{A\rightarrow B\rightarrow D\rightarrow E}= \displaystyle \frac{ k_\mathrm{eff}^{A\rightarrow B\rightarrow D} k_4}{ k_\mathrm{eff}^{A\rightarrow B\rightarrow D} +  k_4}   \\
    k_\mathrm{eff}^{A\rightarrow C\rightarrow E}= \displaystyle \frac{R^{A\rightarrow C} k_1 k_3}{R^{A\rightarrow C}k_1 +  k_3}   \\
    k_\mathrm{eff}^{A \cdots \rightarrow F}=  \displaystyle \frac{\left(k_\mathrm{eff}^{A\rightarrow B\rightarrow D\rightarrow E} + k_\mathrm{eff}^{A\rightarrow C\rightarrow E} \right) \times k_5 }{k_\mathrm{eff}^{A\rightarrow B\rightarrow D\rightarrow E} + k_\mathrm{eff}^{A\rightarrow C\rightarrow E} + k_5}   \\    
    \end{array}
\right.}
\end{equation}

\begin{equation}\label{app:eq2}
\Rightarrow
\left\{
    \begin{array}{c}
    K_\mathrm{tot}^{H-\mathrm{loss}} = R^{A\rightarrow B} k_1 +  k_\mathrm{eff}^{A\rightarrow B\rightarrow D} + k_\mathrm{eff}^{A\rightarrow C\rightarrow E} + k_\mathrm{eff}^{A \cdots \rightarrow F} \\
     K_\mathrm{tot}^{C_{2} H_{x}-\mathrm{loss}} = R^{A\rightarrow C} k_1 + k_\mathrm{eff}^{A\rightarrow B\rightarrow D\rightarrow E} \\
    \end{array}
\right.~.
\end{equation}

\begin{adjustbox}{center,caption={Example of a fragmentation map for an initial species ($A$) that dissociates into several fragments $B$, $C$, etc. The fragmentation rates ($k_i$) and the branching ratios ($R^{X \rightarrow Y}$) permit us to compute the effective fragmentation rates ($k_\mathrm{eff}^{A \cdots \rightarrow M}$) and the total loss rates ($K_\mathrm{tot}^{X-\mathrm{loss}}$) corresponding to the loss of a given X neutral fragment (atom or molecular group).}, label={app:fig_explanation_keff},nofloat=figure,vspace=\bigskipamount}
 \includegraphics[width=0.35\textwidth]{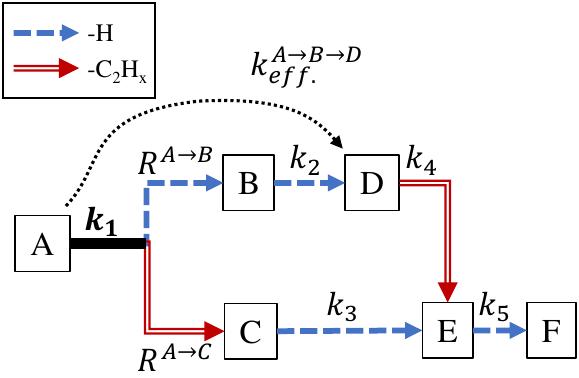}
\end{adjustbox}

\section{Absorption of the relevant species}\label{app:sigma_PAHs}

\subsection{Computation of the photoabsorption cross sections}

\begin{figure*}[htbp]
    {\centering
    \includegraphics[width=0.4\textwidth,valign=c]{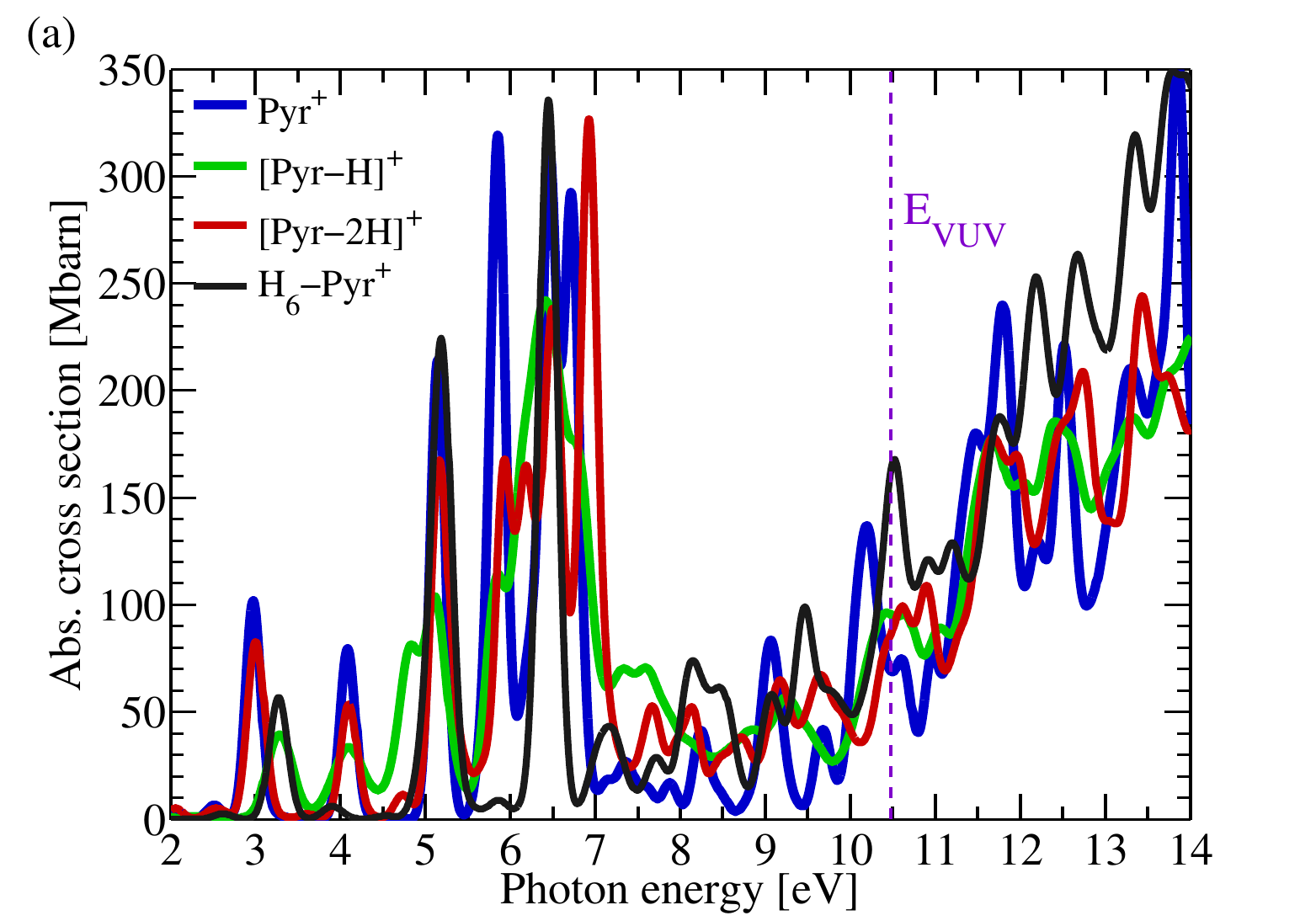} \includegraphics[width=0.4\textwidth,valign=c]{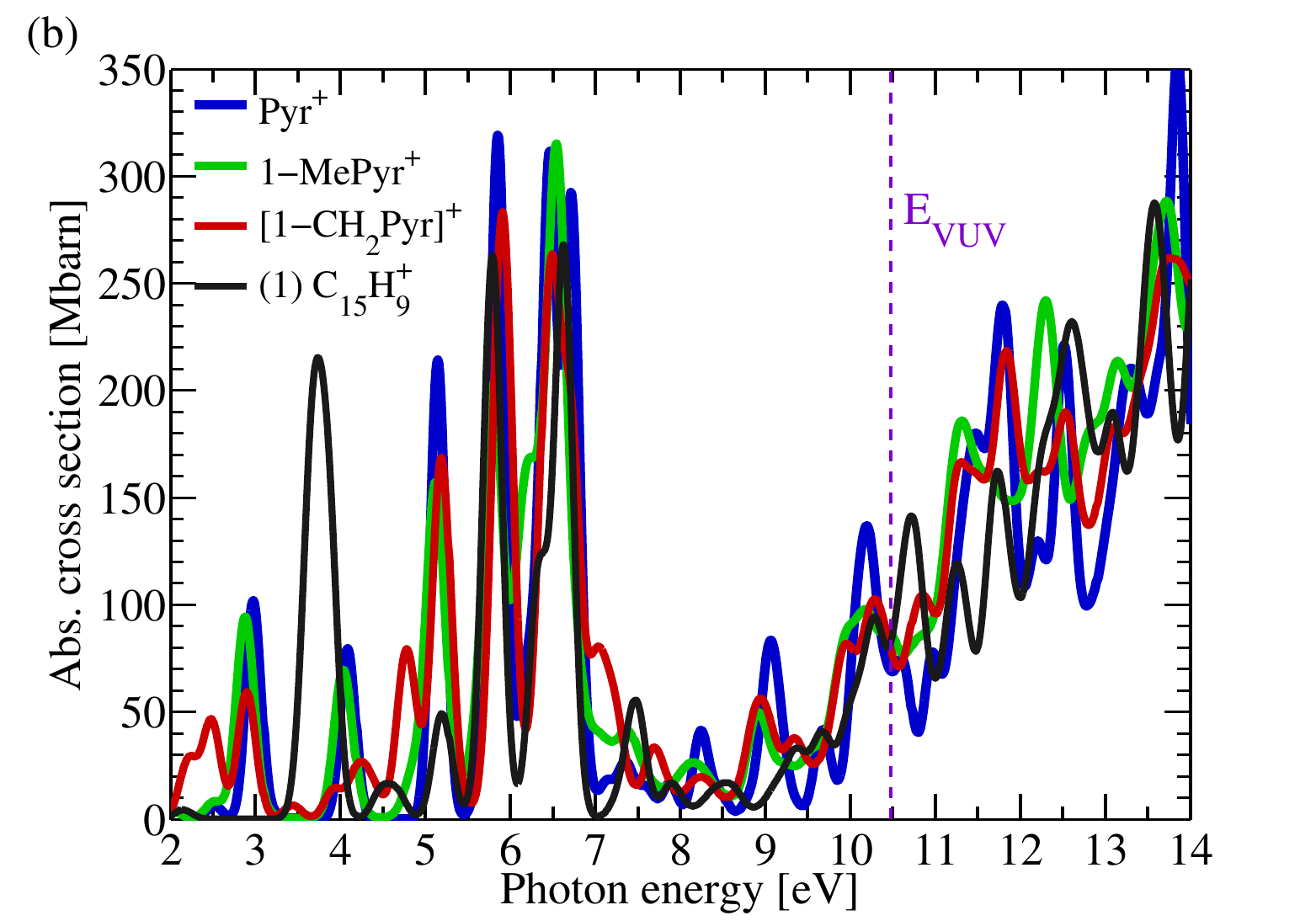} \\
    \includegraphics[width=0.4\textwidth,valign=c]{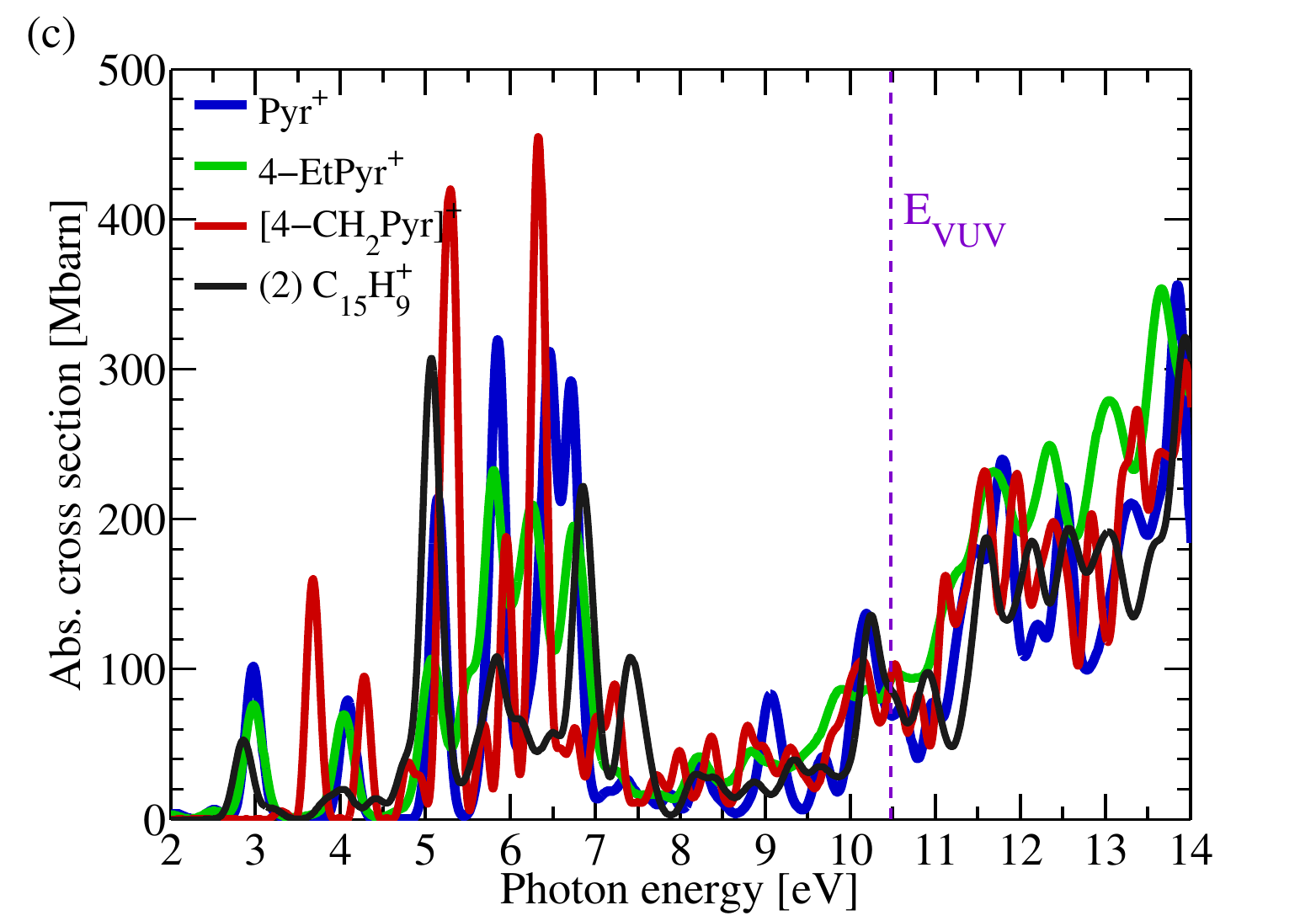} \\
    \includegraphics[width=0.4\textwidth,valign=c]{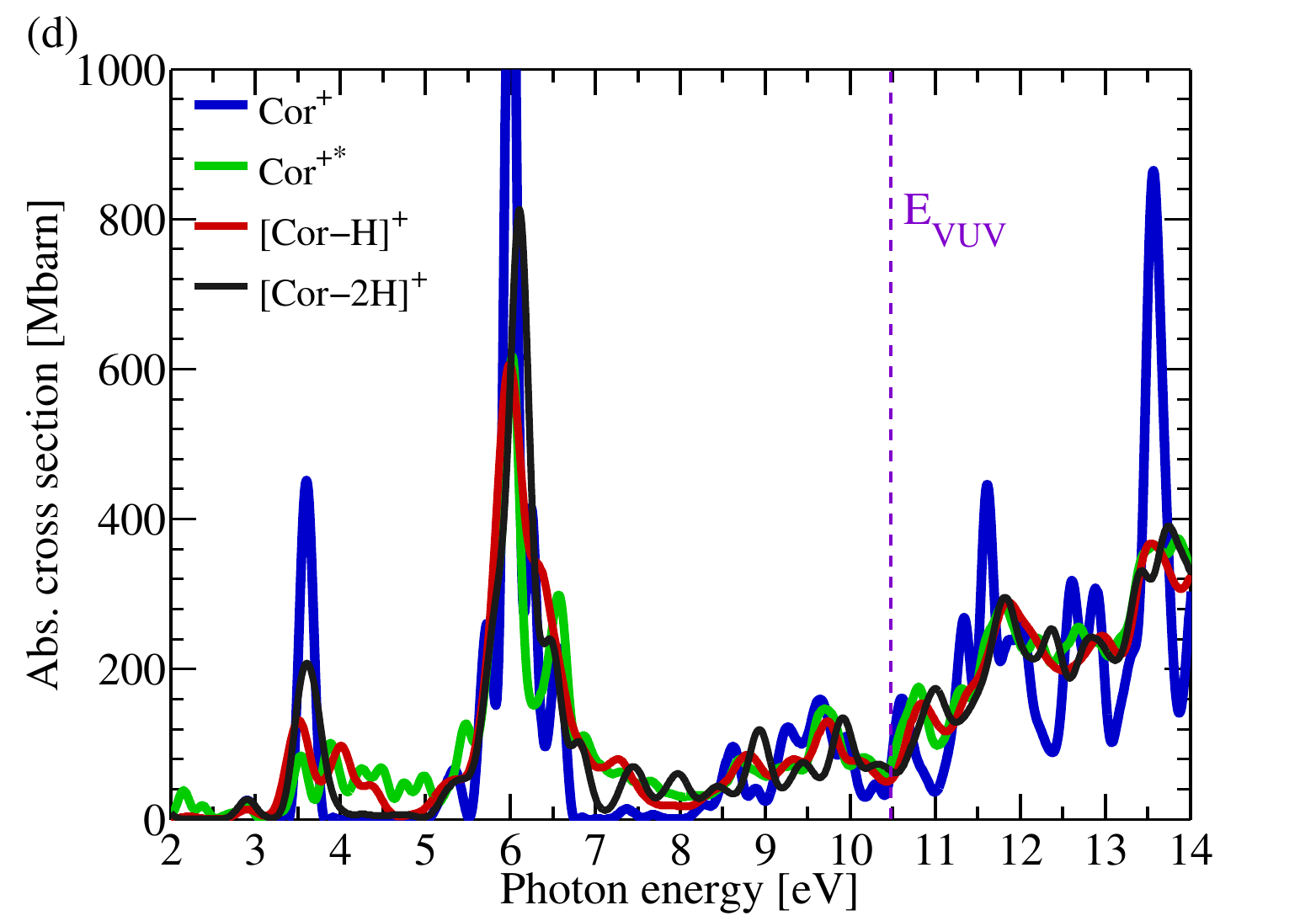} 
    \includegraphics[width=0.4\textwidth,valign=c]{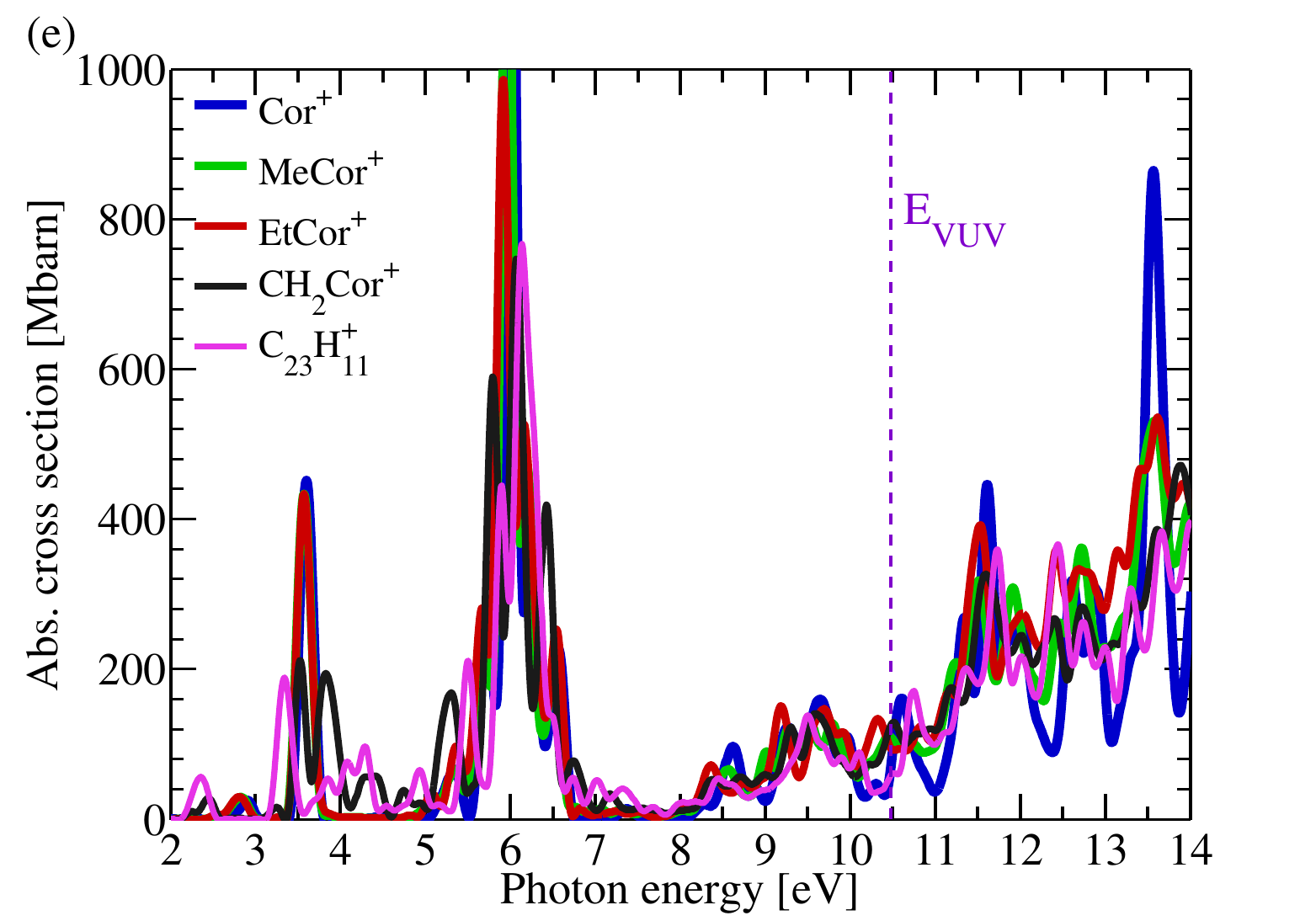} \\}
    \caption{Photoabsorption cross sections calculated using TD-DFT for the studied parent cations and some of their relevant fragments: (a) Pyr$^+$ (H$_6$-Pyr$^+$ also included), (b) 1-MePyr$^+$, (c) 4-EtPyr$^+$, (d) Cor$^+$,
    and, (e) MeCor$^+$ and EtCor$^+$.}
    \label{app:fig_sigma_abs}
\end{figure*}

The photoabsorption cross sections calculated using TD-DFT (cf. Sect. \ref{sec:meth_(TD-)DFT calculations}) are displayed in Fig.~\ref{app:fig_sigma_abs}.
For some species, we considered several isomers and only show here the averaged cross sections. This includes the case of [Pyr-H]$^+$ for which H loss can occur from three carbon atoms that are not equivalent: C$_1$ (1-dPyr$^+$), C$_2$ (2-dPyr$^+$), and C$_4$ (4-dPyr$^+$). We weighted their contribution to obtain the [Pyr-H]$^+$ averaged cross section such as $\sigma_\mathrm{abs}^\mathrm{[Pyr-H]^+}=(4\sigma_\mathrm{abs}^{1-\mathrm{dPyr}^+}  + 2\sigma_\mathrm{abs}^{2-\mathrm{dPyr}^+} + 4\sigma_\mathrm{abs}^{4-\mathrm{dPyr}^+})/10$. For [Pyr-2H]$^+$, we considered two isomers for which the 2H losses occur from adjacent carbons (C$_{1,2}$ or C$_{4,5}$). The cross sections of these two isomers have been equally weighted as suggested by the study of \cite{panchagnula2020}, which compares the IR spectrum of the doubly dehydrogenated pyrene with DFT computed harmonic spectra of the fourteen possible isomers and concludes that a 1:1 mixture of 1,2-ddPyr$^+$ and 4,5-ddPyr$^+$ reproduces fairly well the measurement. For C$_{15}$H$_9^+$, we considered two isomers with a five-membered ring either on the long (1) or the short (2) axis of the initial pyrene structure. They are respectively attributed to the VUV photoproducts of 1-MePyr$^+$ and 4-EtPyr$^+$. For Cor$^{+*}$, we considered the three isomers depicted in Fig.~\ref{fig:maps_bare_PAHs}~(b), which correspond to structures where the H atom has migrated toward the closest carbons. For [Cor-2H]$^+$ we considered only the isomer for which the 2H comes from the same ring consistently with previous studies \citep[e.g.,][]{montillaud2013, west2018, castellanos2018_lab}. We considered only one isomer in the case of [Cor-H]$^+$, MeCor$^+$, EtCor$^+$, C$_{25}$H$_{13}^+$, or C$_{23}$H$_{11}^+$, due to the symmetry of these species.

\subsection{Computation of the absorption rate}

In order to estimate the corrective factor $\gamma$ that is used in Sect.~\ref{sec:cooling} and Fig.~\ref{fig:rescaled_cross_sections}, we determined the average absorption rate for each species. It reads:
  \begin{equation}
      \overline{k_\mathrm{abs}^{M_i}} = \overline{\sigma _\mathrm{abs}^{M_i}} \times \phi_0~~,
 \end{equation}
 where $\overline{\sigma _\mathrm{abs}^{M_i}}$ is the spectrally averaged theoretical photoabsorption cross section over the energy range from 10\,eV to 11\,eV giving the value around the VUV photon energy at $\pm~0.5$\,eV. The values of $\overline{\sigma_\mathrm{abs}^{M_i}}$ and $\overline{k_\mathrm{abs}^{M_i}}$ are reported in Table~\ref{app:table_averaged_sigma_kabs}.

\begin{table*}[h!]
\caption{\label{app:table_averaged_sigma_kabs} Averaged photoabsorption cross sections and absorption rates (over the [10~-~11]\,eV range) of relevant species in this work. The error (standard deviation) on $\overline{\sigma}$ quantifies the variability of the cross sections over the considered energy range. No error bar is provided for $\overline{k_\mathrm{abs}}$ since most of the uncertainty would come from the experimental error on the VUV photon flux, which is globally included in the $\gamma$ correction factor as explained in Sect.~\ref{sec:cooling}.}
    \centering \small{
        {\renewcommand{\arraystretch}{1.2}
\begin{tabular}{*{6}{M{1.7cm}}} \hline\hline
\textbf{Species}      & \textbf{$\overline{\sigma}$ [Mbarn]} & \textbf{$\overline{k_\mathrm{abs}}$  [10$^{-3}$s$^{-1}$]} & \textbf{Species}   & \textbf{$\overline{\sigma}$   [Mbarn]} & \textbf{$\overline{k_\mathrm{abs}}$  [10$^{-3}$s$^{-1}$]} \\ \hline
Pyr$^+$               & $83 \pm 28$                            & 1.17                                               & Cor$^+$            & $72 \pm 41$                            & 1.02                                               \\

[Pyr-H]$^+$           & $82 \pm 15$                            & 1.16                                               & Cor$^{+*}$        & $103 \pm  38$                            & 1.46                                              \\

[Pyr-2H]$^+$          & $76 \pm 26$                            & 1.08                                               & [Cor-H]$^+$        & $92 \pm 37$                            & 1.31                                               \\
                      &                                        &                                                    & [Cor-2H]$^+$       & $93 \pm 36$                            & 1.32                                              \\
1-MePyr$^+$           & $88 \pm 6$                             & 1.25                                               &                    &                                        &                                                    \\
1-CH$_2$Pyr$^+$       & $90 \pm 11$                            & 1.27                                               & MeCor$^+$          & $88 \pm 16$                            & 1.25                                               \\
(1) C$_{15}$H$_{9}^+$ & $94 \pm 24$                            & 1.33                                               & C$_{25}$H$_{13}^+$ & $97 \pm 24$                            & 1.37                                              \\
                      &                                        &                                                    & C$_{23}$H$_{11}^+$ & $84 \pm 40$                            & 1.19                                                    \\
4-EtPyr$^+$           & $91 \pm 9$                             & 1.29                                               &                    &                                        &                                               \\
4-CH$_2$Pyr$^+$       & $81 \pm 17$                            & 1.14                                               & EtCor$^+$          & $103 \pm 17$                           & 1.46                                              \\
(2) C$_{15}$H$_{9}^+$ & $89 \pm 23$                            & 1.27                                               & C$_{25}$H$_{13}^+$ & $97 \pm 24$                            & 1.37                                                \\
                      &                                        &                                                    & C$_{23}$H$_{11}^+$ & $84 \pm 40$                            & 1.19                                                   \\
H$_6$-Pyr$^+$         & $109 \pm 36$                           & 1.54                                               &                    &                                        &                                            \\ \hline       
\end{tabular}           }}
\end{table*}

\end{appendix}
\end{document}